\documentclass[aps,prd,preprintnumbers,nofootinbib,twocolumn,superscriptaddress]{revtex4-2}

\usepackage[utf8]{inputenc}
\usepackage[T1]{fontenc}
\usepackage[english]{babel}
\usepackage[left=2cm,right=2cm,top=2cm,bottom=2cm]{geometry}
\usepackage{amssymb}
\usepackage{amsmath}
\usepackage{amsfonts}
\usepackage{braket}
\usepackage{mathrsfs}
\usepackage{physics}
\usepackage{slashed}
\usepackage{xcolor}
\usepackage{hyperref}
\usepackage{graphicx}
\usepackage{rotating}
\usepackage{bbold}
\usepackage[toc,page]{appendix}
\usepackage{ulem}
\usepackage{csquotes}

\newcommand{\Eqref}[1]{Eq.~\eqref{#1}}

\newcommand{\psiLB}[2][]{\bar{\psi}_{\mathrm{L}#1}^{#2}}
\newcommand{\psiL}[2][]{\psi_{\mathrm{L}#1}^{#2}}
\newcommand{\psiRB}[2][]{\bar{\psi}_{\mathrm{R}#1}^{#2}}

\newcommand{\tRB}[1][]{\bar{t}_{\mathrm{R}#1}}
\newcommand{\tR}[1][]{t_{\mathrm{R}#1}}
\newcommand{\bRB}[1][]{\bar{b}_{\mathrm{R}#1}}
\newcommand{\bR}[1][]{b_{\mathrm{R}#1}}

\newcommand{\tL}[1][]{t_{\mathrm{L}#1}}

\newcommand{\bL}[1][]{b_{\mathrm{L}#1}}

\newcommand{\hbot}{h_\mathrm{b}}
\newcommand{\hbott}{\tilde{h}_\mathrm{b}}

\newcommand{\htop}{h_\mathrm{t}}
\newcommand{\htopt}{\tilde{h}_\mathrm{t}}
\newcommand{\topmassr}{\mu_\mathrm{t}^2} 
\newcommand{\botmassr}{\mu_\mathrm{b}^2} 

\newcommand{\Nf}{N_\mathrm{f}}
\newcommand{\Nc}{N_\mathrm{c}}

\newcommand{\e}{\epsilon}

\newcommand{\vp}{\varphi}

\newcommand{\etaF}{\eta_\mathrm{F}}
\newcommand{\lB}[4]{l^#1_#2\left(#3;#4\right)}
\newcommand{\lF}[4]{l^{\left(\mathrm{F}\right)#1}_#2\left(#3;#4\right)}
\newcommand{\lFB}[7]{l^{\left(\mathrm{FB}\right)#1}_{#2,#3}\left(#4,#5;#6,#7\right)}
\newcommand{\lFBB}[9]{l^{\left(\mathrm{FBB}\right)d}_{#1,#2,#3}\left(#4,#5,#6;#7,#8,#9\right)}

\newcommand{\m}[6]{m^#1_{#2,#3}\left(#4,#5;#6\right)}
\newcommand{\mFF}[3]{m^{\left(\mathrm{F}\right)#1}_4\left(#2;#3\right)}
\newcommand{\mTF}[3]{m^{\left(\mathrm{F}\right)#1}_2\left(#2;#3\right)}
\newcommand{\mFB}[7]{m^{\left(\mathrm{FB}\right)#1}_{#2,#3}\left(#4,#5;#6,#7\right)}
\newcommand{\mFBtilde}[7]{\tilde{m}^{\left(\mathrm{FB}\right)#1}_{#2,#3}\left(#4,#5;#6,#7\right)}

\definecolor{darkpastelgreen}{rgb}{0.01, 0.75, 0.24}
\definecolor{shiraz}{rgb}{0.6,0,0.4}
\definecolor{nileblue}{RGB}{53,87,155}

\begin{document}

\title{Interplay of Chiral Transitions in the Standard Model}

%
%
%
\author{Holger Gies}
\email{holger.gies@uni-jena.de}
\affiliation{\mbox{\it Theoretisch-Physikalisches Institut, Friedrich-Schiller-Universit{\"a}t Jena,}
	\mbox{\it D-07743 Jena, Germany}}
\affiliation{\mbox{\it Abbe Center of Photonics, Friedrich-Schiller-Universit{\"a}t Jena,}
	\mbox{\it D-07743 Jena, Germany}}
\affiliation{Helmholtz-Institut Jena, Fr\"obelstieg 3, D-07743 Jena, Germany}
\affiliation{GSI Helmholtzzentrum für Schwerionenforschung, Planckstr. 1, 
D-64291 Darmstadt, Germany}
\author{Richard Schmieden}
\email{richard.schmieden@uni-jena.de}
\affiliation{\mbox{\it Theoretisch-Physikalisches Institut, Friedrich-Schiller-Universit{\"a}t Jena,}
	\mbox{\it D-07743 Jena, Germany}}
\author{Luca Zambelli}
\email{luca.zambelli@bo.infn.it}
\affiliation{\mbox{\it INFN-Sezione di Bologna, via Irnerio 46, I-40126 Bologna, Italy}}
\begin{abstract}
We investigate nonperturbative aspects
of the interplay of chiral transitions in the standard model in the 
course of the renormalization flow. We focus on the chiral symmetry breaking 
mechanisms provided by the QCD and the electroweak sectors, the latter of which 
we model by a Higgs-top-bottom Yukawa theory. The interplay becomes 
quantitatively accessible by accounting for the fluctuation-induced mixing of 
the electroweak Higgs field with the mesonic composite fields of QCD. In fact, 
our approach uses dynamical bosonization and treats these scalar fields on the 
same 
footing. Varying the QCD scale relative to the Fermi scale we
quantify the mutual impact of the symmetry-breaking mechanisms, specifically 
the departure from the second order quantum phase transition of the pure Yukawa 
sector in favor of a crossover upon the inclusion of the gauge interactions. This allows to 
discuss the ``naturalness'' of the standard model in terms of a 
pseudo-critical exponent which we determine as a function of the 
ratio of the QCD and the Fermi scale. We also estimate the minimum value of the $W$ 
boson mass in absence of the Higgs mechanism. 

%
%
%
%
\end{abstract}
\maketitle
\pagenumbering{arabic}
\section{Introduction} \label{s1}

In the standard model of particle physics, chiral symmetry breaking generates 
the masses of fermionic matter in the visible universe 
\cite{Nambu:1961tp,Weinberg:1967tq,Wilczek:2012sb}. Two 
sectors of the standard model contribute 
apparently independently to fermion mass generation: at the Fermi scale, 
electroweak symmetry breaking triggered by the parameters of the Higgs potential 
induce current quark and charged lepton masses \cite{Weinberg:1967tq} through 
the 
Yukawa interactions to the Higgs field. Subsequently at the QCD scale, the 
intricate dynamics of the strong interactions generate the baryon masses that 
ultimately dominate the visible masses in the 
universe 
\cite{BMW:2008jgk,ParticleDataGroup:2022pth}. 
With the Fermi scale corresponding to the vacuum expectation value of the Higgs 
field $v\simeq 246$~GeV and the QCD scale $\Lambda_\mathrm{QCD}\simeq 
\mathcal{O}(100)$~MeV, the two relevant scales are 2-3 orders of magnitude apart. 

The symmetry-breaking mechanisms are rather different in 
the two sectors: mass generation in QCD is driven by the gluonic interactions 
growing strong towards lower energies, thus being an intrinsically 
nonperturbative phenomenon. Mass generation in the electroweak sector -- 
despite also being a gauge theory -- is described by the Yukawa interactions 
with the Higgs field and appears accessible to perturbation theory. Also, all 
involved quantities including the two different scales given above can be 
traced back to different fundamental parameters of the model. Similar comments 
apply to typical embeddings of the standard model in overarching models, where 
the two scales are parametrized by the breaking mechanisms towards the 
standard-model symmetries. 

The qualitative difference between the two mechanisms of chiral symmetry 
breaking becomes prominent when studying the symmetry transitions as a function 
of control parameters within reduced model sectors. 
For instance, reducing the 
electroweak sector to a pure Yukawa model involving the Higgs field and the 
most strongly coupled top and bottom quarks, the reduced model exhibits a 
chiral quantum phase transition of second order as a function of the mass 
parameter of the Higgs potential \cite{Gerhold:2007yb,Gerhold:2007gx,Gies:14}. 
By 
contrast, the pure QCD sector is essentially governed by the strong coupling 
constant, and the long-range physics is always in the symmetry broken phase.
The combination of the two sectors therefore suggests that the standard model 
is also always in the symmetry broken phase, but exhibits a rapid crossover as 
a function of a suitable control parameter. 

This rapid crossover is also a manifestation of the so-called 
\textit{naturalness problem} 
\cite{tHooft:1979rat,Giudice:2008bi,Grinbaum:2009sk,Dine:2015xga,%
Hossenfelder:2018ikr}: considering the standard 
model as a function of microscopic (bare) parameters at some high-energy scale 
$\Lambda$ (an ultraviolet (UV) cutoff), one of its parameters serving as a 
control parameter has to be tuned rather finely to put the model rather close 
to the rapid crossover. This fine-tuning is necessary to separate the high 
scale $\Lambda$ from the Fermi scale and the subsequent QCD scale, $\Lambda 
\ggg v > \Lambda_{\mathrm{QCD}}$. It should be emphasized that the naturalness 
problem is not a consistency problem of the standard model, but rather a 
peculiar feature that may or may not find an explanation within an embedding in 
a more fundamental theory. 

The present work is devoted to a study of the interplay of the two chiral 
symmetry breaking sectors of the standard model. This interplay is visible in 
the details of how the two transitions merge into the expected rapid crossover. 
Moreover, it becomes apparent by the fact that the relevant degrees of freedom 
partly share the same quantum numbers: in the electroweak sector, we encounter 
the presumambly fundamental Higgs field, whereas a composite mesonic scalar 
field serves as a useful effective degree of freedom for the description of the 
chiral QCD transition. We show that these scalar fields can
be dealt with on the same footing yielding only a single effective field for 
the description of both chiral transitions. 

As an advantage, we can study the region near the rapid crossover as a function 
of the microscopic parameters, allowing to quantify the strength of the 
crossover in terms of a pseudo-critical exponent. We argue that this 
exponent is a measure for the amount of fine-tuning needed on the microscopic 
level, thereby quantifying the naturalness problem. In our approach, we can 
thus also compute how much the naturalness problem in the electroweak sector is 
``alleviated'' by the presence of the QCD sector. 

Finally, our approach can be 
used to estimate the properties of a non-fine-tuned version of the standard 
model. For the case that chiral-symmetry breaking of QCD would dominate fermion 
mass generation and condensate formation, QCD would also generate masses for 
the electroweak gauge bosons \cite{Quigg:2009xr,Quigg:2009vq,Iso:2017uuu} the 
value of which we estimate 
in the present work. 

A special class of such models
	are those that would solve the hierarchy problem by requiring an enhanced 
	symmetry, scale invariance, which is then dynamically broken by
	means of QCD-like dimensional transmutation.
	Typically, the mass spectrum generated by this Coleman-Weinberg
	mechanism is sensitive to the details of the bare Lagrangian 
	at the UV scale $\Lambda$, that is to say to the 
	marginal couplings such as the quartic Higgs coupling.
	Extracting an unambiguous prediction for the mass spectrum
	in these scale invariant Coleman-Weinberg scenarios 
	then requires a UV completion being able to remove
	the UV cutoff and to reduce the freedom in the choice of these
	bare couplings. 
	Still, the details of the translation between the 
    marginal couplings and the IR mass spectrum and interactions
    remain nonperturbative.
    Thus, the study presented in this article
    can be conceived as a first step in this direction.

Rather than focusing straight ahead on the critical region of approximate
scale invariance, whose precise definition at sizable gauge coupling
is already a nontrivial problem, we address a more general
survey of the phase diagram, aiming to provide
an overall picture of how the two extremal scenarios,
the pure-Yukawa sharp second-order phase transition and
the pure QCD universal chiral symmetry breaking,
can be reconciled and connected with an intermediate
and less understood regime, featuring the interplay 
of several sources of symmetry breaking.

Our work uses methods of functional renormalization in order to deal with the 
nonperturbative aspects of the problem. Moreover, these methods give access to 
treat fields as effective degrees of freedom in a scale-dependent manner which 
is crucial to focus on the relevant fields and remove redundant field 
variables. 

In the following, we introduce the various relevant sectors of the standard 
model in Sect.~\ref{sec:model} in order to motivate the model subset which we 
include in our quantitative analysis. Section \ref{sec:RGfloweqs} briefly 
summarizes the central renormalization group flow equations including the 
scale-dependent field transformations that facilitate a study of the interplay 
of the chiral transitions. The nature of the transitions is quantitatively 
studied in Sect.~\ref{sec:results}, containing most of our results. We 
conclude in Sect.~\ref{sec:conc}. Technical details and additional 
quantitative analyses are contained in the appendices.

\section{Chiral symmetry breaking sectors of the standard 
model}

\label{sec:model}

We are mainly interested in the
interplay between 
the two sectors of the standard model which are responsible for the 
fermion mass formation through chiral symmetry breaking. For the electroweak 
sector, the driving mechanism is controlled by the RG-relevant mass parameter of 
the Higgs potential, whereas the QCD sector is controlled by the marginally 
relevant non-Abelian gauge coupling.
Also,
	we neglect the long-range effects 
	of the electroweak gauge 
	bosons associated
	to the
	U(1)${}_\mathrm{Y}\times$SU(2)${}_\mathrm{L}$
	symmetry.
	The latter are
	subleading with respect to
	gluons, even though the limit of vanishing weak and hypercharge gauge 
couplings is presumably not smooth; in the full system, the 
construction of observables therefore requires a more careful discussion 
\cite{Frohlich:1980gj,Maas:2017wzi,Maas:2017xzh,Sondenheimer:2019idq,
Maas:2020kda}.
By dropping the electroweak gauge sector, our study differs from 
those of other deformations of the standard model where the coupling strengths 
of the weak and strong sectors are shifted relative to each other 
\cite{Quigg:2009xr,Lohitsiri:2019wpq,Berger:2019yxb}.
	We also ignore most of the flavor asymmetry 
of the quarks and leptons,
accounting specifically for the third generation
of quarks which are most strongly coupled to the Higgs field,
 and for the
split between the top
and bottom quarks.

Finally, we deal with the strong-coupling IR regime with the help of a 
somewhat minimalistic description that neverlessless gives us a qualitative and 
semi-quantitative access to the chiral transition. For this, we start from 
microscopic QCD and study the renormalization flow towards
the quark-meson model~\cite{Schwinger:1957em,Gell-Mann:1960mvl}
along the lines familiar from functional RG studies 
\cite{Jungnickel:1995fp,Schaefer:1999em,Braun:2003ii,Resch:2017vjs}. 
In 
fact, the quark-meson model can quantitatively be matched to chiral 
perturbation theory \cite{Eser:2018jqo,Divotgey:2019xea} as the low-energy 
limit of QCD.
As argued below, the QCD-induced mesonic 
scalar field
can
in fact account also for the standard-model Higgs
field by means of an appropriate identification
of degrees of freedom and couplings.

\subsection{From QCD to the quark-meson model}
The QCD sector of the standard model consists of a 
non-Abelian SU$(\Nc)$ gauge theory minimally coupled to $\Nf$ massless Dirac 
fermions (quarks) defined by the (Euclidean) action
\begin{equation}
S=\int_x \bar{\psi}^a_i \mathrm{i}\slashed{D}_{ij}\psi^a_j 
+\frac{1}{4}F^{z}_{\mu\nu}F^{\mu\nu}_z.
\label{eq:SQCD}
\end{equation}
The fermions are described by $\psi^a_i$, where $a,b=1,...,\Nf$ denote the 
flavor indices and $i,j=1,...,\Nc$ the fundamental color indices.
The covariant derivative $D^\mu_{ij}$ in the fundamental representation of the 
group SU($N_\mathrm{c}$) couples quarks to the non-Abelian gauge bosons 
$(A^\mu)_{ij}$ and is given by 
\begin{equation}
	D^\mu_{ij}=\partial^\mu\delta_{ij}-ig(A^\mu)_{ij}.
\end{equation}
In general, we work with the physical values of $N_\mathrm{c}=3$ 
colors and $N_\mathrm{f}=6$ flavors. However, when focusing on the top-bottom 
family, we use $N_\mathrm{f}=2$ as is appropriate for a single generation.

For 
the approach to the low-energy phase of QCD, we map the microscopic QCD sector 
defined by \Eqref{eq:SQCD} onto the quark-meson model which we consider as a 
useful effective field theory for our purposes. We emphasize that we do not 
introduce any new independent parameter but extract the IR values of the 
quark-meson model parameters from the RG flow. The translation from the high- 
to the low-energy description proceeds most conveniently by studying the 
four-fermion interactions which are induced during the RG flow through quark 
and gluon fluctuations. For the sake of simplicity, we focus solely on the 
scalar-pseudoscalar channel in a point-like approximation. This is familiar 
from the Nambu-Jona-Lasinio model which can become critical at strong coupling 
and mediates chiral symmetry breaking \cite{Braun:2019aow}. 
At intermediate scales, we therefore work with the effective action
\begin{equation}
	\begin{aligned}
		\Gamma=\int_x  
&\phantom{+}\bigg\{\bar{\psi}^a_i\mathrm{i}\slashed{D}_{ij}\psi^a_j+\frac{1}{4}
F_{\mu\nu}^z F_z^{\mu\nu}
\\
&\phantom{+}+\frac{1}{2}\bar{\lambda}_\sigma\left[(\bar{
\psi}_i^a\psi_i^b)^2-\left(\bar{\psi}_i^a\gamma_5\psi_i^b\right)^2\right]\bigg\}
, 
	\end{aligned}
	\label{eq:QCDAction}
\end{equation}
amended by a suitable gauge-fixing and ghost sector. Again, the four-fermion 
interaction $\bar{\lambda}_\sigma$ is not used as a model parameter, but 
computed from the RG flow of the microscopic action. 

Starting from the asymptotically free high energy domain, the gauge 
coupling grows towards the low 
energy regime \cite{Gross:73,Politzer:73}, subsequently also driving the 
four-fermion 
coupling $\bar{\lambda}_\sigma$ to large values. This is indicative for the 
approach to chiral symmetry breaking and condensate formation. The 
corresponding composite scalar degrees of freedom appearing in the effective 
low-energy theory can be traced by means of a Hubbard-Stratonovich 
transformation. Introducing the auxiliary complex scalar field,
\begin{equation}
\varphi^{ab}=-\mathrm{i}\frac{\bar{h}}{m^2}\bar{\psi}^{b}_iP_\mathrm{L}\psi^{a}
_i,\qquad\varphi^{\ast 
ab}=-\mathrm{i}\frac{\bar{h}}{m^2}\bar{\psi}^{a}_iP_\mathrm{R}\psi^{b}_i, 
	\label{eq:hubbardstratonoviceom}
\end{equation}
we can translate the four-fermion interaction into a Yukawa interaction of the 
fermions with the new scalar degrees of freedom. 
The resulting effective action includes the quark-meson effective theory,
\begin{equation}
	\begin{aligned}
		\Gamma=\int_x &\partial_\mu\varphi^{ab}\partial^\mu\varphi^{\ast 
ab}+U(\vp,\vp^\dagger)+ 
\bar{\psi}^a_i\mathrm{i}\slashed{D}_{ij}\psi^a_j+\frac{1}{4}F_{\mu\nu}^z 
F_z^{\mu\nu}\\
&
+\mathrm{i}\bar{h}\bar{\psi}_i^a\left[P_\mathrm{R}\varphi^{ab 
}+P_\mathrm{L}\varphi^{\ast ba}\right]\psi_i^b . 
	\end{aligned}
	\label{eq:QMAction}
\end{equation}
Again, the Yukawa coupling $\bar{h}$ as well as the mesonic effective 
potential $U$ do not contain free parameters, but are determined by the RG flow 
of QCD and thus governed by the gauge coupling. 
Clearly, this action still contains 
massless gluons, and will do so even after an IR massive decoupling of the 
quarks and mesons. Since we are interested in the chiral transition, the 
presence of free massless gluons in the deep IR of the model is not relevant. 
Of course, confinement and the Yang-Mills mass gap removes gluons from the 
physical IR spectrum.

This model is invariant under 
SU$(N_\mathrm{f})_\mathrm{L}\times$SU$(N_\mathrm{f})_\mathrm{R}$ 
transformations, where these transformations act independently on the left- and 
right-handed spinor
\begin{equation}
\begin{aligned}	
\psi_\mathrm{R}\quad &\rightarrow &&\mathrm{U}_\mathrm{R}\psi_\mathrm{R},\\
\psi_\mathrm{L}\quad &\rightarrow &&\mathrm{U}_\mathrm{L}\psi_\mathrm{L},
\end{aligned}
\end{equation}
as well as on their conjugate transposed as
\begin{equation}
\begin{aligned}
\bar{\psi}_\mathrm{R}\quad &\rightarrow &&\bar{\psi}_\mathrm{R}\mathrm{U}_\mathrm{R}^\dagger,\\
\bar{\psi}_\mathrm{L}\quad &\rightarrow &&\bar{\psi}_\mathrm{L}\mathrm{U}_\mathrm{L}^\dagger.
\end{aligned}
\end{equation}
The scalar field transforms as
\begin{equation}
\begin{aligned}
\varphi \quad&\rightarrow &&\mathrm{U}_\mathrm{R}\vp\mathrm{U}_\mathrm{L}^\dagger,\\
\varphi^\dagger \quad&\rightarrow &&\mathrm{U}_\mathrm{L}\vp^\dagger\mathrm{U}_\mathrm{R}^\dagger.
\end{aligned}
\end{equation}
Additionally the model is invariant under a U$(1)_\mathrm{B}$ symmetry, 
corresponding to baryon number conservation. It acts only on the spinor, since 
the scalar fields are made up of quark-antiquark pairs (c.f.~\Eqref{eq:hubbardstratonoviceom}) and thus carry no baryon number 
\begin{equation}
	\label{eq:baryonsym}
\begin{aligned}
\psi \quad&\rightarrow && \exp(\frac{\mathrm{i}\vartheta_\mathrm{B}}{3})\psi\\
\bar{\psi}\quad&\rightarrow&& \exp(-\frac{\mathrm{i}\vartheta_\mathrm{B}}{3})\bar{\psi}.
\end{aligned}
\end{equation}
The U$(1)_\mathrm{A}$ anomaly could be included in the present framework 
\cite{Jungnickel:1995fp,Pawlowski:1996ch,Gies:2006nz,Pisarski:2019upw,
Fejos:2020pcg, Braun:2020mhk}, but is not relevant for our purposes.
Chiral symmetry breaking in this model induces a non-vanishing vacuum 
expectation value for the scalar field.
The phenomenologically relevant breaking pattern
corresponds to the vacuum configuration
%
%
%
\begin{equation}
\varphi_0=
\begin{pmatrix}
\sigma_0& 0& \dots  &0 \\
0&\sigma_0 &\dots &0 \\
\vdots&\vdots & \ddots & \vdots \\
0&0 &\dots & \sigma_0
\end{pmatrix} ,
\label{eq:QMVacuum}
\end{equation}
which breaks the 
SU$(N_\mathrm{f})_\mathrm{L}\times$SU$(N_\mathrm{f})_\mathrm{R}$ symmetry of the 
model down to the diagonal SU$(N_\mathrm{f})_\mathrm{V}$ subgroup.\\
In this case we consider a quartic approximation for the scalar potential, in 
the symmetric (SYM) and in the spontaneous-symmetry-breaking (SSB) regimes, 
respectively given by
\begin{equation}
	\begin{aligned}
		U(\vp,\vp^\dagger)&=m^2\varrho+\frac{1}{2}\lambda_1 \varrho^2+\frac{1}{4}\tau,\qquad \text{SYM},\\
		U(\vp,\vp^\dagger)&=\frac{1}{2}\lambda_1\left(\varrho-\varrho_0\right)^2+\frac{1}{4}\tau,\qquad \text{SSB},
	\end{aligned}
	\label{eq:QMScalarPotential}
\end{equation}
where we have introduced the invariants
\begin{equation}
	\begin{aligned}
		\varrho&=\tr(\vp^\dagger\vp),\\
		\tau&=
		2
		\tr(\vp^\dagger\vp)^2-
		\varrho^2.
	\end{aligned}
	\label{eq:rhotaudef}
\end{equation}
Here $\varrho_0$ denotes the scale-dependent minimum of the scalar potential 
in the SSB regime. The scalar  mass spectrum obtained by spontaneous chiral 
symmetry breaking with the vacuum configuration given by \Eqref{eq:QMVacuum} is 
given in appendix \ref{app:ScalarSpectrum},
and
has been discussed in detail 
for general $\Nf$ in ref.~\cite{Jungnickel:95,Schreyer:2020}

\subsection{Indentifying the overlap of the Higgs and meson degrees of freedom}

The auxiliary meson field $\vp$
introduced in the low-energy effective 
description of strong interactions,
and the fundamental Higgs doublet $\phi$
of the standard model play similar roles, as their vacuum expectation value parametrizes
mass generation and
chiral symmetry breaking.
This similarity poses fundamental
questions, which are ultimately connected
with the challenge of bridging between 
effective field theories 
valid at very different scales.
See for instance refs.~\cite{Tornqvist:2005yp,Schumacher:2011gi} for  discussions
of the possible interplay of these two fields.

The quark meson model 
is generically conceived as an effective description of the strong interactions
below energies of the order of a few GeV,
with a characteristic scale of
breaking of the approximate chiral symmetry 
given by the vacuum expectation value $v$ of the 
$\sigma$ field, \Eqref{eq:QMVacuum}, of
the order of $f_\pi\simeq 93$~MeV.
An effective UV cutoff scale -- if it exists --
of the Higgs model is
instead unknown. In the present work, we parametrize it by a high-energy scale 
$\Lambda$ which we take as sufficiently large. Towards lower scales, the 
parameters of the Higgs potential generate the electroweak scale, characterized 
by the
breaking of the global SU$(2)_\mathrm{L}\times$SU$(2)_\mathrm{R}$
custodial symmetry 
and of the exact chiral symmetry
at the characteristic 
Fermi scale of $\Lambda_\mathrm{F}=246$~GeV.
 The Fermi scale at the same time 
corresponds to the vacuum expectation value $v$ of the Higgs field.  
From this point of view there seems to be no reason
to aim at the simultaneous description of the two phenomena by means of a single linear sigma model.

Following this perspective,
a first investigation of the 
electroweak
phase transition
in a chiral Higgs-top-bottom model
has been studied in Ref.~\cite{Schmieden:Master} 
and is summarized in appendix \ref{app:ComparisonToMaster}. 
This study adopts RG methods, 
along the lines for instance of 
Ref.~\cite{Gies:14}, and supplements
the quark-meson model with an
independent elementary Higgs field.

As a matter of fact however,
the effective descriptions of symmetry breaking
by the two sigma models describing the strong
and the 
Higgs sectors are indistinguishable, as the relevant corresponding 
components of the scalar fields carry the same conserved quantum numbers (also in 
presence of the full electroweak gauge sectors).
Therefore, we focus here on the 
simplest toy model, where we remove
the redundant double-counting of
the parity-even scalar symmetry-breaking channel.
In the present Higgs-QCD model it is
	particularly clear that the overlap between the QCD meson field and the
Higgs sectors extends even beyond the single degrees of freedom
describing the sigma and Higgs particles,
and can include the whole Higgs
doublet.
Of course with doublet here we mean 
a fundamental representation of the global 
SU$(2)_\mathrm{L}$ group only,
which is a subgroup
of the QCD chiral symmetry,
as the electroweak gauge sector
of the standard model is not taken into consideration
in the present analysis.

In order to make this matching of the scalar degrees of freedom explicit, 
we focus in the following on the 
3rd generation of quarks, the top and bottom family for which 
we reduce our equations to $N_\mathrm{f}=2$.
The identification of 
the Higgs doublet 
with some of the components 
of the collective QCD-meson field becomes transparent from an
analysis of the Yukawa interactions.
Using a reparametrization of the meson field as
\begin{equation}
	\begin{aligned}
\Phi^{ab}&=\frac{1}{\sqrt{2}}\left(\varphi^{ab}+\epsilon^{ac}\epsilon^{bd}{
\varphi^\ast}^{cd}\right)\\ 
\tilde{\Phi}^{ab}&=\frac{1}{\sqrt{2}}\left({\varphi^\ast}^{ab}-\epsilon^{ac}
\epsilon^{bd}{\varphi}^{cd}\right),
	\end{aligned}
\end{equation}
we find that each scalar field $\Phi$ and $\tilde{\Phi}$ 
contains 4 real degrees of freedom for the present case of $\Nf=2$ , and the 
various components can be related under complex conjugation through 
\begin{equation}
	\begin{aligned}
		{\Phi^\ast}^{ab}&=\epsilon^{ac}\epsilon^{bd}\Phi^{cd}\\
		\tilde{\Phi}^{\ast ab}&=-\epsilon^{ac}\epsilon^{bd}\tilde{\Phi}^{cd}.
	\end{aligned}
	\label{eq:fieldTransformation}
\end{equation}
These fields transform under the
$(\boldsymbol{2},\bar{\boldsymbol{2}})$
 representation of the SU$(2)_\mathrm{L}\times$SU$(2)_\mathrm{R}$ symmetry group.
\begin{equation}
	\begin{aligned}
		\Phi &\rightarrow  U_\mathrm{L}\Phi U^\dagger_\mathrm{R}\\
		\tilde{\Phi} &\rightarrow U^\dagger_\mathrm{R}\tilde{\Phi}U_\mathrm{L}.
	\end{aligned}
	\label{eq:FullModelSymmetryTransforms}
\end{equation}
The conventional electroweak Higgs field $\phi^a$ as a complex doublet can be 
identified by
\begin{equation}
	\begin{aligned}
		\phi^a\equiv\Phi^{a2}&\text{, implying}\quad 
\phi_\mathcal{C}^a=\mathrm{i}\left(\sigma_2\right)^{ab}{\phi^\ast}^b.
  \end{aligned}
\end{equation}
As required, $\phi^a$ transforms under SU$(2)_\mathrm{L}$. In addition, we can 
define $\tilde{\phi}^a$ as a complex doublet under 
SU$(2)_\mathrm{R}$ by
\begin{equation}
	\begin{aligned}
		\tilde{\phi}^a=\tilde{\Phi}^{2a}&\text{, implying}\quad
\tilde{\phi}_\mathcal{C}^a=-\mathrm{i}\left(\sigma_2\right)^{ab}\tilde{\phi}^{ 
\ast b},
	\end{aligned}
	\label{eq:fieldDefinition}
\end{equation}
respectively. This allows us to rewrite the Yukawa interaction in the 
quark-meson model as
\begin{equation}
	\begin{aligned}
		\mathscr{L}^\mathrm{QM}_\mathrm{Yuk}=
		\frac{\mathrm{i}\bar{h}}{\sqrt{2}}& \left(\psiLB[,i]{a} {\phi^a} \bR[,i]  
		+ \psiLB[,i]{a} {\phi_\mathcal{C}^{a}} \tR[,i] + \text{h.c.}\right.\\
		&\left. + \psiRB[,i]{a}\tilde{\phi}^{a}\bL[,i]+\psiRB[,i]{a}{\tilde{\phi}_\mathcal{C}}^{a}\tL[,i]+\text{h.c.} 
		\right).
	\end{aligned}
	\label{eq:QMYukawaInteraction}
\end{equation}
Here we have used the following standard notation for
the flavor components of the fermion field
\begin{equation}
	\psi_{\mathrm{L/R,i}}=\begin{pmatrix}
		t_{\mathrm{L/R,i}}\\
		b_{\mathrm{L/R,i}}
	\end{pmatrix}.
\end{equation}
Comparing this Yukawa interaction with the one of the Higgs-top-bottom 
model for the special case $\Nf=2$, we observe the scalar field $\phi$ to 
have the same quantum numbers as the Higgs doublet, while the $\tilde{\phi}$ 
field contains the remaining four real degrees of freedom present in the 
mesonic field of QCD (for $\Nf=2$). The latter has no analogue in the 
electroweak sector, and here describes a doublet
in the fundamental representation of SU$(2)_\mathrm{R}$.

As quarks in the standard model have nonvanishing	electroweak charges, four-Fermi couplings
exist in several channels	with a complicated charge assignment. The 
channels which are related to the Higgs field upon a Hubbard-Stratonovich 
transformation are well known from top quark condensation models 
\cite{Miransky:1988xi,Miransky:1989ds,Bardeen:1989ds,Hill:2002ap}.
 Independently of the 
details, the corresponding channels always have some overlap with the QCD 
induced channels. After a Hubbard-Stratonovich transformation, we match those 
parts of the mesonic field that share the same quantum numbers with the Higgs 
field.

Although meson Yukawa couplings emerge as an effective description of the quark 
four-fermion couplings induced by QCD, the standard model features also fundamental Yukawa 
couplings to the Higgs.	As a consequence, we introduce new coupling constants 
for the top and bottom Yukawa interaction and arrive at the final form  
\begin{equation}
	\begin{aligned}
		\mathscr{L}_\mathrm{Yuk}=&
		\phantom{+}\frac{\mathrm{i}\tilde{h}_b}{\sqrt{2}}\left( \psiLB[,i]{a} \phi^a \bR[,i]
		 + \bRB[,i] \phi^{\ast a} \psiL[,i]{a}\right)\\ &+\frac{\mathrm{i}\tilde{h}_t}{\sqrt{2}}
		 \left( \psiLB[,i]{a} \phi_\mathcal{C}^{a} \tR[,i] + \tRB[,i]\phi_\mathcal{C}^{\ast a}\psiL[,i]{a}\right)\\
		 &+\frac{\mathrm{i}\bar{h}}{\sqrt{2}}
		 \left(\psiRB[,i]{a}\tilde{\phi}^{a}\bL[,i]+\psiRB[,i]{a}{\tilde{\phi}_\mathcal{C}}^{a}\tL[,i]+ \text{h.c.}
		\right),
	\end{aligned}
\label{eq:Yukawaaction}
\end{equation}
where
\begin{equation}
\tilde{h}_{\mathrm{t/b}} = 
\bar{h}_{\mathrm{t/b}} + \bar{h}.
\label{eq:linearh}
\end{equation}
Despite 
the simple new parametrization of \Eqref{eq:linearh}, it is important to 
stress
that the flow of the couplings $\bar{h}_{\mathrm{t/b}}$ is primarily driven by 
the electroweak sector, whereas $\bar{h}$ is driven by QCD fluctuations.
Notice also that the interaction of the SU$(2)_\text{R}$-doublet $\tilde{\phi}$ with 
the quarks remains fully parametrized by the QCD-induced coupling $\bar{h}$.

The model inherits the U$(1)_\mathrm{B}$ symmetry
of \Eqref{eq:baryonsym} from the quark meson model,
corresponding to baryon number conservation in the theory. 
(NB: we 
ignore the U$(1)_\mathrm{B}$ violating sphaleron processes of the standard model 
in our analysis.)

While we have focused here on the top/bottom family, the matching of 
the scalar fields also applies to the lighter quark families as the 
corresponding sectors in the mesonic field carries the same quantum numbers. In 
this manner, the current quark masses are generated at the Fermi scale 
from the corresponding fundamental Yukawa couplings. The main difference, 
however, is that the lighter quarks do not decouple near the Fermi scale and 
thus the corresponding mesonic fields are predominantly driven more and more by 
QCD. In addition, the full mesonic field also contains components that link the 
different families to one another. These inter-family 
components do not correspond 
to any part of the electroweak Higgs field.

\subsection{From the meson potential to the Higgs phase}
We now want to construct a scalar potential out of the invariants of the fields $\phi$ and $\tilde{\phi}$,
\begin{equation}
	\begin{aligned}
		\rho&=\phi^\dagger\phi,\\
		\tilde{\rho}&=\tilde{\phi}^\dagger\tilde{\phi}.
	\end{aligned}
	\label{eq:fieldInvariants}
\end{equation}
The SU$(2)_\text{L}\times$SU$(2)_\text{R}$ invariant $\varrho$ defined in 
\Eqref{eq:rhotaudef} then yields
\begin{equation}
    \varrho=\rho+\tilde{\rho}.
	\label{eq:threerhos}
\end{equation}
A particular quartic potential constructed out of these invariants
and inspired by \Eqref{eq:QMScalarPotential} is given by
\begin{equation}
	U(\rho ,\tilde{\rho})=m^2\left(\rho +\tilde{\rho}\right)+\frac{\lambda_{1}}{2}\left(\rho +\tilde{\rho}\right)^2+\lambda_{2}\rho\tilde{\rho}.
	\label{eq:myScalarPotential}
\end{equation}
This potential, however, cannot be mapped one-to-one to the scalar potential 
given in \eqref{eq:QMScalarPotential}, since the $\tau$ invariant cannot be 
expressed by $\rho$ and $\tilde{\rho}$ alone. 
This mismatch between \Eqref{eq:myScalarPotential}
and \Eqref{eq:QMScalarPotential} has a subdominant effect on the scalar 
spectrum, which slightly differs in the two cases
 as is discussed in appendix \ref{app:ScalarSpectrum}.

In the symmetric case where $m^2>0$, the minimum of the potential has a 
vanishing field, $\rho=0$ and $\tilde{\rho}=0$. In the  
symmetry-breaking regime, we parametrize the potential by
\begin{equation}
	U(\rho, 
\tilde{\rho})=\frac{\lambda_{1}}{2}\left(\rho+\tilde{\rho}
-\kappa\right)^2+\lambda_{2}\rho\tilde{\rho}. 
	\label{eq:potentialbroken}
\end{equation}
For being able to describe the standard-model-like Higgs phase, we specify our vacuum to be 
$\rho=\kappa$ and $\tilde{\rho}=0$.
The vacuum expectation value in the $\phi$ field spontaneously breaks the 
SU$(2)_\mathrm{L}\times$SU$(2)_\mathrm{R}$ symmetry of the QCD part 
of the model down to diagonal SU$(2)$ and induces Dirac masses for the 
fermions through the Yukawa interaction. Choosing our vacuum to be 
\begin{equation}
	\begin{aligned}
		\left.\phi\right|_\mathrm{vac}=
			\begin{pmatrix}
				\phantom{i}0\phantom{i}\\
				\sqrt{\kappa}
			\end{pmatrix},
		\qquad\qquad
		\left.\tilde{\phi}\right|_\mathrm{vac}=
			\begin{pmatrix}
				\phantom{i}0\phantom{i}\\
				0
			\end{pmatrix},
	\end{aligned}
\end{equation}
the induced fermion masses read
\begin{equation}
	\begin{aligned}
		m^2_\mathrm{top}&=\frac{\tilde{h}_\mathrm{t}^2}{2}\kappa,\qquad 
    m^2_\mathrm{bot}&=\frac{\tilde{h}_\mathrm{b}^2}{2}\kappa. 
	\end{aligned}
\end{equation}
The scalar spectrum contains $\Nf^2 -1$ massless degrees of freedom (Goldstone 
bosons), $\Nf^2$ degrees of freedom with mass $\sqrt{\lambda_{2}\kappa}$
and one radial mode corresponding to the Higgs particle with mass 
$\sqrt{2\lambda_{1}\kappa}$. For conventional appliations 
in the quark-meson model with $\Nf=2$ light (first generation) quark flavors, 
the three Goldstone modes are identified with the pions. In the present 
setting, where we consider the third generation, the $\phi$ 
field corresponds to the Higgs instead. Upon embedding into the full standard 
model, the gauged SU$(2)_\mathrm{L}$ provides 
mass to the corresponding gauge bosons through 
the Brout-Englert-Higgs mechanism and the Goldstone bosons disappear.
In the present simplified model, these massless Goldstone bosons would 
naively remain in the spectrum. In order to avoid artifacts caused by them, we 
decouple them from the flow by giving them a mass of the order of the 
longitudinal gauge bosons, once the Higgs field develops a vacuum expecation 
value near the Fermi scale.
%
%
\section{RG flow equations}
\label{sec:RGfloweqs}
%
Effective field theories are
a perfect tool for the description of
phase transitions, as is well established since
the early work of Ginzburg and Landau.
However, to address issues
such as decoupling of degrees of freedom,
hierarchy of scales, or even the origin
of symmetry breaking, they have to be
complemented with additional information.
RG methods are often the preferred 
source of this information.
To be able to follow the RG flow of
our effective toy model
from its weak-coupling standard-model-like
UV behavior down to the chiral-symmetry-breaking
 regime, we
need nonperturbative methods.
The precise running of the strong gauge
coupling and the details of the QCD IR dynamics are not essential for our
main case study of the interplay of the chiral transitions.
We therefore model the gauge flow with a
simple ansatz for the running gauge coupling
described in the following.
On the other hand, the influence that 
this strongly coupled gauge sector has
on the matter fields and especially on
the Higgs-meson sector is of crucial
relevance. For the latter, we resort
to functional RG methods whose 
description and implementation is the main body of this section.

\subsection{Gauge sector}
In the gauge sector, we use a perturbative two-loop equation for the running of 
the gauge coupling. 
We improve this flow such that we obtain an IR fixed point $\alpha_\ast=g^2_\ast/(4\pi)$ for that coupling,
whose presence e.g.~in Landau gauge has been put forward by a number of studies, 
see for instance 
refs.~\cite{vonSmekal:1997ohs,Gies:02a,Fischer:2003rp,Pawlowski:2003hq,%
Fischer:2008uz,Weber:2011nw,Reinosa:2017qtf}. In the FRG scheme that we are 
using, such a fixed point is rather generic as it is compatible with the 
decoupling of IR modes due to confinement and mass gap generation. 
This fixed point has to be chosen above the critical gauge 
coupling needed to induce chiral symmetry breaking 
\cite{Gies:04,Miransky:1984ef,Gies:2005as,Braun:2006jd,Braun:2010qs}.  
The precise choice of $\alpha_*$ is then irrelevant for
	our analysis, since massive hadronic freeze-out occurs 
	before the fixed-point regime.
Although capturing chiral symmetry breaking, this
\textit{ad hoc} modification of the beta function
together with the restriction to pointlike fermionic self-interactions 
is a somewhat crude approximation which can be substantially improved within a 
systematic propagator and vertex expansion 
\cite{Mitter:2014wpa,Braun:2014ata,Eichmann:2016yit,Binosi:2016wcx,
Cyrol:2017ewj}. 

We describe the
scheme-dependent intermediate flow between the perturbative
weakly coupled regime and the nonperturbative fixed point
by means of a smooth exponential function \cite{Gies:04}.
Other parametrizations, for instance
using a Heaviside step function as the opposite extreme,
would not significantly change our results.
The beta function for the gauge coupling then reads
\begin{equation}
	\begin{aligned}	
		\partial_tg^2=\etaF g^2
=-2&\Bigg(c_1\frac{g^4}{16\pi^2}+c_2\frac{g^6}{\left(16\pi^2\right)^2}
\Bigg)\\
&\times\left(1-\exp(\frac{1}{\alpha_\ast}-\frac{1}{\frac{g^2}{4\pi
}})\right) ,	
	\end{aligned}
\end{equation}
where we have introduced the constants
\begin{equation}
	\begin{aligned}
		c_1&=\left(\frac{11}{3}\Nc-\frac{2}{3}\Nf\right),\\
		c_2&=\left(\frac{34}{3}\Nc^2-\frac{10}{3}\Nc\Nf-2C_2(\Nc)\Nf\right).
	\end{aligned}
\end{equation}
At high energies, we use $\Nf=6$ active flavors within 
the beta function but dynamically account for their decoupling across their 
mass thresholds for the heavy quarks.\\
After crossing into the broken regime at $k=k_\mathrm{trans}$, we gradually switch off the top and bottom quark in the running of the strong gauge coupling by decreasing the effective number of active flavors to 5 at the top mass threshold, and then to 4 at the bottom threshold. The thresholds of both quarks are determined by their respective Yukawa coupling at the transition scale
\begin{equation}
m^2_t \approx \frac{1}{2}\tilde{h}^2_t\big\vert_{k=k_\mathrm{trans}} k^2_\mathrm{trans}, \qquad m^2_b \approx \frac{1}{2}\tilde{h}^2_b\big\vert_{k=k_\mathrm{trans}} k^2_\mathrm{trans}.
\end{equation}
Expressed through RG time, we decouple the quarks from the running of the strong gauge coupling at
\begin{equation}
t_\mathrm{top}=\ln(\frac{\tilde{h}_t}{\sqrt{2}})\qquad\mathrm{and}\qquad t_\mathrm{bottom}=\ln(\frac{\tilde{h}_b}{\sqrt{2}})
\end{equation}
after we entered the broken regime.

\subsection{Matter sector}
We study the matter sector of our model using the functional renormalization 
group (FRG) as a method that allows us to address strong-coupling regimes as 
well as the perturbative accessible weak-coupling limit. The starting point is 
a (bare) microscopic action $S$ defined at an UV cutoff $\Lambda$ and the RG 
flow of a scale-dependent effective action $\Gamma_k$ given by the 
Wetterich 
equation~\cite{Wetterich:93,Ellwanger:1993mw,Morris:1993qb,Bonini:1992vh}
\begin{equation}
	\partial_t\Gamma_k=\frac{1}{2}\mathrm{STr}\left(\frac{\partial_t R_k}{\Gamma_k^{(2)}+R_k}\right).
	\label{eq:Wetterich}
\end{equation}
This scale-dependent effective action interpolates between the bare action 
$\Gamma_{k=\Lambda}=S$, and the full quantum effective action 
$\Gamma_{k=0}=\Gamma$, where all quantum corrections have been integrated out. 
In \Eqref{eq:Wetterich}
$R_k$ denotes the regulator, which
is to some extent an arbitrary function
of momentum.
Its additive insertion in the denominator 
(i.e.~in the exact propagator) 
serves as an IR cutoff for momenta smaller than $k$.
On the other hand,
 the derivative $\partial_t R_k$
 in the numerator is chosen to provide UV regularisation
 of the trace integral. 
Despite its simple closed one-loop form,
\Eqref{eq:Wetterich} can be proved to
be exactly equivalent to the standard 
functional-integral definition of $\Gamma$
in presence of a mass-like deformation
$R_k$ in the bare action.
 This exactness, combined with
 systematic
 approximation schemes like
 the derivative expansion or the vertex expansion,
turns \Eqref{eq:Wetterich}
into a useful nonperturbative tool,
which has been extensively applied
to a wide selection of problems
in quantum, statistical, or many body 
physics.
 For more details we direct the interested
 reader to the rich secondary literature 
\cite{Berges:02,Gies:06,Pawlowski:2005xe,Braun:2011pp,Kopietz:2010zz,Dupuis:2020fhh}.

For the application of the FRG equation, we project
the exact equation \eqref{eq:Wetterich}
onto the ansatz derived in the previous section,
specifically \Eqref{eq:QMAction} with the
field redefinition that
leads to the standard-model-like Yukawa interactions
of \Eqref{eq:Yukawaaction}.
This can be understood as 
a 
leading order of the derivative expansion for the quantum effective action. 
Additionally we allow for a scale-dependent kinetic term of the fields by 
introducing a wave-function-renormalization factor for each field.
The gauge coupling, the scalar potential $U$, the Yukawa couplings, and the 
wave function renormalizations for the various fields $Z_i$ are all scale 
dependent; we suppress the index $k$ often used to denote scale-dependent 
quantities for notational clarity. 
Using this truncation in the Wetterich equation and projecting onto the 
respective operators yields the $\beta$ functions for their couplings. 
 
 We account for the running of the wave-function renormalizations through the 
so-called anomalous dimensions of the respective fields,
\begin{equation}
	\eta_i = -\partial_t \log Z_i\,.
	\label{eq:AnomalousDimension}
\end{equation}
Furthermore it is useful to introduce dimensionless, renormalized quantities as
\begin{equation}
	\begin{aligned}
		&\rho=Z_\phi 
k^{2-d}\phi^\dagger\phi,\quad\tilde{\rho}=Z_{\tilde{\phi}}k^{2-d}\tilde{\phi}
^\dagger\tilde{\phi},\\
&\htop=Z_\phi^{-1}{Z_\mathrm{L}^{\mathrm{t}}}^{-1}{
Z_\mathrm{R}^{\mathrm{t}}}^{-1}\htopt,\quad
\hbot=Z_\phi^{-1}{Z_\mathrm{L}^{
\mathrm{b}}}^{-1}{Z_\mathrm{R}^{\mathrm{b}}}^{-1}\hbott,\\
&h^2=Z_{\tilde{\phi}}^{-1}{Z_\mathrm{L}^{\mathrm{t}}}^{-1/2}{Z_\mathrm{L}^{
\mathrm{b}}}^{-1/2}{ 
Z_\mathrm{R}^{\mathrm{t}}}^{-1/2}{Z_\mathrm{R}^{\mathrm{t}}}^{-1/2}\bar{h}^{2} .
	\end{aligned}
	\label{eq:DimlessQuantities}
\end{equation}
For concrete computations, we work in $d=4$ later on.
We also employ the Landau gauge, where the gauge parameter is set to zero. This 
gauge choice is know to be a fixed point of the RG flow of the gauge-fixing 
parameter	\cite{Caswell:1974cj,Ellwanger:95,Litim:98}.
Using standard calculation techniques~\cite{Berges:02}, we extract the flow 
equations and obtain for the scalar potential 
\begin{equation}
	\begin{aligned}
		\partial_t u = &-d u + (d-2+\eta_{\phi})\rho 
u_{\rho}+(d-2+\eta_{\tilde{\phi}})\tilde{\rho}u_{\tilde{\rho}}\\
&+v_d\Big\{3\lB{
d}{0}{\mu^2_\mathrm{G}}{\eta_{\phi}}+\lB{d}{0}{\mu^2_\mathrm{H}}{\eta_{\phi}}
\\
&\phantom{v_d\Big\{{}}+3\lB{d}{0}{\mu^2_\mathrm{m}}{\eta_{\tilde{\phi}}}+\lB{d
}{0}{\mu^2_\mathrm{m,r}}{\eta_{\tilde{\phi}}}\Big\}\\
&-2 d_\gamma \Nc v_d 
\left\{\lF{d}{0}{\mu^2_\mathrm{t}}{\eta_\mathrm{t}}+\lF{d}{0}{\mu^2_\mathrm{b}}{
\eta_\mathrm{b}}\right\} ,
	\end{aligned}
	\label{eq:potentialFlow}
\end{equation}
where subscripts in $u$ denote derivatives with respect to the corresponding 
variable. The various threshold functions $l_0^d$, $l^{\mathrm{(F)}d}_0$ etc. 
responsible for the decoupling of massive modes can be found in the appendix 
\ref{app:threshold} for the special choice of the linear regulator 
\cite{Litim1}. In \Eqref{eq:potentialFlow}, $d_\gamma$ is the dimension of the 
spinor representations of the fermions, and 
$v_d^{-1}=2^{d+1}\pi^{d/2}\Gamma(d/2)$. We have also introduce the abbreviations
\begin{equation}
	\begin{aligned}
		\mu_\mathrm{t}^2&=\frac{\htop^2}{2}\rho+\frac{h^2}{2}\tilde{\rho},\\
		\mu_\mathrm{b}^2&=\frac{\hbot^2}{2}\rho+\frac{h^2}{2}\tilde{\rho},\\
		\mu^2_\mathrm{G}&=u_\rho,\\
		\mu^2_\mathrm{H}&=u_\rho+2\rho u_{\rho,\rho},\\
		\mu^2_\mathrm{m}&=u_{\tilde{\rho}},\\
		\mu^2_\mathrm{m,r}&=u_{\tilde{\rho}}+2\tilde{\rho}u_{\tilde{\rho},\tilde{\rho}}.
	\end{aligned}
	\label{eq:massAbbreviations}
\end{equation}
We simplify the task of following the RG evolution of the full effective 
potential by means of a quartic polynomial truncation, adopting the 
parametrizations of \Eqref{eq:myScalarPotential} and \Eqref{eq:potentialbroken} 
in the SYM and in the standard-model-like SSB regimes respectively. A more general 
functional analysis of the scalar potential would be able to fully describe all 
possible SSB regimes~\cite{Borchardt:2016kco,Beekman:2019sof}. The remaining 
flow equations of the model can be found in the App.~\ref{app:FlowEquations}. 
\subsection{Partial Bosonization}
For our scale-dependent analysis, the Hubbard-Stratonovich transformation 
\eqref{eq:hubbardstratonoviceom} which translates our four-fermion interaction 
into a Yukawa interaction between quarks and mesons needs to be treated 
dynamically: If we translated the full four-fermion interaction into 
the scalar sector at a fixed matching scale $k_\mathrm{m}$, gluonic 
fluctuations would generate a nonzero four-fermion coupling again at scales 
$k<k_\mathrm{m}$. To fully account for these radiatively generated interactions, 
we apply the bosonization prescription continuously, i.e. at every 
scale $k$. To this end we promote the scalar field to be scale dependent, 
reabsorbing the four-fermion interaction into the scalar sector for all $k$. 
This yields additional contributions to the flow equation of the effective 
average action \cite{Gies:2001nw,Gies:2002kd,Pawlowski:2005xe,%
Floerchinger:2009uf,Baldazzi:2021ydj,Ihssen:2023nqd}
\begin{equation}
D_t\Gamma_k[\varphi_k]=\partial_t\Gamma_k[\varphi_k]\Bigr\rvert_{\varphi_k}-\int_x\frac{\delta\Gamma_k[\varphi_k]}{\delta\varphi_k}\partial_t\varphi_k,
\label{eq:modifiedflow}
\end{equation}
The first term on the right-hand side accounts for the RG flow of the model of 
\Eqref{eq:QCDAction}, without dynamical re-bosonization. The second term takes 
the new scale dependence of the fields into account and ensures a vanishing 
fermionic self-interaction on all scales. In order to keep the concrete 
computations simple, we use the approximative scheme of 
\cite{Gies:2001nw,Gies:04}.
 See refs.~\cite{Gies:06,Braun:2011pp} for an 
introduction to this method.

We choose the following ansatz for the scale dependence of the scalar field
\begin{equation}
\begin{aligned}
\partial_t\varphi^{ab}_k(q)&=-\mathrm{i}\left(\bar{\psi}^bP_\mathrm{L}\psi^a\right)(q) \partial_t \alpha_k(q) +\varphi^{ab}\partial_t \beta_k(q),\\
\partial_t\varphi^{\ast ab}(q)&=-\mathrm{i}\left(\bar{\psi}^aP_\mathrm{R}\psi^b\right)(-q)\partial_t\alpha_k(q)+\varphi^{\ast ab}\partial_t\beta_k(q),
\end{aligned}
\label{eq:HSfieldflow}
\end{equation}
with $\alpha_k(q)$ and $\beta_k(q)$ being functions to be chosen such that our 
four fermion coupling stays zero for the full flow 
\cite{Gies:2001nw,Gies:04,Gies:06}. 
The scale dependence of the meson scalar fields, descending from 
\Eqref{eq:HSfieldflow}, entails a corresponding scale dependence of the Higgs 
(left-charged) and dual-Higgs (right-charged) scalar doublets, which can be 
straightforwardly computed from the linear field transformation 
\eqref{eq:fieldTransformation}. 

The scale dependence of the scalar field, together with the second term in 
\Eqref{eq:modifiedflow}, yields additional contributions to the flow equations, 
specifically to the flow of the Yukawa coupling $h$ and the scalar potential 
$u$. Recall that our parametrization of the Yukawa couplings shown in 
\Eqref{eq:Yukawaaction} assumes a simple linear superposition of the meson-like 
and fundamental Higgs-like couplings, 
see \Eqref{eq:linearh}.
Correspondingly, we capture the effects of the scale-dependent partial 
bosonization solely in the meson-like Yukawa coupling $h$. This is of course a 
simple modeling of a more intricate structure of higher-dimensional operators 
mixing the quarks, the Higgs and the mesons. 

In the symmetric regime, the additional contribution to the (dimensionless) mass 
parameter in the scalar potential is conveniently expressed in terms of the 
ratio
 \begin{equation}
 	\e=\frac{m^2}{k^2h^2},
 	\label{eq:defe}
 \end{equation}
and given by \cite{Gies:04} 
\begin{equation}
\partial_t \e\big\rvert_{\mathrm{d.b.}} = 
2\frac{\e}{h^2}(1+\e)(1+(1+\e)Q_\sigma)(\beta_{\lambda_\sigma}^{g^4}g^4
+\beta^{g^2h^2}_{\lambda_\sigma}g^2h^2), 
\end{equation}
where $\#\big\rvert_{\mathrm{d.b.}}$ denotes the additional contibutions 
stemming from dynamical bosonization (the second term in 
\eqref{eq:modifiedflow}), while the Yukawa contribution reads
\begin{equation}
\partial_t 
h^2\big\rvert_{\mathrm{d.b.}}=2(1+2\e+Q_\sigma(1+\e)^2)(\beta_{\lambda_\sigma}^{
g^4}g^4+\beta^{g^2h^2}_{\lambda_\sigma}g^2h^2).
\end{equation}
In the broken regime, we similarly find
\begin{align}
\partial_t \kappa\big\rvert_{\mathrm{d.b.}} =& 2\frac{\kappa}{h^2}(1-\kappa 
\lambda_1)(1+(1-\kappa\lambda_1)Q_\sigma)
\notag\\
&\times(\beta_{\lambda_\sigma}^{g^4}g^4+\beta^
{g^2h^2}_{\lambda_\sigma}g^2h^2)
\end{align}
as the additional contribution to the running of the (dimensionless) minimum of 
the potential, and
\begin{equation}
\partial_t h^2\big\rvert_{\mathrm{d.b.}} = 2(1-2\kappa\lambda_1 + Q_\sigma 
(1-\kappa\lambda_1)^2)(\beta_{\lambda_\sigma}^{g^4}g^4+\beta^{g^2h^2}_{
\lambda_\sigma}g^2h^2)
\end{equation}
for the Yukawa coupling in the SSB regime. Here we have used
\begin{equation}
\begin{aligned}
\beta_{\lambda_\sigma}^{g^4}&=-\frac{1}{4}\frac{9\Nc^2-24}{\Nc}v_d\lFB{d}{1}{2}{0}{0}{\eta_\psi}{\eta_\mathrm{F}},\\
\beta^{g^2h^2}_{\lambda_\sigma}&=\frac{48}{\Nc}v_d\lFBB{1}{1}{1}{0}{\e}{0}{\eta_\psi}{\eta_\phi}{\etaF},
\end{aligned}
\end{equation}
for brevity.

The quantity $Q_\sigma = \partial_t\left(\bar{\lambda}_\sigma 
(k^2)-\bar{\lambda}_\sigma (0)\right)/\partial_t\bar{\lambda}_\sigma (0)$ 
measures the suppression of 
the four-fermion coupling $\bar{\lambda}_\sigma$ for large external momenta 
and is, in principle, straightforwardly computable. A suppression implies 
$Q_\sigma < 0$. In the SSB regime, where fermions are massive, 
non-pointlike four-fermion interactions are suppressed by the inverse fermion 
mass squared coming from the two internal propagators \cite{Gies:03}. Therefore 
we choose the ansatz 
\begin{equation}
Q_\sigma \!=\! \frac{Q_\sigma^0 }{2}\!\left(
\mFB{d}{1}{2}{\mu^2_\mathrm{t}}{0}{\eta^\mathrm{t}}{\etaF}\!+\mFB{d}{1}{2}{\mu^2_\mathrm{b}}{0}{\eta^\mathrm{b}}{\etaF}\!\right)
\end{equation}
which has the right decoupling properties but introduces a free parameter 
$Q_\sigma^0$. Tests for various different values of $Q_\sigma^0 \in [-0.5,0]$ 
show that results are qualitatively independent of the precise choice of this 
parameter \cite{Schmieden:Master,Gies:04}. 
\subsection{Evaluation of the RG flows}

As a check of the reliability of our aproximations,
we compare 
the so-called local potential approximation (LPA),
where all anomalous dimensions of the fields
are neglected, to the generalization 
including the running wave function renormalizations of the 
fields (sometimes called LPA${}^\prime$).
This check confirms the stability of our main results,
that we present in the next section.

Furthermore and as mentioned above, our reduction of the standard model comes 
with some artifacts such as unphysical massless modes. As argued above, the 
massless gluons do not affect the chiral transitions and can therefore 
safely be ignored. By contrast, the massless would-be Goldstone bosons of the 
Higgs field in the SSB regime would induce an unphysical logarithmic flow. We 
cure this artifact by introducing masses in the SSB regime for these modes in 
order to model the Brout-Englert-Higgs (BEH) mechanism in the full model following 
the prescription of \cite{Gies:14}. 
As studied in more detail in Ref.~\cite{Schmieden:Master} and substantiated for 
the present work in App.~\ref{app:ScalarSpectrum}, this procedure 
has little to no influence on the IR values of interest in this study.

Finally, we comment on a subtlety related to the different parametrizations of 
the scalar potential elucidated in Sect.~\ref{sec:model}. The parametrization 
\Eqref{eq:myScalarPotential} of the potential differs from that of 
\Eqref{eq:QMScalarPotential}, as the invariant $\tau$ cannot be expressed in 
terms of $\rho$ and $\tilde{\rho}$. This leads to a slightly different scalar 
mass spectrum in the two cases. Because of the different spectrum, the RG flow 
equation for the coupling $\lambda_{2}$ in the potential differs from the one 
found in the quark meson model (see for instance 
refs.~\cite{Jungnickel:95,Berges:02}).
For simplicity, we adopt the beta function of $\lambda_2$ 
stemming from the quark-meson model and the
potential parametrization \eqref{eq:QMScalarPotential}, 
with the correct  thresholds corresponding to the scalar mass spectrum
arising from \Eqref{eq:myScalarPotential}.
The consistency of this approach is confirmed by the stability of the results
shown below.
In fact, implementing the RG equation coming directly from the form
of \Eqref{eq:myScalarPotential} would just change the IR value of the 
$\lambda_2$ coupling by a small amount, affecting only the masses of the 
SU$(2)_\text{R}$ doublet. By contrast, the effects
we are interested in do not depend on $\lambda_2$.

%
%
\section{Nature of the chiral transition} 
\label{sec:results}

Let us now investigate the nature of the chiral transitions within our 
parametrization of the essential standard model sectors. More specifically, we 
are interested in studying the transitions as a function of the microscopic 
(bare) parameters. From an RG perspective, the most RG-relevant parameter that 
gives access to both sides of the transition can serve as a control parameter 
to explore the near critical regime.

Let us start from the limiting case, where we consider only the 
Higgs-top-bottom Yukawa sector. This sector exhibits a behavior 
reminiscent to a second-order chiral quantum phase transition with the mass 
parameter of the Higgs potential serving as a control parameter \cite{Gies:14}.
In the language of critical phenomena~\cite{Zinn-Justin:2002ecy}, the 
chiral order parameter corresponding to the Higgs expectation value scales 
in the ordered (broken) phase near criticality according to a power law
\begin{equation}
	\langle \phi \rangle \propto \abs{t}^{\beta},
	\label{eq:orderparameterscaling}
\end{equation}
where $\beta$ denotes a critical exponent, and $t$ corresponds to the reduced 
temperature in a thermodynamic setting, (for quantum phase transitions, the 
corresponding control parameter is often called $\delta$). In the present model, 
the control parameter can be identified as
\begin{equation}
	t=\e -\e_\ast \equiv \delta\e_\Lambda,
	\label{eq:reducedTemperature}
\end{equation}
where $\e_\ast$ denotes the critical value of the dimensionless mass parameter 
of the Higgs potential at the UV cutoff $\Lambda$ separating the different 
phases. The precise value depends on all other bare couplings as well as on 
the RG scheme, but the deviation $\delta\e_\Lambda$ is a suitable choice as an 
analogue for the reduced temperature.

By means of so-called scaling and hyperscaling relations 
\cite{Zinn-Justin:2002ecy}, the critical exponent $\beta$ can be related to 
two other thermodynamic exponents used to describe the power law behaviour of 
the correlation length, as well as the anomalous dimension exponent $\eta$
\begin{equation}
	\beta=\frac{\nu}{2}\left(d-2+\eta\right).
	\label{eq:betaExponent}
\end{equation}
Here $\eta$ corresponds to the anomalous dimension coming from the wave function renormalization evaluated at the critical point. Thus, it describes the long-range 
behavior of the correlation function 
($\langle\phi(0)\phi(r)\rangle \propto r^{-d+2-\eta}$) when the 
correlation length diverges. In addition, we encounter the correlation length 
exponent $\nu$. Using the RG approach to the theory of critical phenomena, we 
can relate this critical exponent $\nu$ to the RG scaling exponent $\Theta$ 
of the most relevant perturbation at the critical point by
\begin{equation}
	\nu=\frac{1}{\Theta}.
	\label{eq:NutoTheta}
\end{equation}
At small interactions, i.e., near the Gaussian fixed point, the mass parameter 
of the Higgs potential is the most relevant perturbation, with its 
dimensionless version $\epsilon$ scaling as $\epsilon \sim k^{-2} =: 
k^{-\Theta}$, cf. \Eqref{eq:defe}, such that $\Theta=2$ characterizes the 
non-interacting limit. Hence, $\Theta$ denotes the power-counting dimension of 
the scalar mass parameter, and reflects its quadratic running tied to the 
naturalness problem:  in presence of a high energy scale such as a UV cutoff 
$\Lambda$ and assuming all bare parameters to be of order 1 at this 
scale, physical dimensionful long-range observables, like the vacuum 
expectation value  of the scalar field (the Fermi scale), are expected to be of 
the same order of magnitude of $\Lambda$.  Much smaller values are possible only 
at the expense of an exceptional amount of fine-tuning of the bare 
parameters at the scale $\Lambda$ such that bare parameters and 
contributions from quantum fluctuations cancel to a high degree.

In the vicinity of the Gaussian fixed point, perturbation theory predicts 
corrections to the canonical scaling yielding
\begin{equation}
	\Theta = 2-\eta,
	\label{eq:Theta}
\end{equation}
which -- upon expansion -- leads to at most logarithmic corrections to 
canonical scaling in accordance with Weinberg's theorem. Upon insertion of 
\Eqref{eq:Theta} into \Eqref{eq:betaExponent}, we find
\begin{equation}
	\beta=\frac{1}{2}\left(1+\eta+\frac{1}{2}\eta^2+\mathcal{O}\left(\eta^3\right)\right).
	\label{eq:betaToeta}
\end{equation}
for the scaling of the order parameter. 

\begin{figure}[htbp]
	\includegraphics[width=0.5\textwidth]{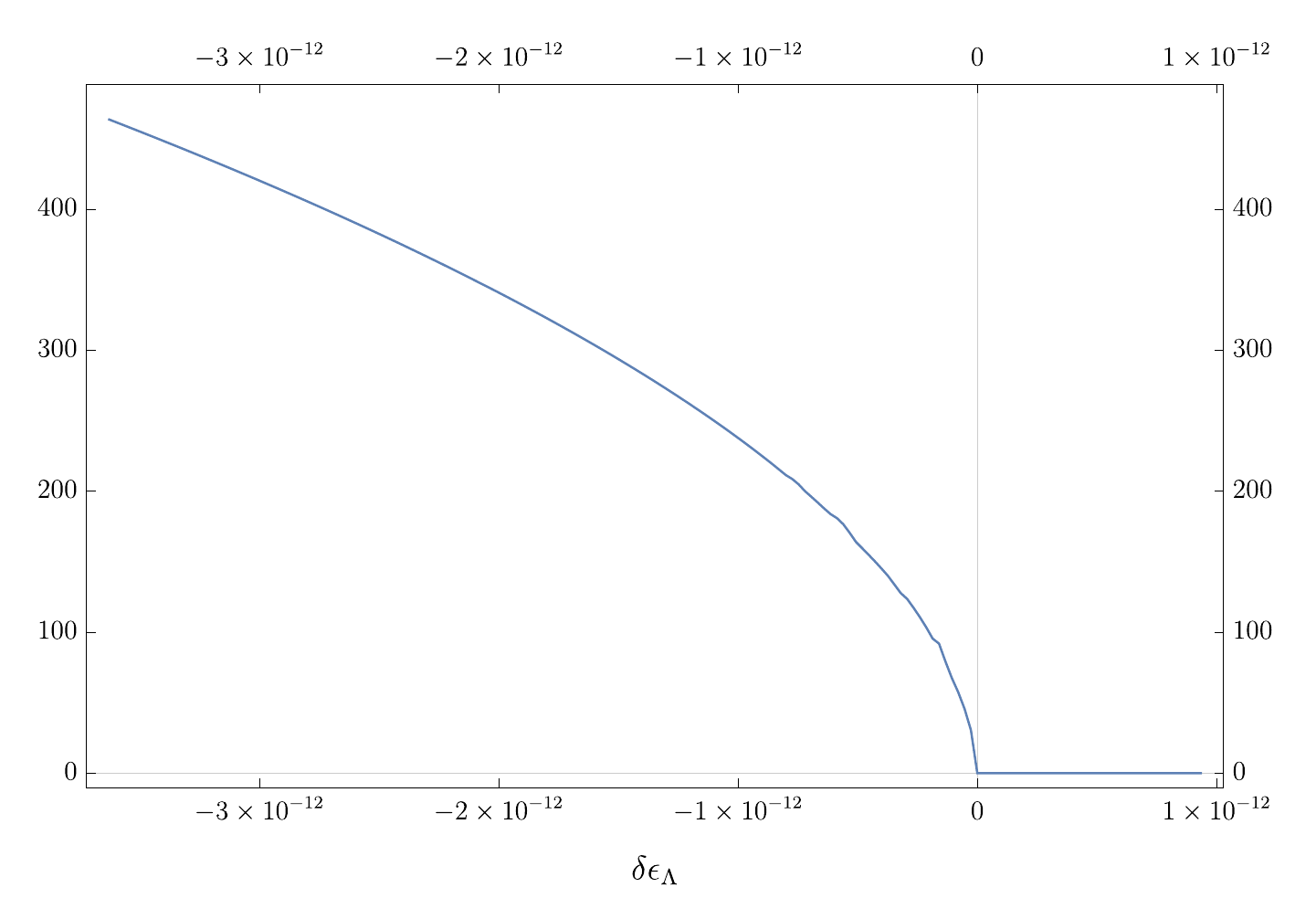}
	\caption{Second order quantum phase transition in the 
Higgs-top-bottom model \cite{Gies:14}: the chiral order parameter (vacuum 
expectation value of the Higgs field) $v$ is shown as a function of the control 
parameter $\delta \epsilon_\Lambda$. In the positive half plane we stay in the 
symmetric regime for the full flow, i.e. there is no vacuum expectation value 
$v=0$ in the long-range limit. 
For negative values of $\delta\e_\Lambda$, 
we end up in the broken phase, as is indicated by a non-zero vev $v$. 
%
%
Near the critical point $\delta\e_\Lambda=0$, the increase of the order 
parameter is governed by a power law, \Eqref{eq:orderparameterscaling}, with 
critical exponent close to $\beta\simeq 1/2$. A physical value for the Fermi 
scale 
$v\simeq 246$~GeV requires to fine-tune the control parameter rather closely to 
the phase transition.}
	\label{fig:PhaseTransitionHiggs}
\end{figure}

Let us illustrate this using the quantum phase transition of the reduced 
Higgs-top-bottom model. For otherwise fixed couplings, the vacuum expectation 
value of the Higgs field in this ungauged model is shown in 
Fig.~\ref{fig:PhaseTransitionHiggs} as a function of the control parameter 
$\delta \epsilon_\Lambda$. For positive values of the control parameter, the 
model is in the symmetric phase $v=0$. For negative values, the order parameter 
$v$ increases rapidly according to the scaling 
law~\eqref{eq:orderparameterscaling}. Since the model is comparatively weakly 
coupled, we expect the critical exponents to be close to their 
\textit{mean-field} values 
$\beta\simeq 1/2$ and $\eta\simeq 0$. (More precisely, the top-Yukawa coupling 
as the largest coupling introduces sizable quantum corrections such that 
$\eta\simeq 0.07$, see below.)

Correspondingly, the model needs to be 
fine-tuned severely in order to achieve a large scale separation. For instance 
for Fig.~\ref{fig:PhaseTransitionHiggs}, we have initiated the flow at 
$\Lambda=10^8$~GeV . In order to obtain a 
vacuum expectation value at the Fermi scale $v\simeq246$~GeV, we have to tune 
the control parameter to about $\delta\epsilon_\Lambda\simeq -1\times 
10^{-12}$ in units of the cutoff $\Lambda$. A ``natural choice'' of  
$\delta\epsilon_\Lambda \sim - \mathcal{O}(1)$ would have lead to 
a much larger order parameter and thus to unphysically large quark and Higgs 
masses near the cutoff scale. The required amount of fine tuning is governed by 
the critical exponent $\Theta\simeq2$. Therefore, the naturalness problem could 
be reduced, if $\Theta$ received large corrections to power-law scaling through 
a finite $\eta$.

The above reasoning is typically also applied to the full standard model. 
This, however, relies on some assumptions,
which might not be realized in nature.
A first assumption is that the massless Gaussian fixed point
be the relevant fixed point for the understanding
of the properties of the system on all relevant scales.
In presence of other fixed points with different 
properties, the sensitivity of the mass spectrum on
bare parameters might be completely different.
The second assumption is that -- even at the
Gaussian fixed point -- the relevant operator related to
a scalar bare mass parameter be the only/most significant
deformation determining the mass spectrum of particles
and in particular the physical value of the Higgs and
W/Z bosons masses.
In presence of more than one scale-breaking relevant
deformation, this might not be the case.
In particular, in asymptotically free non-Abelian gauge theories
there is always one marginally-relevant deformation for each gauge
coupling, with associated power-counting critical exponent $\Theta=0$.
The interplay of the chiral transition driving mechanisms and the understanding 
of the naturalness problem in
presence of such couplings is however much less
advanced, and at the heart of our following studies.

Upon inclusion of a QCD sector, the nature of the transition 
changes qualitatively rather independently of the initial conditions: as QCD is 
always in the chirally broken phase, the symmetric phase of the model 
disappears completely as does the critical point. The second-order quantum 
phase transition of the ungauged model is washed out and becomes a 
smooth crossover. Nevertheless, if the QCD scale is much smaller than the Fermi 
scale, we expect the comparatively sharp increase of the Yukawa model to remain 
a prominent feature of $v$ as a function of $\delta \e_\Lambda$. If this is the 
case, we refer to the large-$v$ regime to the left of this sharp increase as 
the \textit{Higgs regime} in contradistinction to the \textit{QCD regime} to 
the right of the rapid transition where $v$ is much smaller.

\begin{figure}[htbp]
	\includegraphics[width=0.5\textwidth]{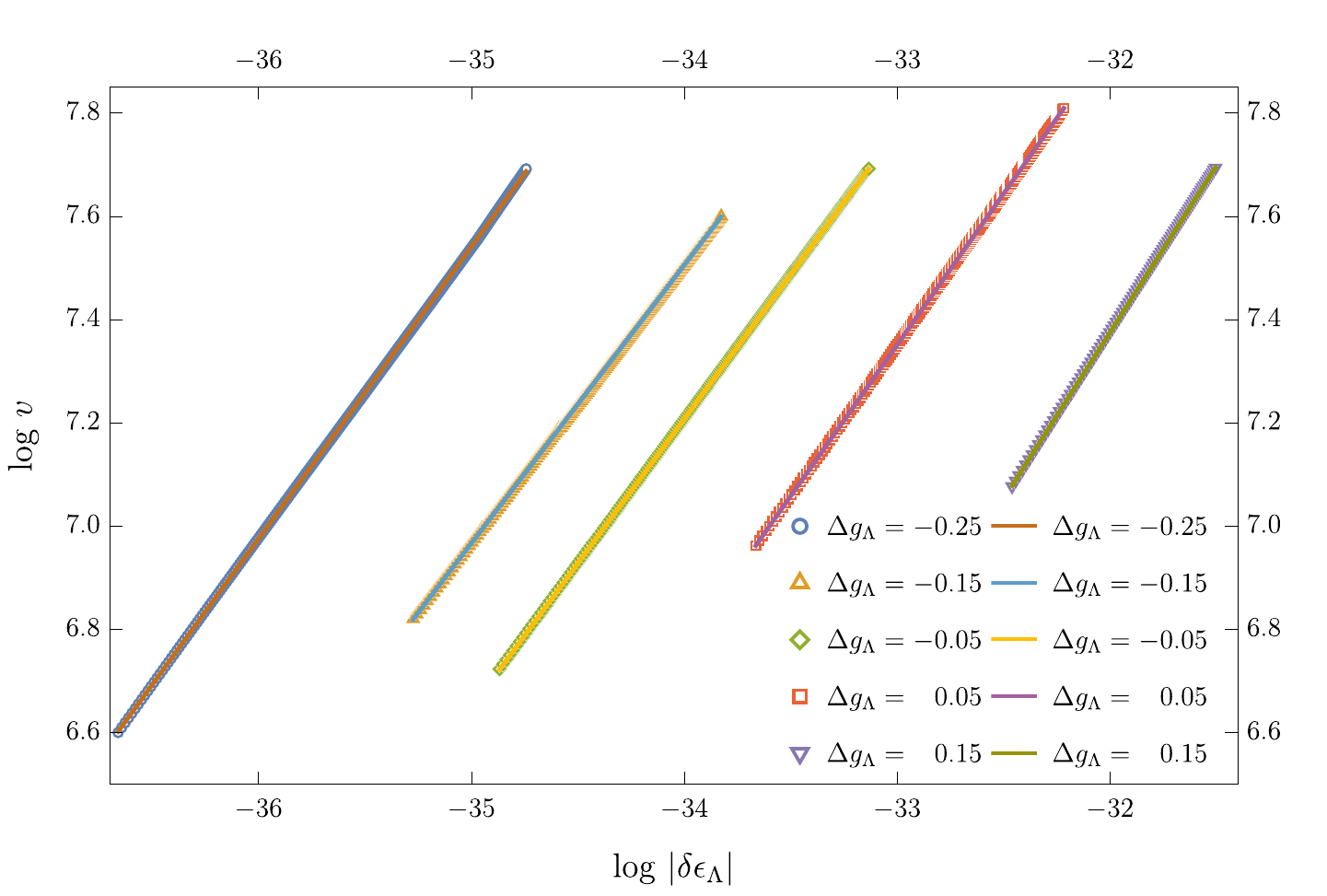}
	\caption{The phase transition of 
the full model in a logarithmic plot: the increase of the order parameter 
$v$ (a.u.) is shown as a function of the control parameter $\delta \e_\Lambda$ 
in the regime of a strong increase of $v$, similarly to the left half plane of 
Fig. \ref{fig:PhaseTransitionHiggs}. The pseudo-critical exponent $\beta$ for 
each choice of initial value of the gauge coupling can be extracted from the 
slope in this double-logarithmic plot (c.f.~Eq.~\eqref{eq:orderparameterscaling}).}
	\label{fig:FitExponent}
\end{figure}

If this is the case, we can still try to quantify this increase by a 
power-law scaling similar to \Eqref{eq:orderparameterscaling} with a 
corresponding \textit{pseudo-critical} exponent $\beta$. More precisely, we 
extract the pseudo-critical exponent of the chiral transition by 
fitting the vacuum expectation value in the Higgs regime  to the power law 
ansatz (Eq. \eqref{eq:orderparameterscaling}). 
This procedure is illustrated in Fig. 
\ref{fig:FitExponent}, where the chiral order parameter $v$ (in arbitrary 
units) is plotted doubly-logarithmically as a function of the control parameter 
for different initializations of the gauge sector; here, $\Delta g_\Lambda$ 
measures the deviation of the initial conditions for the gauge 
coupling from the physical point). Each case displays a clear power-law for the 
order-parameter scaling that can be fitted by \Eqref{eq:orderparameterscaling}. 

In this way, we get an estimate for the pseudo-critical exponent 
$\beta$, even though the transition is actually a crossover. Using Eq. 
\eqref{eq:betaToeta}, we can re-express this exponent through $\eta$. Upon 
insertion of $\eta$ into \Eqref{eq:Theta}, we obtain a correction to scaling 
and can thus quantify the impact of the gauge sector on the amount of 
fine-tuning necessary to separate the Fermi scale from the UV scale $\Lambda$. 
If $\eta$ is positive and sufficiently large, the naturalness problem of the 
model could be alleviated.

Let us finally point out two aspects that are not fully accounted for 
here: first, the SU$(2)_\text{L}$ gauge group is ignored here but participates 
also as a potential source for chiral symmetry breaking in its strong-coupling 
regime. In the Higgs regime, this source is always screened by the finite 
gauge-boson masses. In the QCD regime, it is potentially active. However, since 
the QCD sector is much more strongly coupled, it dominates the chiral 
observables quantitatively. We expect our results to remain essentially 
unaffected, as long as we preserve the hierarchy of the two non-Abelian gauge 
sectors of the standard model.

Second, the picture of universality developed for the Higgs-top-bottom 
model is not exact since the model does presumably not feature a UV-complete 
limit. We therefore have to work with a finite UV cutoff $\Lambda$, implying 
that slight violations of universality can occur, but are generically 
suppressed by powers of $1/\Lambda$. We find that $\Lambda=10^8$ GeV is 
sufficient for our purposes. We have tested for violations of universality by 
varying the initital conditions of the marginal couplings on the 
$\mathcal{O}(1)$ level. The residual non-universal effects in the QCD regime 
remain below the permille level, see App. \ref{app:universalityviolation}.

\subsection{The QCD regime}\label{sec:DeepQCD}
In the QCD regime, 
one expects the induced vacuum expecation value in the IR to be of the 
order of the scale of 
strong interactions $\Lambda_\mathrm{QCD}$. In this work, we use a simple but 
useful definition of this scale by identifying it with the location of the 
IR Landau pole of the strong gauge coupling obtained from the one-loop 
beta function. In the deep QCD regime, the phase transition is 
triggered by the gauge sector (more specifically via a strongly increasing 
four-fermion coupling in the effective action \Eqref{eq:QCDAction} before 
performing the partial bosonization), in contrast to quark and Higgs 
fluctuations in the deep Higgs regime.
This corresponds to a choice of initial values 
where the scalar mass $m^2>0$ 
is of the order of the UV scale $\Lambda$, making the scalar field non-dynamical 
and thus reducing our model to a low-energy effective theory of the strong 
interactions.
In other words, a limiting case of our reduced standard model is the pure quark 
meson model \cite{Jungnickel:95}, however with $\Nf=6$ quark flavors.
 
The vacuum expectation value induced in the deep QCD regime depends 
solely on the value of the gauge coupling at the initialization scale 
$g_\Lambda$. Since there is no remnant of the electroweak Higgs mechanism 
in this regime, all quarks have zero current quark mass.

As it should be, the other marginal parameters of the model (scalar 
self-interactions and Yukawa couplings) have hardly any influence on the 
resulting 
vacuum expectation value. More precisely, their influence arises only 
because of the fact that we work with a finite UV cutoff $\Lambda$; this 
introduces violations of universality which vanish exactly in the limit 
$\Lambda\to\infty$. For completeness, we quantify these universality violations 
in App.~\ref{app:universalityviolation} for our practical choice 
$\Lambda=10^8$ GeV and find them to be on the sub-permille level for the 
quantities of interest.

This feature is expected since the gauge coupling $g$ is the only parameter 
present in fundamental massless
QCD. Through the scale anomaly encoded in the RG running 
manifesting in \textit{dimensional transmutation}, it 
is linked to the dimensionful quantity $\Lambda_\mathrm{QCD}$ setting the scale 
for all long-range observables such as the vacuum expectation value $v$. 
The chiral condensate $v$  and $\Lambda_\mathrm{QCD}$ are intimately connected; 
using our simple definition of $\Lambda_\mathrm{QCD}$ mentioned above, we list 
some quantitative results for varying initial values of the gauge coupling 
$g_\Lambda$ in Tab.~\ref{tab:QCDvev}.
\begin{table}[htbp]
	\begin{tabular}{cccc}
		$g_\Lambda$ & $\Lambda_\mathrm{QCD}$ in GeV & $v_\mathrm{QCD}$ in GeV & $v_\mathrm{QCD}/\Lambda_\mathrm{QCD}$ \\ \hline
		0.519       & $6.4 \times 10^{-11}$                         & $4.8 \times 10^{-12}$   & 0.075                                                 \\
		0.619       & $1.6 \times 10^{-5}$                           & $1.1\times 10^{-6}$     & 0.068                                                 \\
		0.719       & $3.3\times 10^{-2}$                           & $2.1\times 10^{-3}$     & 0.063                                                 \\
		0.819       & 5.0                                           & 0.29                    & 0.059                                                 \\
		0.919       & 158                                           & 8.8                     & 0.055         \\
		1.019 & 1911 &101&0.053                                       
	\end{tabular}
\caption{Comparison of the QCD scale $\Lambda_\mathrm{QCD}$ (for simplicity 
defined by the location of the one-loop IR Landau pole for $\Nf=6$ active quark flavors) and the  
vacuum expectation value in the deep QCD regime ($v_\mathrm{QCD}$).  The 
\textit{physical} initial condition $g_\Lambda=0.719$ is chosen at 
$\Lambda=10^8$ GeV such that we obtain the experimentally measured value of 
$\alpha_\mathrm{s}(m_\mathrm{Z})=\frac{g^2(m_\mathrm{Z})}{4\pi}= 0.117$ at the Z 
boson mass scale. The other bare couplings are set to zero at the high energy scale (scalar self interactions) or by IR physics (top and bottom Yukawa couplings). The chiral condensate however is (almost) independent of the values of these other couplings, as is checked in appendix \ref{app:universalityviolation}.}
\label{tab:QCDvev}
\end{table}
We see that while $v$ and $\Lambda_{\mathrm{QCD}}$ vary a lot under shifts in $g_\Lambda$, the ratio of the two quantities stays approximately constant over ranges of the bare strong gauge coupling considered in this work.
We consider the choice $g_\Lambda=0.719$ as the physical initial 
condition at $\Lambda=10^8$ GeV, since it corresponds to the experimentally 
measured value of 
$\alpha_\mathrm{s}(m_\mathrm{Z})=\frac{g^2(m_\mathrm{Z})}{4\pi}\simeq 0.117$ at 
the Z boson mass scale. This yields $\Lambda_\mathrm{QCD}\simeq 33$~MeV and 
$v\simeq 2.1$~MeV for the IR QCD scales. Naively, this appears to be rather 
small compared to the physical values, e.g. $v=f_\pi\simeq 93$~MeV. This, 
however, is a consequence of the fact that our IR flow proceeds with $\Nf=6$ 
massless quarks down to the scale of symmetry breaking. Since the quarks induce 
screening, the gauge coupling grows strong at lower scales compared to the 
standard model case that includes the decoupling of the ``heavy'' quarks as a 
consequence of the Higgs mechanism.

Notice that the 
value of the chiral condensate $v$ is consistently smaller than the scale 
obtained by the one-loop calculation.
Both vary exponentially as a function 
of $g_\Lambda$. 
The results show that the ratio
between the QCD scale and the
would-be-Fermi scale $v$, far 
from being a free parameter,
actually features a maximum value
of $\mathcal{O}(10)$. Furthermore we observe that the ratio of the two scales varies rather slowly over a wide 
range of initial conditions $g_\Lambda$.

\subsection{Pseudo-critical exponent}
\begin{figure}[htbp]
	\includegraphics[width=0.48\textwidth]{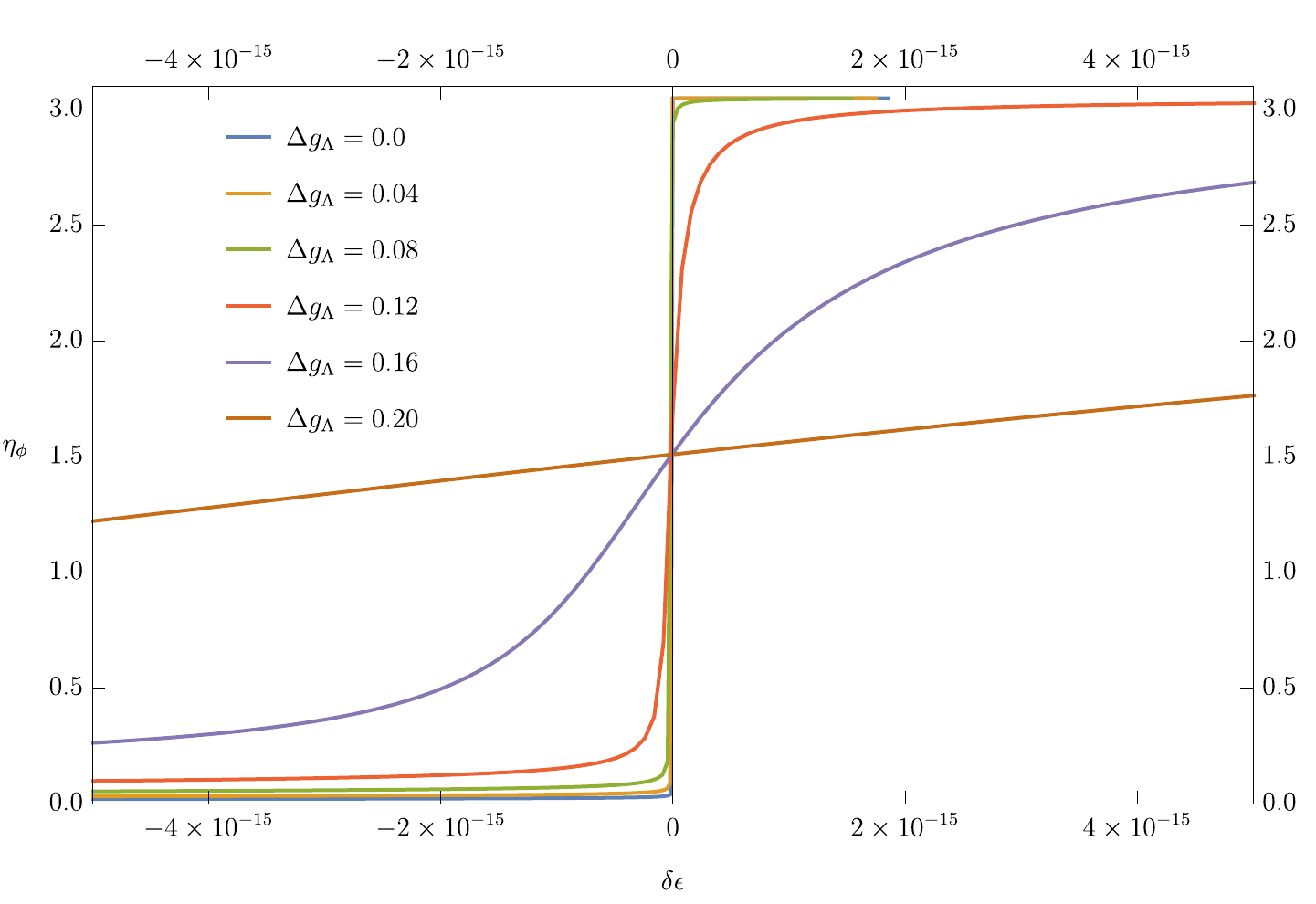}
	\caption{The scalar anomalous dimension $\eta_{\phi}$ at the transition scale. A small $\eta_\phi$ corresponds to a Higgs-like regime ($\delta\e<0$) whereas a large one can be identified with a QCD-like regime. While for small gauge couplings $g_\Lambda$ the scalar anomalous dimension jumps at the critical point (characterized by $\delta\epsilon=0$), indicating that a critical point for a second order phase transition can be straightforwardly defined, and thus extracting a pseudo-critical exponent is sensible. For large gauge couplings, this region grows, making it hard to define a critical point as well as extract a meaningful pseudo-critical exponent.}
	\label{fig:TransitionVsCrossover}
\end{figure}

Let us now analyze the interplay between the chiral transitions in the full 
model. For this, we first choose initial values for the marginal couplings at 
the initialization scale $\Lambda$, and then vary the relevant parameter of the 
scalar potential, i.e.~the control parameter $\delta \e_\Lambda$. Integrating 
the flow equations yields the vacuum expectation value $v$ as a function of 
$\delta \e_\Lambda$ similar to the chiral Higgs-top-bottom model, c.f.~Fig.~\ref{fig:PhaseTransitionHiggs}. As expected, the second order phase 
transition of the latter becomes a crossover for any finite value of 
the initial gauge coupling $g_\Lambda$, meaning that the discontinuity
at $\delta \e_\Lambda=0$ is smoothed
in a neighbourhood of the 
would-be-critical point. The size of this critical region grows for increasing
 $g_\Lambda$, 
 as illustrated in
 Fig.~	\ref{fig:TransitionVsCrossover}. 
Still, we generically observe a rather sharp transition between a 
Higgs-like and a QCD-like regime. In this regime, we identify the physical point 
for the Fermi scale where $v=246$ GeV. The required choice for the control 
parameter $\delta \e_\Lambda$ lies typically at small negative values; we call 
the deviation from this value $\delta \e_\Lambda^\mathrm{phys}$.  Near the 
physical point,
i.e.~for $|\delta \e_\Lambda^\mathrm{phys}|\ll 1$, we fit the 
results for the vacuum expectation value to the scaling hypothesis coming from the theory of second order 
phase transitions,
that is \Eqref{eq:orderparameterscaling}. 
Example results of such fits have 
been shown in Fig.~\ref{fig:FitExponent}. Deviations from canonical 
power-counting scaling occurs as departures of the pseudo-critical exponent 
$\beta$ from the value $\beta=1/2$,
 which we measure in terms of the 
pseudo-critical correction-to-scaling exponent $\eta$
according to \Eqref{eq:betaToeta}. 
As discussed above, 
finite positive values of $\eta$ are a direct quantitative measure of how the 
interplay of the chiral transitions in the standard model alleviates the 
naturalness problem.

Let us investigate how this quantity depends on the microscopic
parameters of the Lagrangian at the scale $\Lambda$. In general, changing 
a microscopic parameter changes the physical quantities of the theory. For 
instance, changing the top-Yukawa coupling in the Higgs regime corresponds to 
changing the top mass relative to the vacuum expectation value. However, this 
direct influence on the long-range observables becomes less pronounced -- and 
even vanishes ultimately -- in the deep QCD regime where the full model is 
dominated by the gauge sector. Since the physical point is near the transition 
from the Higgs to the QCD regime, the precise dependence of the 
pseudo-critical exponent on the microscopic couplings is a priori unclear.  

\begin{figure}[htbp]
	\includegraphics[width=0.5\textwidth]{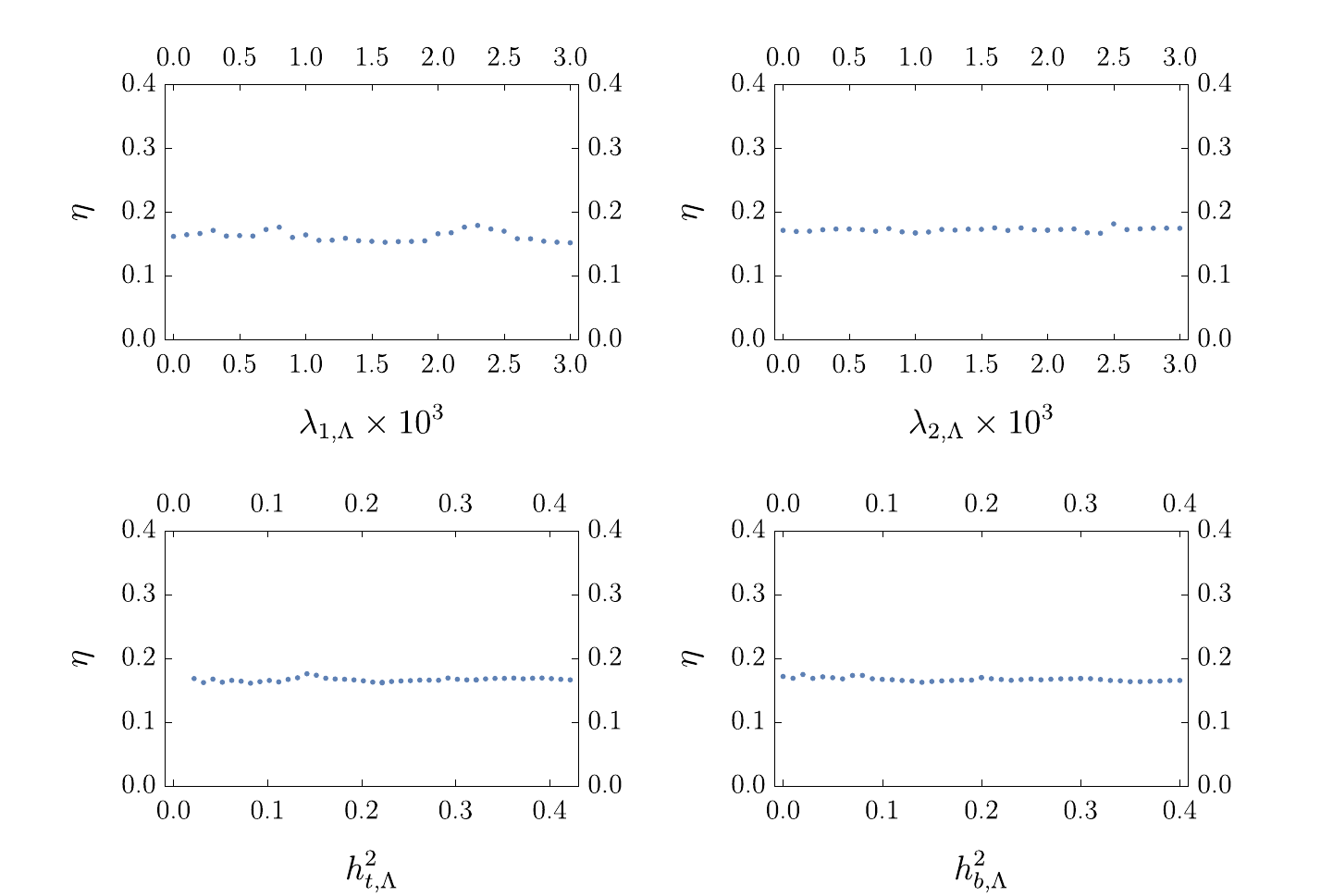}
	\caption{The critical exponent $\eta$ for different couplings at the high energy 
scale $\Lambda$: the quartic couplings $\lambda_1$  
(upper left panel) and $\lambda_2$ (upper right), and the Yukawa couplings
 of the top (lower left) and bottom (lower right) quarks. 
The gauge coupling $g_\Lambda$
at the initialization
scale is kept 
fixed at the physical value. 
The size of the fluctuations of $\eta$
 is a measure for the numerical 
precision.}
	\label{fig:TotalPlotPhaseTrans}
\end{figure}

The computational procedure is the same for all parameters of interest: we keep 
all couplings fixed at the initialization scale $\Lambda$ while varying the 
control parameter 
$\delta \e_\Lambda^\mathrm{phys}$ to extract the behavior of the phase 
transition and to measure $\eta$ for a given set of parameters. We repeat this 
for different values of the coupling of interest at $\Lambda$, while keeping 
all other values fixed. Results are shown in Fig.~\ref{fig:TotalPlotPhaseTrans} 
for all couplings in the model except for the gauge coupling.
We find that the pseudo-critical exponent $\eta$ remains basically unaffected 
by the initial value of the Yukawa and scalar couplings in the explored regime. 
The data in Fig.~\ref{fig:TotalPlotPhaseTrans} shows only some glitches 
on top of a constant value which reflect the numerical precision. The 
fine-tuning procedure necessary to arrive at $v=246$ GeV together with the large 
integration range starting from $\Lambda=10^8$ GeV leaves room for an accumulation 
of numerical errors. For practical reasons, we have restricted the scalar 
couplings to relatively small values, since stronger
scalar self-interactions would require the flow to already start in the broken 
regime. This choice is also
justified by the observation
that the Higgs quartic coupling in the full
standard model appears to
be nearly vanishing at the Planck scale.
\begin{figure}[htpb]
	\includegraphics[width=0.5\textwidth]{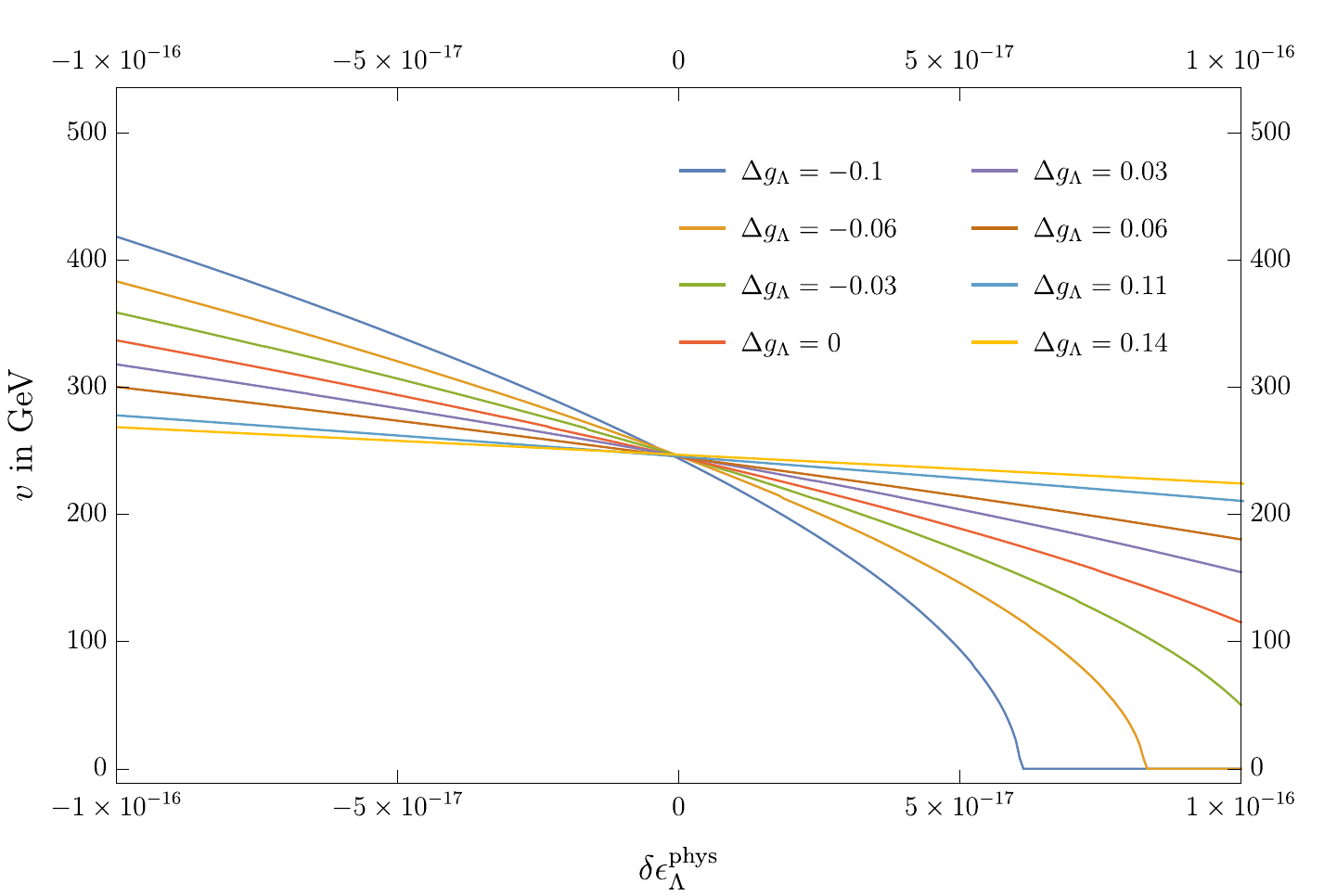}
	\caption{Vacuum expectation value in vicinity of the Fermi scale 
$v=246\,\text{GeV}$ for different values of the gauge coupling $g_\Lambda$ at 
the UV scale. For larger gauge couplings we observe less change in the vacuum 
expectation value under variation of the relevant direction of the scalar 
potential at the high energy scale $\delta\e_\Lambda$, indicating a 
stronger interplay of the phase transitions and a corresponding mitigation of 
the naturalness problem.}
	\label{fig:WindowPlot}
\end{figure}

By contrast, varying the initial value of the gauge coupling has a strong 
influence on the transition from the Higgs to the QCD regime.
This is visible in 
Fig.~\ref{fig:WindowPlot}, where the slope of the order parameter field $v$ as 
a function of the control parameter flattens for increasing values of the gauge 
coupling; the latter is expressed in terms of the deviation $\Delta g_\Lambda$ 
of the initial condition from the physical point. This has a direct impact on 
the naturalness problem: the flatter the curve is when it crosses the physical 
point, the less fine-tuning is needed to put the system in the vicinity of the 
Fermi scale. We emphasize that the physical point $\Delta g_\Lambda=0$ already 
exhibits a flatter curve than the pure Higgs-top-bottom model with canonical 
power-counting. This demonstrates that the QCD sector in the standard model 
already alleviates the naturalness problem compared to a pure Higgs-Yukawa 
model which is often used to illustrate the naturalness problem.

\begin{figure}[t]
	\includegraphics[width=0.48\textwidth]{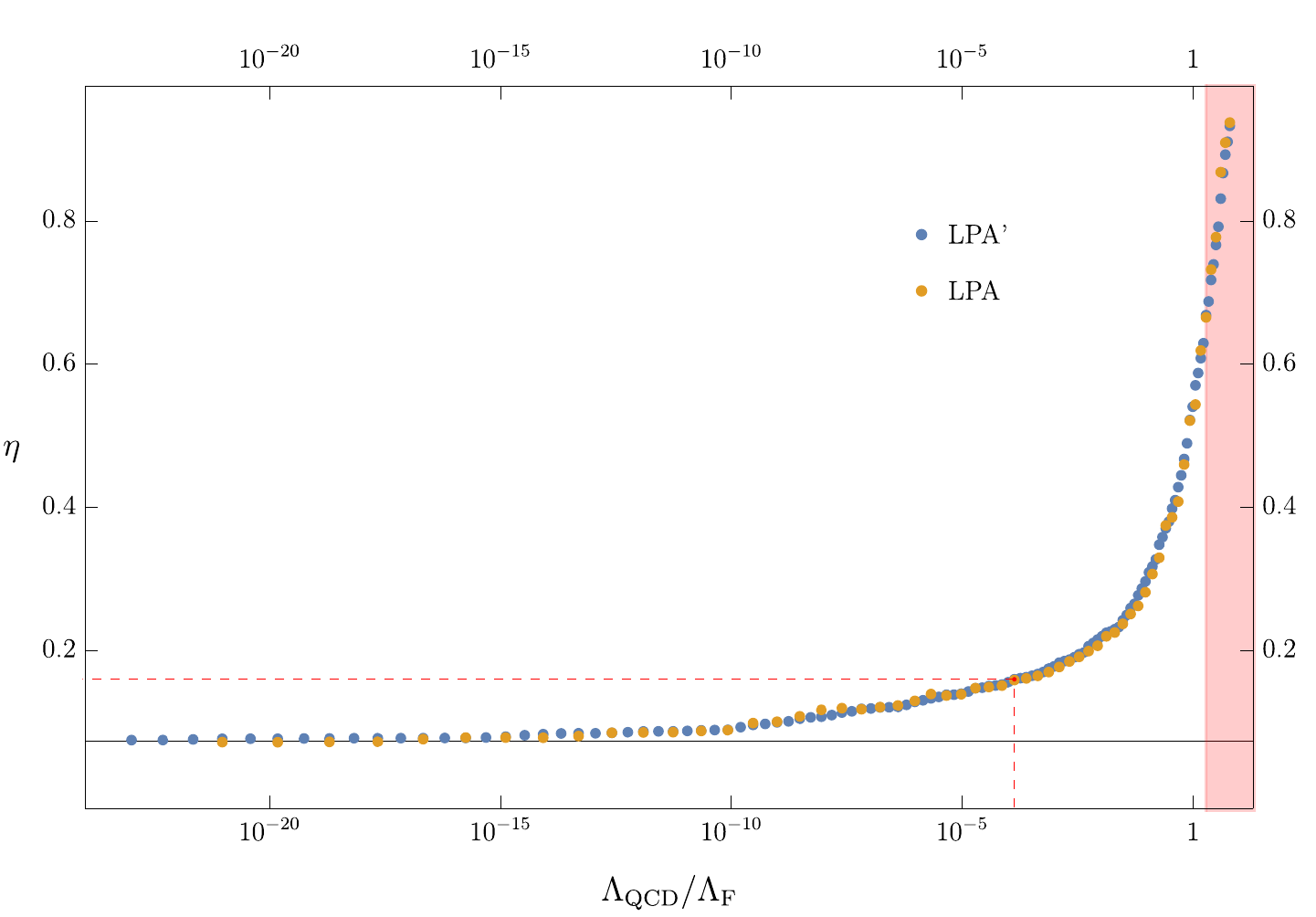}
	\caption{Pseudo-critical exponent $\eta$ for different initial conditions 
of $g_\Lambda$. By dimensional transmutation, we express $g_\Lambda$ as the 
QCD scale $\Lambda_\mathrm{QCD}$ (for $\Nf=6$). The Fermi scale 
$\Lambda_\mathrm{F}$ is fixed to $246\,\mathrm{GeV}$ for this work. The 
interplay between the chiral transitions in the standard model becomes the more 
pronounced the closer the two scales are. For $\Lambda_\mathrm{QCD}\to 
\Lambda_\mathrm{F}$, the naturalness problem weakens significantly. The physical 
point corresponding to a standard-model-like running of the gauge coupling is denoted by the 
red dotted line. The shaded region on the right marks the regime where 
the chiral transition gets so smooth that the  
pseudo-critical exponent becomes less meaningful.
} 
	\label{fig:etaVSQCD}
\end{figure}

Quantifying these results analogously to the previous analysis, we again 
extract the pseudo-critical exponent $\eta$ for these transitions. The 
interplay of the chiral transitions and the alleviation of the naturalness 
problem is visible from the monotonic increase of $\eta$ as a function of 
$g_\Lambda$. In Fig.~\ref{fig:etaVSQCD}, we plot the corresponding critical 
exponent against the dimensionless 
ratio $\Lambda_\mathrm{QCD}/\Lambda_\mathrm{F}$; here, we have translated 
$g_\Lambda$ into $\Lambda_\mathrm{QCD}$, denoting the QCD scale coming from 
dimensional transmutation, $\Lambda_\mathrm{F}$ denotes the Fermi scale which is 
set to $246\,\mathrm{GeV}$ for this work.

Deep inside the Higgs regime, i.e.~for 
$\Lambda_\mathrm{QCD}/\Lambda_\mathrm{F}\to 0$, we observe that the 
pseudo-critical exponent approaches a constant value $\eta\simeq0.07$. This is the value of 
the pure Higgs-top-bottom model  
arising from the fluctuations in the Yukawa sector. Surprisingly, already in the limit
of negligible QCD corrections,
the critical exponent differs
from the pure Gaussian
value.
This is an effect of having fixed
the IR observables, such
as the Fermi scale $v$ and the top mass $m_\mathrm{t}^2$,
a requirement that forces the theory off the Gaussian
fixed point. A more explicit argument 
supporting this interpretation is presented in the next section.

  For an increasing scale 
ratio $\Lambda_\mathrm{QCD}/\Lambda_\mathrm{F}$ also $\eta$ increases, 
exemplifying a stronger interplay of the chiral transitions. At the physical 
point which in our definition is characterized by $\Lambda_\mathrm{QCD}=33$~MeV, 
we observe that the pseudo-critical exponent takes the value $\eta=0.16$. 
 We take this as a manifestation that the QCD sector 
exerts a non-negligible influence on the chiral transition. Quantitatively, the naturalness 
problem of the standard model is mildly alleviated by this interplay, even 
though the reduction of the power-law scaling from the canonical 
value $\Theta=2$ to $\Theta\simeq 1.84$ remains on the $10\%$ level and thus 
does not make a qualitative difference.

A ``solution'' of the naturalness problem would require a pseudo-critical 
exponent $\eta\sim \mathcal{O}(1)$. For values of  $\Lambda_\mathrm{QCD}$ 
larger than the physical point, we observe the onset of a strong increase of 
$\eta$ near $\Lambda_\mathrm{QCD}/\Lambda_\mathrm{F}\simeq 10^{-2}$. Physically 
this implies that an increasing part of the chiral order parameter is generated 
by the QCD sector. Correspondingly, the separation of $v$ from the cutoff scale 
$\Lambda$ requires less and less fine-tuning of the bare parameters in the 
Higgs sector, but is taken care of by chiral symmetry breaking of QCD. The 
scale of the latter depends only logarithmically on the high scale and a large 
scale separation thus becomes natural.

However recall that
	the ratio
	$\Lambda_\mathrm{QCD}/\Lambda_\mathrm{F}$ cannot attain arbitrarily large values,
	as we argued in the previous section.
Furthermore, the simplistic
description of
Fig.~\ref{fig:WindowPlot} of chiral-symmetry
breaking in terms of a pseudo-critical exponent $\eta$
breaks down well below
that upper bound. This is because
the behavior of
$v$ across the transition
between the symmetric and
the SSB regime  ceases to 
be comparable to the
power-law ansatz of \Eqref{eq:orderparameterscaling}
for too large $g_\Lambda$; in 
Fig.~\ref{fig:etaVSQCD} this is indicated by the shaded region.

\subsection{Scaling behaviour near the QCD-dominated regime}

As observed in the preceding section, the pseudo-critical exponent
exhibits a strong increase beyond
$\Lambda_\mathrm{QCD}/\Lambda_\mathrm{F}\gtrsim 10^{-2}$. We argue in the 
following that this can be understood within a simple approximation of the flow 
equations. Approaching the QCD regime from the side of the Higgs regime, the 
RG flow transition between the symmetric and SSB  regimes starts to occur 
closer and closer to $\Lambda_\mathrm{QCD}$. Near the latter, the running of 
the Yukawa coupling, as well as the scalar self interaction, becomes governed 
by the flow of the strong gauge coupling $g^2$. This can be seen by looking at 
the $\beta$ functions of these couplings in a specific limit:

Right above the scales before the RG flow enters the SSB regime,  
the dimensionless mass parameter $\e$, and thus the scalar masses, are 
negligible; by contrast, the meson Yukawa coupling $h^2$ near the QCD regime
is much larger than the top and bottom Yukawa couplings, which therefore can be 
set to zero for the present analysis. Looking at the flow equations in this 
limit 
reveals that there are quasi fixed points in the composite couplings $h^2/g^2$ 
and $\lambda_{1}/h^2$, implying that the evolution of $h^2$ and $\lambda_{1}$ 
are 
directly connected to the flow of $g^2$.
This kind of partial fixed point
behavior is well know both as a feature
of IR flows, in which case it is called
a Pendelton-Ross kind of fixed point~\cite{Pendleton:1980as},
as well as a UV signature 
of total asymptotic freedom
or safety,
in which case it has been referred to
with a variety of names:
eigenvalue condition~\cite{Chang:1974,Callaway:1988ya},
reduction of couplings~\cite{Zimmermann:1984sx,Heinemeyer:2014vxa},
nullcline~\cite{Bond:2016dvk},
fixed flow~\cite{Giudice:2014tma}
or quasi fixed point~\cite{Gies:2015lia,Gies:2016kkk}.
In the present model,
the quasi fixed point as
a solution of the
RG equations has
been first described by Cheng, Eichten and Li 
(CEL)~\cite{Cheng:1973nv},
and we therefore call it 
a CEL solution.

In this specific limit and considering for simplicity the one-loop approximation for all beta functions of interest here, we have
\begin{equation}
\begin{aligned}
\partial_t g^2 &= -\frac{7}{8\pi^2}g^4,\\ 
\partial_t h^2&=\frac{h^2}{16\pi^2}\left(\frac{19}{2}h^2-16g^2-12\frac{g^4}{h^2}\right),\\
\partial_t \lambda_1 &= \frac{\lambda_1}{4\pi^2}\left(3h^2+4\lambda_1-\frac{3}{2}\frac{h^4}{\lambda_1}\right).
\end{aligned}
\label{eq:oneloopapprox}
\end{equation}
Note that the last terms of the flow of $h$ and $\lambda_1$ arise from 
dynamical bosonization.
In \Eqref{eq:oneloopapprox}, we have already set $d=4$, $\Nf=2$ in the 
Higgs-top-bottom sector beta functions, $\Nf=6$ for the gauge coupling, and 
$\Nc=3$. It is useful to introduce rescaled couplings $\hat{h}=\frac{h}{g}$, 
as well as $\hat{\lambda}_1=\frac{\lambda_1}{h^2}$. Their flow equations 
read
\begin{eqnarray}
\partial_t\hat{h}^2&=&\frac{1}{32\pi^2}g^2\left(19 
\hat{h}^4-4\hat{h}^2-24\right),\label{eq:rescaledflows}\\
\partial_t\hat{\lambda}_1&=&\frac{1}{32\pi^2}\frac{g^2}{\hat{h}^2}\Big[24 
\hat{\lambda}_1+32\hat{h}^2\hat{\lambda}_1\nonumber\\
&&\qquad\qquad +\hat{h}^4\left(32\hat{
\lambda } _1^2+5\hat{\lambda}_1-12\right)\Big],\nonumber
\end{eqnarray}
each of which exhibit a fixed point. These fixed 
points imply that we have $h^2 \sim g^2$ and $\lambda_{1}\sim h^2 \sim g^2$ 
within the validity regime of these one-loop equations. Correspondingly, 
the scalar anomalous dimension takes the form
\begin{equation}
\eta_\phi = \frac{3}{8\pi^2}h^2.
\end{equation}
\begin{figure}[htbp]
\includegraphics[	width=0.48\textwidth]{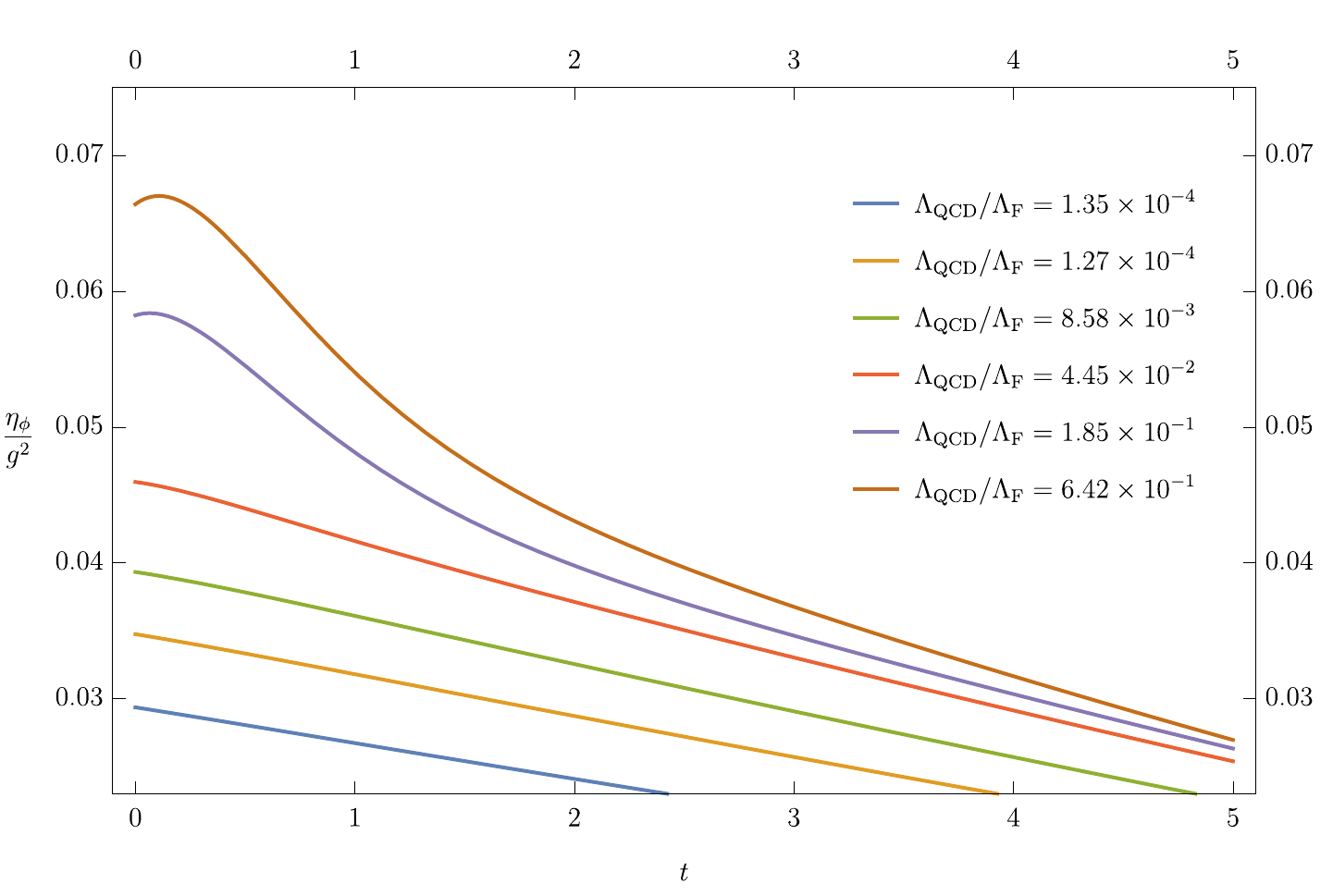}
\caption{The running of $\eta_{\phi}/g^2$ close to the transition from the disordered to the ordered phase for different scale separations $\Lambda_{\mathrm{QCD}}/\Lambda_\mathrm{F}$. The RG time $t$ is normalized such that $t=0$ coincides with the transition scale. We can see that for increasing values of $g_\Lambda$, and thus $\Lambda_{\mathrm{QCD}}/\Lambda_\mathrm{F}\rightarrow 1$, the running of $\eta_{\phi}/g^2$ levels out, corresponding to the quasi fixed-point in $\hat{h}$ and $\hat{\lambda}$.
} 
\label{fig:closeToTransition}
\end{figure}
As a consequence, in an intermediate 
interval of values of 
$g_\Lambda$,
where the latter grows but the
system is not yet fully
dominated by QCD fluctuations, the scalar anomalous dimension scales as $g^2$ close to the 
transition. This
can be seen in Figure \ref{fig:closeToTransition},
where the flattening of the
three upper trajectories
near $t=0$ (the RG time
at which the flows enter into the SSB regime) signals
an effective scaling 
$\eta_\phi\sim g^2$.
When this happens, the quasi-fixed point
value of the field anomalous dimension at the transition
in turn becomes
a good approximation
of the pseudo-critical exponent,
as we now argue.

\begin{figure}[htbp]
	\includegraphics[width=0.48\textwidth]{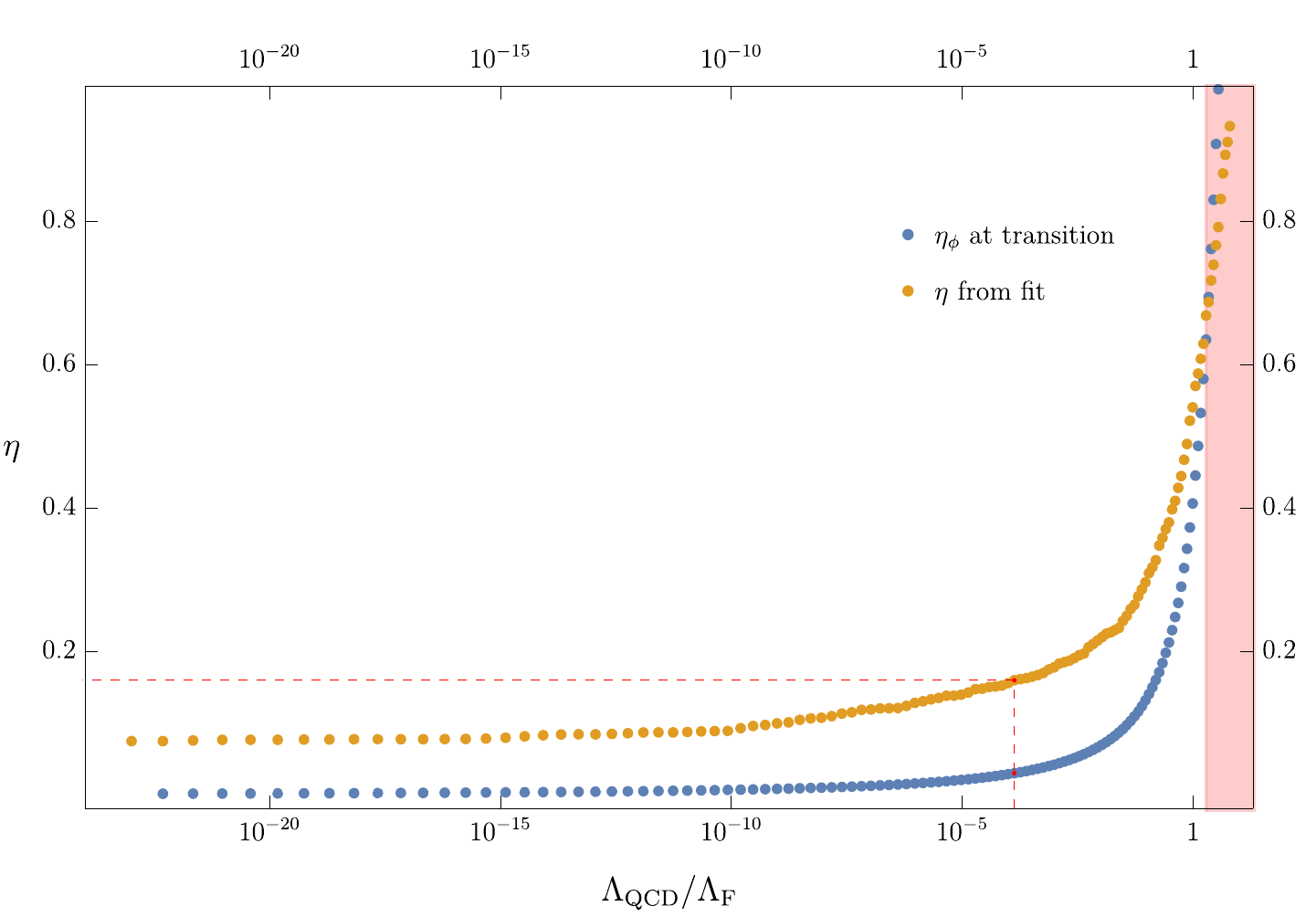}
	\caption{Comparison of the pseudo-critical exponent 
$\eta$ and the scalar anomalous dimensions $\eta_\phi$ evaluated 
near the transition of the RG flow from the symmetric to the broken regime using the full set of flow equations.
Both exhibit a strong increase towards values of $\mathcal{O}(1)$ near 
$\Lambda_\mathrm{QCD}/\Lambda_\mathrm{F}\simeq 10^{-2}$ indicating a strong 
interplay of the chiral transitions. The physical point, characterized by the running of the gauge coupling, is denoted by the red dashed line. 
} 
	\label{fig:CompareEtaFitTransition}
\end{figure}

This phenomenon
can be qualitatively
understood as a 
manifestation of the fact that a strongly growing scalar anomalous dimension has 
has a sizable impact on the running of the (dimensionless) mass parameter of 
the scalar potential:
\begin{equation}
\partial_t \e = 
-(2-\eta_\phi)\e-\frac{5}{8\pi^2}\lambda_{1}+\frac{3}{8\pi^2}h^2.
\label{eq:etaphiapprox}
\end{equation}
An increasing and large value of $\eta_\phi$ modifies 
the canonical quadratic running of $\epsilon$ substantially and eventually 
removes the fine-tuning problem, as is expected in a QCD-dominated model. 
At the transition from the unbroken to the broken phase, the scalar 
anomalous dimension can therefore be expected
to be related to the critical exponent 
$\eta$.
This is highlighted
and quantitatively confirmed by a comparison of the value of the scalar anomalous dimension
$\eta_\phi$ 
(computed within the full set of flow equations)
at the SYM-SSB 
transition scale, and the pseudo-critical exponent $\eta$ obtained from the 
power law behavior of the vacuum expectation value $v$ close to the critical 
point 
(c.f.~Eq.~\eqref{eq:orderparameterscaling}),
which is shown in  
Fig.~\ref{fig:CompareEtaFitTransition}.

Starting from the deep Higgs
regime for very low ratio
$\Lambda_\mathrm{QCD}/\Lambda_\mathrm{F}$,
 one can see in this figure that the
anomalus dimension  at the
transition scale is positive
$\eta_\phi>0$ already
in the pure Higgs-top-bottom model,
since a nonvanishing
top Yukawa coupling
is required by the IR conditions
imposed on the RG trajectories.
In this limit, we observe that
$\eta$ is  bigger than $\eta_\phi$ at
the transition by about an order of magnitude. 
We attribute this to a built-up of fluctuations at the onset of the SSB regime
from the transition scale to the freeze out scale. 
This is a nontrivial result stemming from the
threshold and strong coupling effects 
along the RG flows.
On the other hand, moving towards the intermediate region,
 both quantities
exhibit a strong increase and approach each other towards values of $\eta\sim\mathcal{O}(1)$ near 
$\Lambda_\mathrm{QCD}/\Lambda_\mathrm{F}\simeq 10^{-1}$;
the latter correspond to
$g_\Lambda$
values that trigger the onset of 
the quasi-fixed-point
behavior at the transition.

As a final remark on
	this intermediate scaling
	behavior, let us mention
	another line of reasoning
	that can be used to 
	qualitatively expect
	and understand the results presented above.
Since the region
between the deep Higgs and the
deep QCD regimes  
 corresponds to intermediate values of the mass parameter 
$\delta \epsilon$, it
can be expected to include
an almost-critical point
of minimal breaking 
of scale invariance.
The latter is in fact explicitely and severly broken
in the two opposite extrema of a deep Mexican-hat standard-model-like Higgs potential ($\delta\epsilon\ll-1$)
and of a very massive 
quasi-decoupled QCD-like meson
($\delta\epsilon\gg 1$).
Scale
invariance is a defining property of RG fixed points.
In the present model
the only complete fixed point featured
by our RG equations is the
Gaussian one.
The CEL solution in turn
describes the UV critical 
surface of this fixed point
in its vicinity,
that is, the 
weak-coupling parameterization
of the locus of points
corresponding to a minimal
breaking of scale invariance.
In fact, the relevant, i.e. mass-like, deformation
from the fixed point is zero
on this surface.
Hence, in the weakly coupled perturbative regime, the CEL solution describes a technically natural
theory, where the Higgs mass needs no fine tuning and stays 
small due to the approximate scale symmetry, which is broken only dynamically by QCD through
a Coleman-Weinberg mechanism.
This expectation must be confronted
with the necessity to
exit the mere perturbative
regime and follow the
theory at strong
coupling in the IR.
The results of this section
therefore can be interpreted 
as a first quantitative test
of this perturbative 
understanding, beyond
the weakly coupled regime.
Surprisingly, the CEL solution
appears to provide a fair first
approximation of the
location of the 
``natural'' theory
in the phase space 
of the model,
even if for 
$\Lambda=10^8$~GeV the
latter is close to
$g_\Lambda^2=\mathcal{O}(1)$.

Let us stress that the precise value of $\eta$ at 
the physical point in  Fig.~\ref{fig:CompareEtaFitTransition}
is not only determined by the IR value of
measured quantities, but also depends on
theoretical assumptions, such as
the precise RG flow of the standard model.
Shifting the $\eta$ curve
along the $\Lambda_\mathrm{QCD}/\Lambda_\mathrm{F}\simeq 10^{-2}$
axis is possible by changing the
RG evolution, thus
effectively changing $g_\Lambda$,
while keeping the IR parameters fixed.
Determining the RG flow is not 
only a matter of improving the 
approximations over
the nonperturbative domain of 
the theory, but also and
primarily a question
of \textit{defining}
the quantum
field theory beyond perturbation
theory.
As a possible embodyment of this fact
we can refer to the scenario of Ref.~\cite{Alessandro1,Alessandro2}
where new solutions of the RG equations
were constructed by appropriately
parametrizing the boundary
conditions on the scale-dependent
effective action.
These generalized solutions 
and the new parameters that they involve 
allow to
vary the RG flow between the UV scale
$\Lambda$ and the IR, while preserving the
observables at the latter scale.

\subsection{Electroweak Gauge Bosons}

So far, we have ignored the electroweak gauge sector, as it contributes 
only subdominantly to the interplay of the chiral transitions. In turn, 
however, the chiral transition, of course, affects the electroweak gauge 
sector strongly. This is well known and fully standard in the Higgs 
regime, where the chiral transition at the same time features the BEH 
mechanism, rendering part of the electroweak gauge bosons massive. But 
analogous results also hold in the QCD regime. In other words, the BEH 
mechanism does not need to be triggered by a carefully designed or even 
fine-tuned scalar potential, but is also operative if the vacuum expectation 
value of a composite effective scalar field is generated by other interactions, 
such as the QCD sector.  

As an illustration, let us include ingredients of the electroweak gauge 
sector on an elementary level in order to estimate the mass of the electroweak 
gauge bosons as a consequence of the chiral transition in the QCD regime.

For this, we use the flow equations extracted for the model (c.f.~Sect.~\ref{sec:RGfloweqs}), and amend them with the 
one-loop flow 
equation obtained for the $\mathrm{SU}(2)$ gauge coupling $g_\mathrm{EW}$. 
\begin{equation}
\partial_t g_\mathrm{EW}= -\frac{g_\mathrm{EW}^3}{(4\pi)^2}\left(\frac{22}{3}-\frac{4}{3}n_g -\frac{1}{6}\right).
\label{eq:EWgaugeBosonflow}
\end{equation}
The first term in the parenthesis comes from gauge boson loops, the second one 
from fermionic loops and the last contribution stems from scalar loops. Since in 
our analysis we focus on flows which start in the symmetric regime and end up in 
the broken one, no threshold functions accounting for the masses of the degrees 
of freedom are necessary for the gauge and fermionic loops in the symmetric 
regime. The scalar field however is massive in the SYM regime, and in this regime we will 
account for this effect by introducing the mass threshold corresponding to two 
internal scalar propagators of the Higgs field for this contribution,
\begin{equation}
-\frac{1}{6}\rightarrow -\frac{1}{6}\frac{1-\frac{\eta_{\phi}}{6}}{(1+\e)^2}.
\label{eq:EWgaugeFlowModification}
\end{equation}
This coincides with the standard perturbative contribution in the deep euclidean region $\epsilon\to0$.
In the following, we ignore the hypercharge sector and thus also the mass 
splitting of the $W$ and $Z$ bosons.

As soon as the flow enters the broken regime, the gauge bosons and fermions 
obtain masses, leading to a freeze-out of the flow in this regime. 
As a simple approximation, we use the value of the electroweak gauge coupling at 
the transition scale as the relevant IR value for an estimate of the 
$W$-boson mass.

The $W$ mass can then be read off straightforwardly, since it is fully 
determined by the vacuum expectation value $v$ and the freeze-out value of its 
gauge coupling $g_\mathrm{EW,IR}$ through
\begin{equation}
m_\mathrm{W}=\frac{1}{2}g_\mathrm{EW,IR}v.
\label{eq:WBosonMass}
\end{equation}
The initial conditions for the flow of the electroweak gauge coupling are chosen 
such that they coincide with the running in the standard model, namely its value 
at the $Z$ boson mass scale,
\begin{equation}
\frac{g_\mathrm{EW}^2}{4\pi}(M_Z) \simeq 0.034, \quad \mathrm{where}\quad 
M_Z=90.117\,\mathrm{GeV}.
\label{eq:EWGaugeInitial}
\end{equation}
\begin{figure}[htbp]
\includegraphics[width=0.5\textwidth]{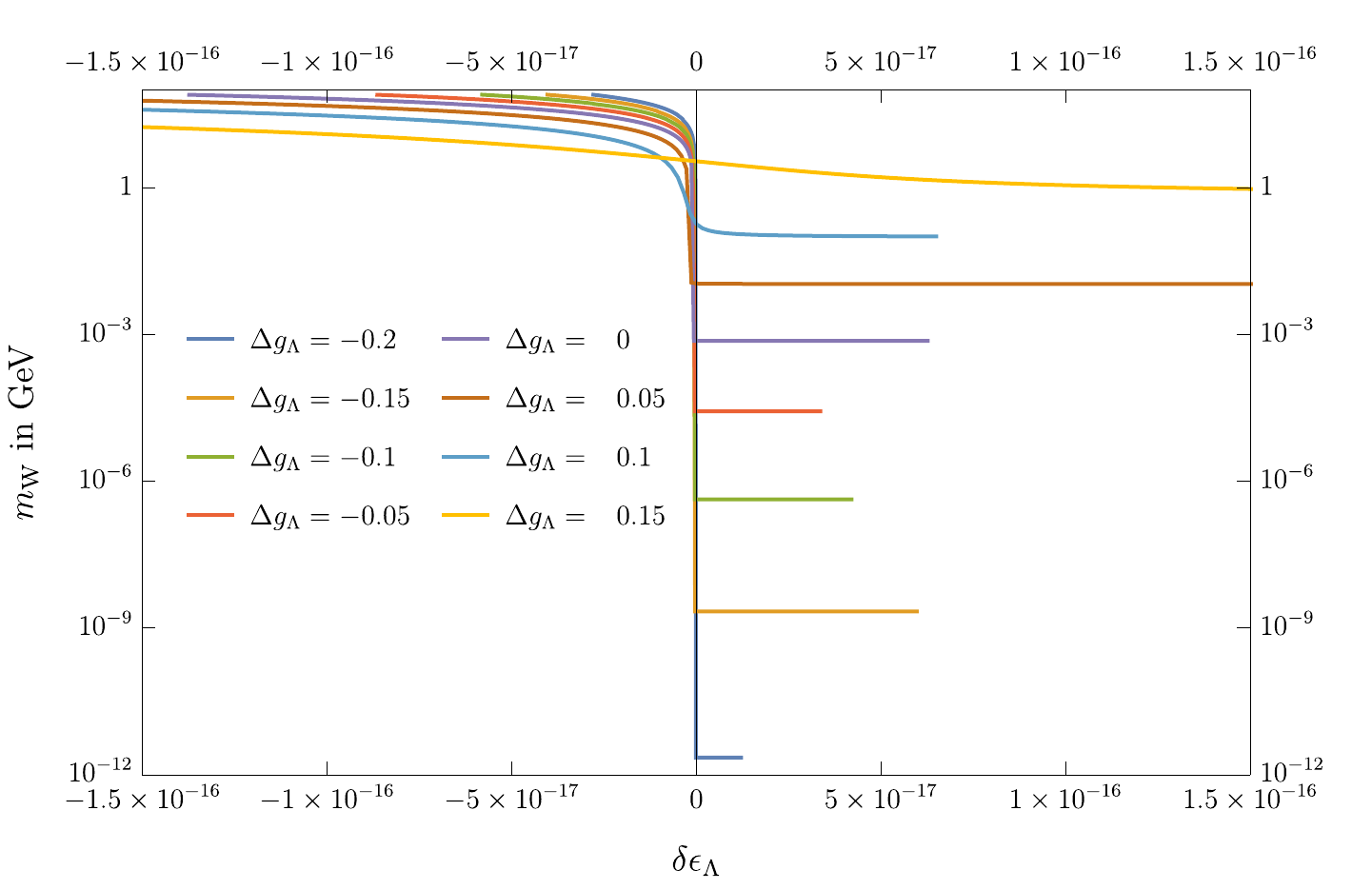}
\caption{$W$ boson mass for different initial values of the strong gauge 
couling at the UV scale $g_\Lambda$ as a function of the initial (dimensionless) 
mass parameter $\delta\e_\Lambda$. Because the gauged system features 
only a crossover, the $W$ boson mass is nowhere zero in the phase diagram. 
Instead, we observe a lower bound for the $W$ boson mass in the QCD regime 
(corresponding to $\delta\e_\Lambda \gg 0$). 
}
\label{fig:WBosonMass}
\end{figure}

In Fig.~\ref{fig:WBosonMass}, we depict the resulting mass values for the 
$W$ boson across the chiral transition region for various initial values of the 
QCD gauge coupling. In the Higgs regime, we observe the conventional strong 
dependence of the gauge boson mass on the control parameter which needs to be 
fine-tuned to obtain the Fermi scale and -- as a consequence -- the physical 
value of the $W$ boson mass. For any 
nonvanishing value of the QCD gauge coupling, 
the transition is actually a crossover, such that the vacuum expectation value 
as well as the $W$ boson mass never vanish. In fact, in the deep QCD regime, 
both dimensionful quantities approach a 
 non-zero value which is 
dominated by the QCD sector.

\begin{figure}[b]
\includegraphics[width=0.5\textwidth]{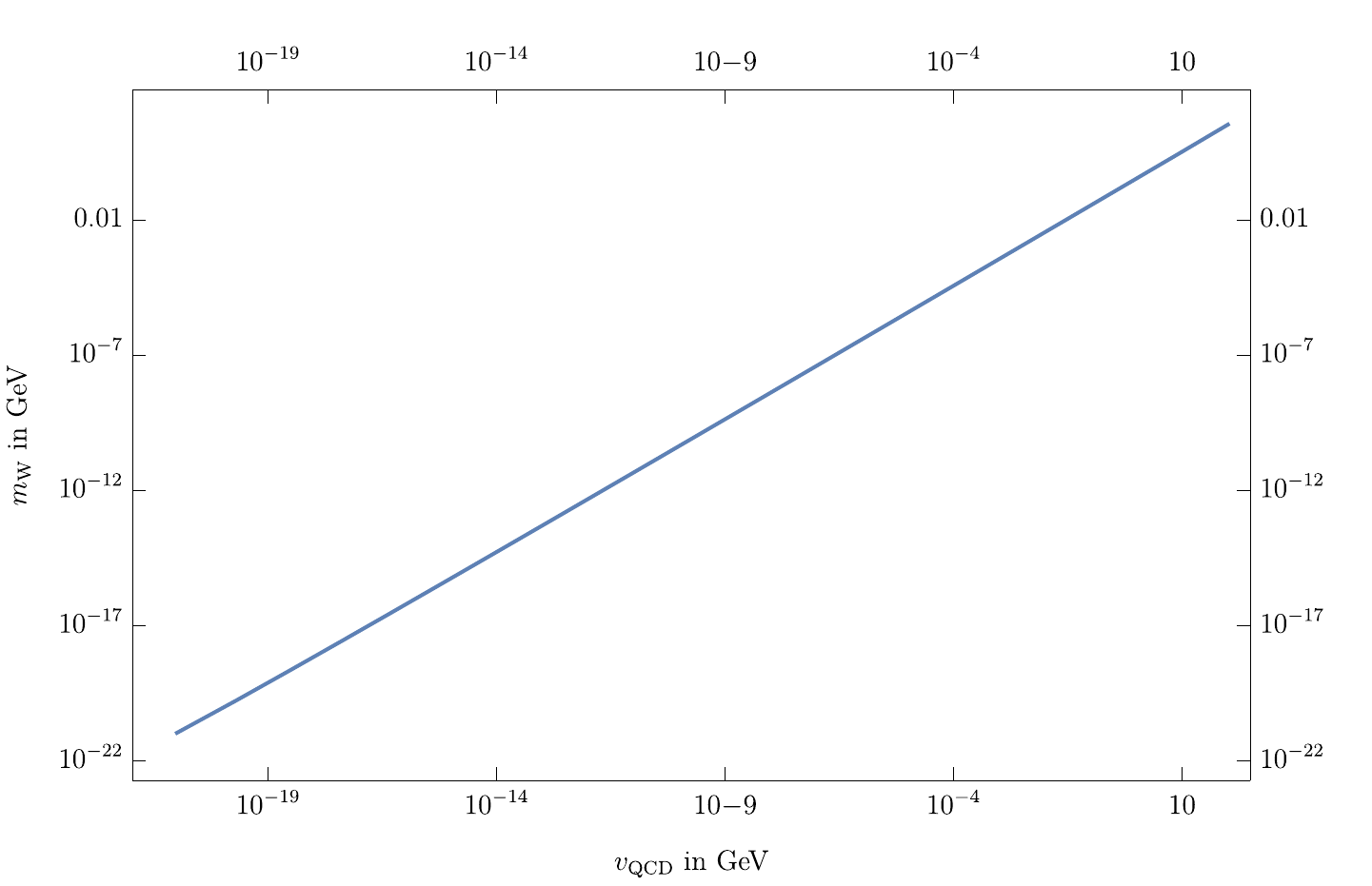}
\caption{The minimal $W$ boson mass in the deep QCD regime for different values 
of $g_\Lambda$, expressed through the chiral condensate $v_\mathrm{QCD}$. We 
extract the mass of the weak vector bosons in the far right half plane of 
Fig.~\ref{fig:WBosonMass} and plot it against the chiral vev obtained in the QCD 
regime. We observe an essentially linear dependence of the mass on the 
QCD-induced vacuum expectation value.
} 
\label{fig:DeepQCDWBosonMass}
\end{figure}

In this regime, the scale is set purely by the gauge coupling, or (after 
dimensional transmutation) by $\Lambda_\mathrm{QCD}$. In order to illustrate 
this dependence, we plot the resulting plateau value for the $W$ boson mass as 
a function of the QCD induced chiral condensate $v=v_\mathrm{QCD}$, see  Fig.~\ref{fig:DeepQCDWBosonMass}. We observe an essentially linear 
dependence.
This plateau value can be interpreted as the minimum possible value of 
the $W$ boson mass for the case that the BEH mechanism is not driven by a 
scalar Higgs potential but fully by the chiral QCD transition. This 
minimum $W$ boson mass value has, for instance, been estimated in
\cite{Quigg:2009xr,Quigg:2009vq}. However, a comparison of different results is 
inflicted by the fact that it involves not only a comparison of approximations 
within a theory, but rather a comparison of different theories that arise from 
fixing quantum field theories differently at different scales.

Taking our approach at face value, the scale in the QCD regime is set by 
fixing the gauge coupling to its physical value at a high scale, before 
evolving it into the IR. As discussed above, this leads to a rather small value 
of the chiral condensate $v\simeq 2.1$~MeV in the QCD regime as a consequence of 
the fact that the QCD sector runs with $\Nf=6$ (screening) massless flavors all 
the way down to the SSB regime. As a consequence, our straightforward estimate 
of the minimum $W$ boson mass value is $m_W= 0.74$~MeV, which 
is a rather small value compared to other estimates.

However, instead of matching our parameters to those of the physical 
standard model at the high scale, we could also perform a low scale matching: 
one possible choice (among many others) is to fix the couplings such that the 
chiral condensate acquires its physical value $v=f_\pi \simeq 93$~MeV also in 
the deep QCD regime. As a consequence, our prediction for the $W$ boson mass 
would be $m_W= 33$~MeV which is similar to that of \cite{Quigg:2009xr}.

\section{Conclusions}
\label{sec:conc}
The naturalness puzzles of the standard
model, e.g.~the gauge or the flavor hierarchy problems or the strong CP problem,
are typically addressed within  
pure model-building arenas.
In this sense, a common attitude is
the expectation that "solving" these 
problems amounts to
creatively devising new models able to
draw a simple straightforward prediction
of parameters that would remain 
otherwise free and
mysterious within the standard model. It is fair to say that this goal often
remains unattained, unfortunately.
Several popular
and largely appreciated proposals
for beyond-the-standard-model physics,
such as for instance technicolor or
axions,
actually require complex nonperturbative
computations to extract quantitative
predictions.

Even within the
standard model itself, it is difficult to quantitatively
assess how \textit{unnatural}
the measured value of some 
key parameter is; in general, this requires quite
nontrivial computations which 
go beyond the weakly-coupled perturbative
regime.
This problem is intrinsic in the
way in which naturalness questions are often posed, that relies on the ability to
connect the input values at some
microscopic UV scale to the output
values extracted from experiments.
The larger the gap between these two scales,
the harder the computations.
In the standard model, the main responsible building block for this difficulty is
the QCD subsector: 
the IR fluctuations of gluons are not
tamed by any BEH mechanism, contrary to the 
electroweak W/Z bosons, and their interactions grow stronger
at lower energies.

The task of properly
defining and quantifying  \textit{unnaturalness} is further aggravated by
the ambiguity in the choice of the 
UV scale at which the microscopic
parameters ultimately determining the measured 
observables are fixed. This second problem
is caused by the fact that the standard model
-- as we understand it -- is not a UV complete theory.
The main troublemaker in this case is the
electroweak sector, and specifically the
U$(1)_{\mathrm{Y}}$ gauge coupling. 
Finally, there is a priori no preferred 
probability distribution for the microscopic parameters with respect to which 
naturalness can be quantified; also in this work, we implicitly use a flat 
distribution as is standard for the discussion of tuning a control parameter to 
a (quantum) phase transition.

In the present work, we have focused on the
gauge-hierarchy problem as a prototypical
naturalness case study.
We have explored the possibility
to quantitatively describe this
feature of the standard model,
by adopting modern techniques 
inspired by effective field theories and the renormalization group,
while pushing them into the nonperturbative
territory.

In this manner, we have been able to study the interplay of the two
main sources of chiral symmetry breaking
of the standard model:
the electroweak Higgs interactions and 
the QCD sector. 
Whereas the breaking scales, i.e., the Fermi and the QCD scale, 
are several orders of magnitude apart, we observe a qualitative and 
quantitative interplay: the second-order transition of a pure Higgs-top-bottom 
sector is turned into a crossover by the presence of the QCD sector,
see Fig.~\ref{fig:TransitionVsCrossover}. 
We suggest to quantitatively characterize
the  deviation from Gaussianity of the Higgs-mass
fine-tuning problem by means of a 
pseudo-critical exponent $\eta$. A result of this study is that the QCD sector 
alleviates the naturalness problem naively quantified by the quadratic running 
of the Higgs mass parameter by a pseudo-critical exponent $\eta$ on the order of 
$\sim10$\% , see 
Fig.~\ref{fig:TotalPlotPhaseTrans}.  
Remarkably, we have observed that $\eta$
is nonvanishing even in case of negligible 
QCD corrections.

These conclusions have been drawn
within the usual
understanding of the standard model
as an effective theory below a suitable
UV cutoff, which in our study was set
at $\Lambda\sim10^8~$ GeV,
and with bare couplings at this scale
fully in the perturbative regime.
The only non-standard ingredient
we used in our investigation is
a non-perturbative improvement of the RG equations
to capture threshold effects in the matter
sector and the expected strong-coupling 
regime of the QCD gauge coupling.

As a useful ingredient of our analysis, we have managed to describe the 
fundamental Higgs field and a mesonic composite scalar field on the same 
footing. This is particularly insightful for the study of deformations of the 
standard model in which the Fermi and the QCD scale are shifted relative to 
each other. Specifically, an increase of the QCD scale demonstrates that the 
standard model can be continuosly connected to a deformed model where 
all scales are set by the QCD sector and the pseudo-critical exponent renders 
the connection between microscopic and macroscopic parameters natural. 

In the latter case, the theory appears
to enter into an almost scale invariant
regime which lies in a narrow intermediate 
window between the deep Higgs and
the deep QCD regimes.
This pseudo-critical region, 
where mass is expected to
be generated
by a Coleman-Weinberg
mechanism, has been shown
to be well approximated by a quasi-fixed point
condition of the Cheng-Eichten-Li 
type~\cite{Cheng:1973nv}, despite
the fact that for standard-model-like
IR observables this happens to
require values of the bare gauge coupling $g_\Lambda$ of order one.

This observation motivates
interest in applying similar methods
to investigate the naturalness problem
of extensions of the standard
model which are able to 
render the Coleman-Weinberg scenario
compatible with the measured values of
IR observables.
Among these, of particular interest 
are UV complete extensions, since
in these models the 
quasi-fixed-point behavior 
is expected to be continuously 
connected with the controlled UV
asymptotics for increasing values 
of $\Lambda$.

Of course, our treatment of the nonperturbative QCD sector is rather 
rudimental and relies on some model input in the pure gauge sector. 
Nevertheless, it is capable of connecting the microscopic description in terms 
of quarks and gluons to a long-range description on terms of a quark-meson 
model that can be matched quantitatively to chiral perturbation theory. Still, 
there is, of course, room for substantial improvement required for claiming a 
full quantitative control of the nonperturbative domain. For instance, in the
deep QCD regime we have estimated 
the radiatively generated mass
of the W boson.
We expect 
this estimate to be inflicted by the approximations 
and assumptions made in this nonperturbative sector. Another subtle issue that we have
not investigated in this work, is the possible
gauge and parametrization dependence
of our results, which we leave for future
developments.

Nevertheless, we believe that our approach offers a useful framework
for addressing the interplay of the chiral transitions in the standard model. 
As the transitions are considered to root in rather different sectors, such a 
treatment in a unified framework is valuable and has been missing so far.

\acknowledgments

We are grateful to Axel Maas,
Daniel Litim,
Roberto Percacci, Simon Schreyer, René Sondenheimer, Jan Pawlowski,
Gian Paolo Vacca, and Christof Wetterich 
for helpful discussions. 
This work has been funded by the Deutsche Forschungsgemeinschaft (DFG) under 
Grant Nos. 398579334 (Gi328/9-1) and 406116891 within the Research Training 
Group RTG 2522/1.
This project has also received funding from the European
Union’s Horizon 2020 research and innovation programme
under the Marie Skłodowska-Curie Grant Agreement
No. 754496.

\appendix
\section{Flow equations of the model}\label{app:FlowEquations}
The flow equations of the model are derived using the functional 
renormalization group flow equation for the effective action~\ref{eq:Wetterich}. 
Here all results are expressed in terms of dimensionless quantities, c.f.~\Eqref{eq:DimlessQuantities}.

For the Yukawa couplings we obtain
\begin{widetext}
		\begin{equation}
		\begin{aligned}
			\partial_t h^2=&\left(d-4+\eta_\mathrm{b}+\eta_\mathrm{t}+\eta_{\tilde{\phi}}\right)h^2+v_d h^4\left\{\lFB{d}{1}{1}{\topmassr}{\mu^2_\mathrm{m}}{\eta_\mathrm{t}}{\eta_{\tilde{\phi}}}+\lFB{d}{1}{1}{\botmassr}{\mu^2_\mathrm{m}}{\eta_\mathrm{b}}{\eta_{\tilde{\phi}}}+\lFB{d}{1}{1}{\topmassr}{\mu^2_\mathrm{m,r}}{\eta_\mathrm{t}}{\eta_{\tilde{\phi}}}\right.\\&\left.+\lFB{d}{1}{1}{\botmassr}{\mu^2_\mathrm{m,r}}{\eta_\mathrm{b}}{\eta_{\tilde{\phi}}}+\lFB{d}{1}{1}{\topmassr}{\mu^2_\mathrm{m}}{\eta_\mathrm{t}}{\eta_{\tilde{\phi}}}+\lFB{d}{1}{1}{\botmassr}{\mu^2_\mathrm{m}}{\eta_\mathrm{b}}{\eta_{\tilde{\phi}}}+\lFB{d}{1}{1}{\topmassr}{\mu^2_\mathrm{m}}{\eta_\mathrm{t}}{\eta_{\tilde{\phi}}}\right.\\&\left.+\lFB{d}{1}{1}{\botmassr}{\mu^2_\mathrm{m}}{\eta_\mathrm{b}}{\eta_{\tilde{\phi}}}+\lFB{d}{1}{1}{\topmassr}{\mu^2_\mathrm{G}}{\eta_\mathrm{t}}{\eta_\phi}+\lFB{d}{1}{1}{\topmassr}{\mu^2_\mathrm{H}}{\eta_\mathrm{t}}{\eta_\phi}+\lFB{d}{1}{1}{\topmassr}{\mu^2_\mathrm{G}}{\eta_\mathrm{t}}{\eta_\phi}\right.\\&\left.+\lFB{d}{1}{1}{\topmassr}{\mu^2_\mathrm{G}}{\eta_\mathrm{t}}{\eta_\phi}+\lFB{d}{1}{1}{\botmassr}{\mu^2_\mathrm{G}}{\eta_\mathrm{b}}{\eta_\phi}+\lFB{d}{1}{1}{\botmassr}{\mu^2_\mathrm{H}}{\eta_\mathrm{b}}{\eta_\phi}+\lFB{d}{1}{1}{\botmassr}{\mu^2_\mathrm{G}}{\eta_\mathrm{b}}{\eta_\phi}\right.\\&\left.+\lFB{d}{1}{1}{\botmassr}{\mu^2_\mathrm{G}}{\eta_\mathrm{b}}{\eta_\phi}\right\}-4(3+\xi)\frac{N_\mathrm{c}^2-1}{2N_\mathrm{c}}g^2h^2\left\{\lFB{d}{1}{1}{\topmassr}{0}{\eta_\mathrm{t}}{\eta_\mathrm{F}}+\lFB{d}{1}{1}{\botmassr}{0}{\eta_\mathrm{b}}{\eta_\mathrm{F}}\right\},
			\label{eq:dth2}
		\end{aligned}
	\end{equation}
\end{widetext}
\begin{equation}
\begin{aligned}
\partial_t h_\mathrm{t}^2=&\left(d-4+2\eta_\mathrm{t}+\eta_{\phi}\right)h_\mathrm{t}^2
\\&
-v_d h_\mathrm{t}^4
\left\{ \lFB{d}{1}{1}{\topmassr}{\mu^2_\mathrm{G}}{\eta_\mathrm{t}}{\eta_{\phi}}\right.\\&\left.-\lFB{d}{1}{1}{\topmassr}{\mu^2_\mathrm{H}}{\eta_\mathrm{t}}{\eta_{\phi}}\right\}\\&-2v_d h_\mathrm{t}^2h_\mathrm{b}^2\ \lFB{d}{1}{1}{\botmassr}{\mu^2_\mathrm{G}}{\eta_\mathrm{b}}{\eta_\phi}\\&-4(3+\xi)\frac{N_\mathrm{c}^2-1}{2N_\mathrm{c}}g^2h_\mathrm{t}^2\ \lFB{d}{1}{1}{\topmassr}{0}{\eta_\mathrm{t}}{\eta_\mathrm{F}},
\label{eq:dtht2}
\end{aligned}
\end{equation}
\begin{equation}
\begin{aligned}
\partial_t h_\mathrm{b}^2=&\left(d-4+2\eta_\mathrm{b}+\eta_{\phi}\right)h_\mathrm{b}^2
\\&-v_d h_\mathrm{b}^4\left\{\lFB{d}{1}{1}{\botmassr}{\mu^2_\mathrm{G}}{\eta_\mathrm{b}}{\eta_{\phi}}\right.
\\&\left.-\lFB{d}{1}{1}{\botmassr}{\mu^2_\mathrm{H}}{\eta_\mathrm{b}}{\eta_{\phi}}\right\}
\\&-2v_d h_\mathrm{b}^2h_\mathrm{t}^2\ 
\lFB{d}{1}{1}{\topmassr}{\mu^2_\mathrm{G}}{\eta_\mathrm{t}}{\eta_\phi}
\\&-4(3+\xi)\frac{N_\mathrm{c}^2-1}{2N_\mathrm{c}}g^2h_\mathrm{b}^2\ 
\lFB{d}{1}{1}{\botmassr}{0}{\eta_\mathrm{b}}{\eta_\mathrm{F}},
\label{eq:dthb2}
\end{aligned}
\end{equation}
where $\kappa$ denotes the minimum of the scalar potential. Here, we have 
assumed $\tilde{h}_t=\htop+h$ and treat $\htop$ and $h$ separately, accounting 
for the rebosonization procedure described above.\\
Finally, the anomalous dimensions of the bosonic and fermionic fields read
	\begin{equation}
	\begin{aligned}
	\eta_{\phi}=&2N_\mathrm{c}\frac{d_\gamma}{d}v_d \left( \htopt^2\left[\mFF{d}{\topmassr}{\eta_\mathrm{t}}- \frac{\htopt^2}{2}\kappa\mTF{d}{\topmassr}{\eta_\mathrm{t}}\right]\right.\\
	&\left. +\hbott^2\left[\mFF{d}{\botmassr}{\eta_\mathrm{b}}- \frac{\hbott^2}{2}\kappa\mTF{d}{\botmassr}{\eta_\mathrm{b}}\right]\right)
	\label{eq:etaphi},
	\end{aligned}
	\end{equation}
	\begin{equation}
	\begin{aligned}
	\eta_{\tilde{\phi}}=&2N_\mathrm{c}\frac{d_\gamma}{d}v_d 
h^2\left(\mFF{d}{\topmassr}{\eta_\mathrm{t}}+\mFF{d}{\botmassr}{\eta_\mathrm{b}
}\right.\\&\left.-\frac{h^2}{2}\kappa\left(\mTF{d}{\topmassr}{\eta_\mathrm{t}}
+\mTF{d}{\botmassr}{\eta_\mathrm{b}}\right)\right),
	\end{aligned}
	\end{equation}
	\begin{equation}
	\begin{aligned}
	\eta^{\mathrm{t}}_\mathrm{R}=&2\frac{v_d}{d}\htopt^2\left\{
	2\mFB{d}{1}{2}{\botmassr}{\mu^2_\mathrm{G}}{\eta_\mathrm{b}}{\eta_\phi}
	\right.\\	
	+&\left.\mFB{d}{1}{2}{\topmassr}{\mu^2_\mathrm{G}}{\eta_\mathrm{t}}{\eta_\phi}
	+\mFB{d}{1}{2}{\topmassr}{\mu^2_\mathrm{H}}{\eta_\mathrm{t}}{\eta_\phi}
	\right\} \\
	+&v_d h^2
	\left\{
	2\mFB{d}{1}{2}{\botmassr}{\mu^2_\mathrm{m}}{\eta_\mathrm{b}}{\eta_{\tilde{\phi}}}
	\right.
	\\
	+&\left.\mFB{d}{1}{2}{\topmassr}{\mu^2_\mathrm{m}}{\eta_\mathrm{t}}{\eta_{\tilde{\phi}}}+\mFB{d}{1}{2}{\topmassr}{\mu^2_\mathrm{m,r}}{\eta_\mathrm{t}}{\eta_{\tilde{\phi}}}
	\right\},
	\end{aligned}
	\end{equation}
	\begin{equation}
	\begin{aligned}
	\eta^{\mathrm{b}}_\mathrm{R}=&2\frac{v_d}{d}\hbott^2
	\left\{
	2\mFB{d}{1}{2}{\topmassr}{\mu^2_\mathrm{G}}{\eta_\mathrm{t}}{\eta_\phi}
	\right.\\
	+&\left.\mFB{d}{1}{2}{\botmassr}{\mu^2_\mathrm{G}}{\eta_\mathrm{b}}{\eta_\phi}+\mFB{d}{1}{2}{\botmassr}{\mu^2_\mathrm{H}}{\eta_\mathrm{b}}{\eta_\phi}
	\right\}\\
	+&v_d h^2
	\left\{
	2\mFB{d}{1}{2}{\topmassr}{\mu^2_\mathrm{m}}{\eta_\mathrm{t}}{\eta_{\tilde{\phi}}}
	\right.\\
	+&\left.\mFB{d}{1}{2}{\botmassr}{\mu^2_\mathrm{m}}{\eta_\mathrm{b}}{\eta_{\tilde{\phi}}}+\mFB{d}{1}{2}{\botmassr}{\mu^2_\mathrm{m,r}}{\eta_\mathrm{b}}{\eta_{\tilde{\phi}}}
	\right\},
	\end{aligned}
	\end{equation}
	\begin{equation}
	\begin{aligned}
	\eta^{\mathrm{t}}_\mathrm{L}=&2\frac{v_d}{d}\htopt^2
	\left(
	\mFB{d}{1}{2}{\topmassr}{\mu^2_\mathrm{G}}{\eta_\mathrm{t}}{\eta_\phi}\right.\\
	+&\left.\mFB{d}{1}{2}{\topmassr}{\mu^2_\mathrm{H}}{\eta_\mathrm{t}}{\eta_\phi}
	\right)
	\\
	+&2v_d \hbot^2
	\mFB{d}{1}{2}{\botmassr}{\mu^2_\mathrm{G}}{\eta_\mathrm{b}}{\eta_\phi}
	\\
	+&v_d h^2\left(\mFB{d}{1}{2}{\topmassr}{\mu^2_\mathrm{m}}{\eta_\mathrm{t}}{\eta_{\tilde{\phi}}}\right.\\
	+&\left.\mFB{d}{1}{2}{\topmassr}{\mu^2_\mathrm{m,r}}{\eta_\mathrm{t}}{\eta_{\tilde{\phi}}}
	+2\mFB{d}{1}{2}{\botmassr}{\mu^2_\mathrm{m}}{\eta_\mathrm{b}}{\eta_{\tilde{\phi}}}
	\right),
	\end{aligned}
	\end{equation}
	\begin{equation}
	\begin{aligned}
	\eta^{\mathrm{b}}_\mathrm{L}=&2\frac{v_d}{d}\hbott^2
	\left(
	\mFB{d}{1}{2}{\botmassr}{\mu^2_\mathrm{G}}{\eta_\mathrm{b}}{\eta_\phi}\right.\\
	+&\left.\mFB{d}{1}{2}{\botmassr}{\mu^2_\mathrm{H}}{\eta_\mathrm{b}}{\eta_\phi}
	\right)
	\\+&2v_d \htop^2 \mFB{d}{1}{2}{\topmassr}{\mu^2_\mathrm{G}}{\eta_\mathrm{t}}{\eta_\phi}
	\\+&v_d h^2\left(\mFB{d}{1}{2}{\botmassr}{\mu^2_\mathrm{m}}{\eta_\mathrm{b}}{\eta_{\tilde{\phi}}}\right.\\
	+&\left.\mFB{d}{1}{2}{\botmassr}{\mu^2_\mathrm{m,r}}{\eta_\mathrm{b}}{\eta_{\tilde{\phi}}}
	\right.\notag\\
	+&\left.2\mFB{d}{1}{2}{\topmassr}{\mu^2_\mathrm{m}}{\eta_\mathrm{t}}{\eta_{\tilde{\phi}}}
	\right).
	\end{aligned}
	\label{eq:anomalousDimensions}
	\end{equation}
	%
%
Here, it is useful to define anomalous dimensions of the top and bottom 
quark as
\begin{equation}
\eta^\mathrm{t}=\frac{1}{2}\left(\eta_\mathrm{L}^\mathrm{t}+\eta_\mathrm{R}
^\mathrm{t}\right),\qquad
\eta^\mathrm{b}=\frac{1}{2}\left(\eta_\mathrm{L}^\mathrm{b}+\eta_\mathrm{R}
^\mathrm{b}\right).
\end{equation}
\section{Scalar Spectrum}\label{app:ScalarSpectrum}
In our quartic approximation, the scalar potential for the different regimes 
are parametrized as
\begin{equation}
	\begin{aligned}
		U(\rho 
,\tilde{\rho})&=m^2\left(\rho+\tilde{\rho}\right)+\frac{\lambda_{1}}{2}
\left(\rho+\tilde{\rho}\right)^2+\lambda_{2}\rho\tilde{\rho},\qquad 
&&\text{SYM}\\ 
		U(\rho 
,\tilde{\rho})&=\frac{\lambda_{1}}{2}\left(\rho+\tilde{\rho}
-\kappa\right)^2+\lambda_{2}\rho\tilde{\rho},\qquad &&\text{SSB} 
	\end{aligned}
	\label{eq:B1}
\end{equation}
where we have used
\begin{equation}
	\begin{aligned}
		\rho &= \phi^{\ast a}\phi^a\quad\text{and}\quad\phi=\frac{1}{\sqrt{2}}
			\begin{pmatrix}
				\phi_1 + \mathrm{i}\phi_2\\
				\phi_4 + \mathrm{i}\phi_3
			\end{pmatrix},\\
		\tilde{\rho}&=\tilde{\phi}^{\ast a}\tilde{\phi}^a\quad\text{and}\quad\tilde{\phi}=\frac{1}{\sqrt{2}}
			\begin{pmatrix}
				\tilde{\phi}_1 + \mathrm{i}\tilde{\phi}_2\\
				\tilde{\phi}_4 + \mathrm{i}\tilde{\phi}_3
			\end{pmatrix},
	\end{aligned}
\end{equation}
and the vacuum configuration of the fields are chosen as
\begin{equation}
	\phi\vert_\text{vac}=
		\begin{pmatrix}
			0\\
			\sqrt{\kappa}
		\end{pmatrix},
	\qquad\tilde{\phi}\vert_\text{vac}=
		\begin{pmatrix}
			0\\0
		\end{pmatrix}.
\end{equation}
In the symmetric regime we have $\kappa=0$, wheras in the broken regime 
$\kappa$ is non-zero.

\begin{figure}
	\centering
	\includegraphics[width=0.49\textwidth]{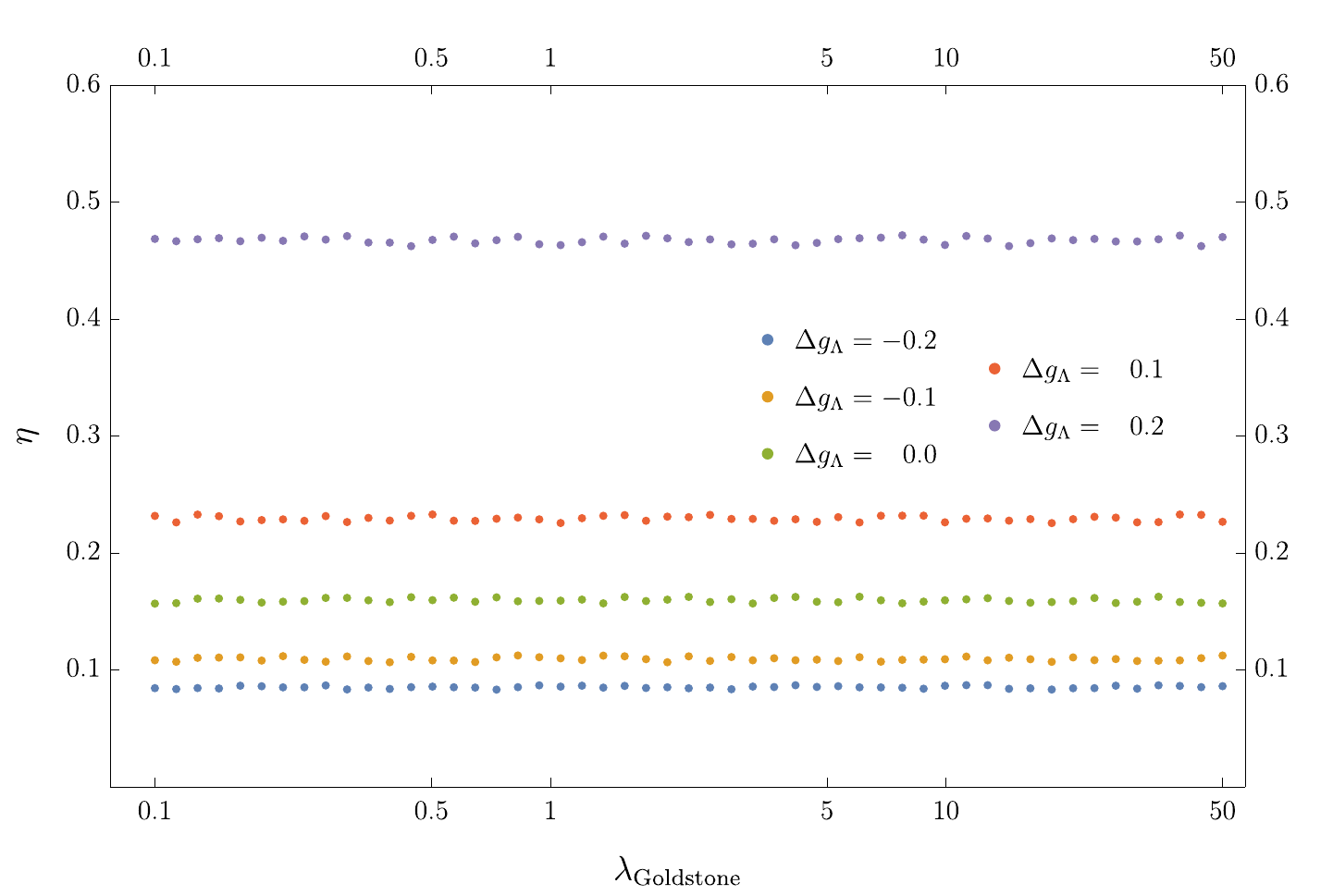}
	\caption{Dependence of the phase transition on the Goldstone 
mass parameter $\lambda_\text{Goldstone}$. We find no significant influence on 
the critical exponent $\eta$. This analysis has been performed for a fixed value 
of $g_\Lambda =0.719$.} \label{fig:GoldstoneParameter} 
\end{figure}

The mass spectrum of the various components of the two scalar fields 
(abbreviated by $f_i$) is given by
\begin{equation}
	M^2_{f_i}=\frac{\partial^2 U(\rho,\tilde{\rho})}{\partial {f_i}^2}\Bigg\vert_\text{vac}.
\end{equation}
In the symmetric regime, we find all components of the fields to have the same 
mass, given by
\begin{equation}
	M^2_{f_i}=m^2.
\end{equation}
In the broken regime, the components obtain different masses. The components of 
the $\tilde{\phi}$ field acquire a mass
\begin{equation}
	M^2_{\tilde{\phi}_i}=\lambda_{2}\kappa.
\end{equation}
For the $\phi$ field we find
\begin{equation}
	\begin{aligned}
		M^2_{\phi_4}&=2\lambda_1\kappa\\
		M^2_{\phi_i}&=0 \quad\qquad\text{for{ }} i\neq 4.
	\end{aligned}
\end{equation}
In comparison to the standard model, these massless Goldstone modes are 
an artifact of not having included the electroweak gauge sector. The BEH 
mechanism renders these modes together with the $W$ and $Z$ bosons massive. For 
our purposes, it suffices to model the BEH mechanism by giving a mass to these 
modes
\begin{equation}
	M^2_{\phi_i}=\lambda_\text{Goldstone}\kappa\quad\text{for{ }}i\neq 4
\end{equation}
proportional to the vacuum expectation value. In this way, they are 
unaffected in the symmetric regime, but decouple in the broken regime. This 
modeling of the BEH mechanism comes at the expense of a new parameter 
$\lambda_\text{Goldstone}$. We study the dependence of the phase transition 
on this  parameter, but find no significant 
influence  on the quantities of 
interest such as the pseudo-critical exponent, see Fig. 
\ref{fig:GoldstoneParameter}. 
For the computational purposes, we simply set this parameter to 
$\lambda_\text{Goldstone}=1$ in this work.

It is instructive to compare the scalar spectrum found here for the 
parametrization \eqref{eq:B1} of the potential 
with that derived from the complete quartic form \eqref{eq:QMScalarPotential}, 
as discussed in the literature \cite{Jungnickel:95}. While the spectrum in the 
symmetric regime is the same for both potentials, exhibiting 8 degrees of 
freedom with squared mass $m^2$, the masses in the broken regime differ for the 
two parametrizations. While we have one radial mode with mass 
$2\lambda_{1}\kappa$ for the SSB potentials in both cases, the full quartic 
form \eqref{eq:QMScalarPotential} gives rise to four massless Goldstone modes, 
and three modes with squared mass $\lambda_{2}\kappa$, whereas there are only 
 three Goldstone modes and four $\lambda_{2}\kappa$ modes for our 
potential \eqref{eq:B1}. The reason for this difference lies in the fact, 
that the structure of the additional trace invariant $\tau$ of 
\Eqref{eq:rhotaudef} is not properly taken into account by our parametrization 
in terms of $\rho$ and $\tilde\rho$. Comparing the resulting flow equations 
for the couplings in the scalar potential shows no difference in 
the mass parameter $m^2$ and quartic coupling $\lambda_{1}$ beta functions. The 
difference in the scalar spectrum, however, yields a slightly different flow 
equation for $\lambda_{2}$ as a consequence. To correct for this 
different parametrization, we use the  
beta function for the $\lambda_{2}$ coupling as it follows from the complete 
quartic parametrization \eqref{eq:QMScalarPotential}. Nevertheless, we have 
checked that the use of the beta function that would follow from \Eqref{eq:B1} 
has no significant influence on the results of this work. 

\begin{figure}[t]
	\includegraphics[width=0.48\textwidth]{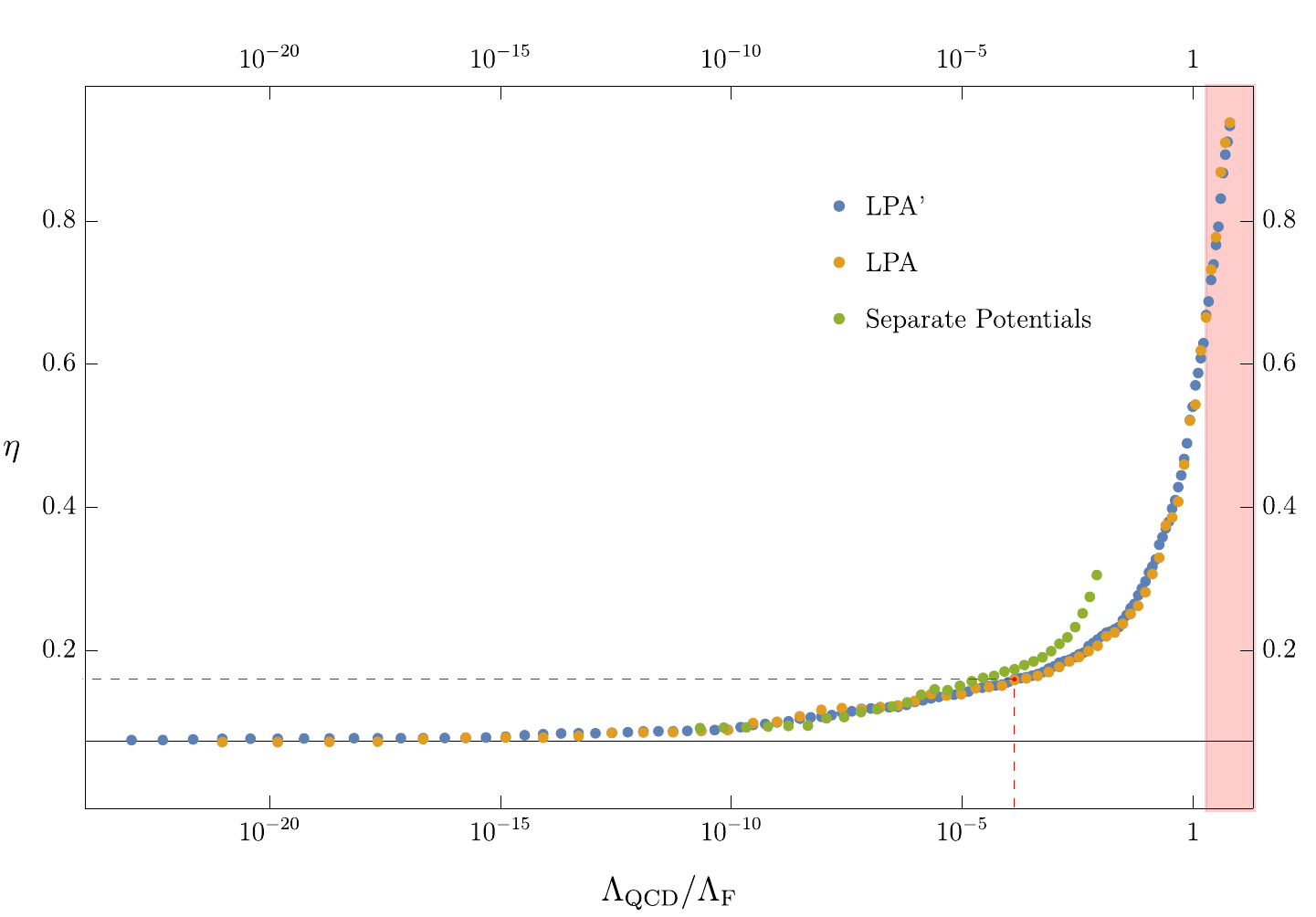}
	\caption{Comparison of the results of this work with the 
		approach used in \cite{Schmieden:Master} based on disentangled scalar 
		fields. Despite quantitative differences, we observe qualitative similarities 
		in the sense that an increasingly strong interacting gauge sector alleviates 
		the fine-tuning necessary in the model. The physical point, where the running of the strong gauge coupling coincides with the standard model is denoted by the red dashed line.}
	\label{fig:compareModels}
\end{figure}
\section{Comparison to an approach with separate 
scalar fields}\label{app:ComparisonToMaster}
In a previous study \cite{Schmieden:Master}, 
we have studied the interplay of the chiral transitions using a 
description where  
the mesonic degrees of freedom are not combined with the Higgs field into one 
collective field, but both the meson and Higgs potential are kept separate and 
disentangled. Whereas two distinct scalar potentials, $V$ for the mesons and 
$U$ for the Higgs field, offer the advantage of straightforwardly identifying 
the degrees of freedom, it is not as easy to indentify the \enquote{correct} 
vacuum.

In the disentangled description, it naively seems possible to have a Higgs potential in the broken regime while the 
meson potential is still symmetric, or vice versa. This is, however, 
modified by the fluctiations, since a non-vanishing 
vacuum expectation value in one of the potentials induces a linear term in the 
other potential through the Yukawa interactions, shifting the minimum away from 
vanishing field amplitude also in the 
other potential. In the study \cite{Schmieden:Master}, we have for 
simplicity assumed that the true vacuum state is given by a linear superposition of the two minima. Subsequently, an analysis 
analogous to the one in this work is performed. Concentrating on the 
chiral transitions and the pseudo-critical exponent, we find qualitatively 
similar results, i.e. the fine-tuning problem gets alleviated for growing 
gauge 
couplings in the UV, as is shown is Fig.~\ref{fig:compareModels}.
For more details and computational subtleties, we refer 
the reader to \cite{Schmieden:Master}.

\section{Threshold Functions}\label{app:threshold}
The various quantities $l^d_n(\cdot)$, $m^d_n(\cdot)$, etc., denote threshold 
functions. These quantities correspond to loop integrals of the functional 
renormalization group and parametrize the decoupling of massive modes from the 
flow equations. These quantities are numerically of order 1, their precise 
form depends on the regulator. Even though the have been frequently discussed 
in the literature \cite{Gies:14,Jungnickel:95,Gies:04,Berges:02}, we list the 
expressions for the threshold functions used in this work explicitly for 
completeness.

We define the regularized kinetic terms in momentum space for bosons and 
fermions as
\begin{equation*}
\begin{aligned}
P(q)&=q^2\left(1+r_{k,\mathrm{B}}(q)\right),\\
P_\mathrm{F}(q)&=q^2\left(1+r_{k,\mathrm{F}}(q)\right)^2 \\
&= q^2\left(1+r_{k,\mathrm{L}}(q)\right)\left(1+r_{k,\mathrm{R}}(q)\right),
\end{aligned}
\end{equation*}
where the last line is used for chiral fermions where  
regulator shape functions are introduced for the left- and right-handed part, 
respectively. The general definitions of the threshold functions can be found 
in the literature \cite{Gies:14,Jungnickel:95,Gies:04,Berges:02,Braun:2011pp}.

For the present work, we use the piecewise linear regulator \cite{Litim1} for 
the bosonic regulator shape function
\begin{equation*}
r_\mathrm{B}=\left(\frac{k^2}{q^2}-1\right)\Theta\left(k^2-q^2\right),
\end{equation*}
and define the fermionic one implicitly by
\begin{equation*}
(1+r_\mathrm{B})=(1+r_\mathrm{L})(1+r_\mathrm{R}), \quad 
r_\mathrm{L}=r_\mathrm{R}.\,
\end{equation*}
This regulator allows to compute the threshold functions 
analytically which is advantageous for the subsequent numerical integration of 
the flow equations. The explicit form reads
\begin{widetext}
\begin{equation*}
\begin{aligned}
\lB{d}{n}{\omega}{\eta_\phi}=&\frac{2(\delta_{n,0}+n)}{d}\frac{1-\frac{\eta_\phi}{d+2}}{\left(1+\omega\right)^{n+1}},\qquad\
\lF{d}{n}{\omega}{\eta_\psi}=\frac{2(\delta_{n,0}+n)}{d}\frac{1-\frac{\eta_\psi}{d+1}}{\left(1+\omega\right)^{n+1}},\\
\lFB{d}{n_1}{n_2}{\omega_1}{\omega_2}{\eta_\psi}{\eta_\phi}=&\frac{2}{d}\frac{1}{\left(1+\omega_1\right)^{n_1}\left(1+\omega_2\right)^{n_2}}\\&\left[\frac{n_1}{1+\omega_1}\left(1-\frac{\eta_\psi}{d+1}\right)+\frac{n_2}{1+\omega_2}\left(1-\frac{\eta_\phi}{d+2}\right)\right],\\
\lFBB{n_1}{n_2}{n_3}{\omega_1}{\omega_2}{\omega_3}{\eta_\psi}{\eta_{\phi_1}}{\eta_{\phi_2}}=&\frac{2}{d}\frac{1}{\left(1+\omega_1\right)^{n_1}\left(1+\omega_2\right)^{n_2}\left(1+\omega_3\right)^{n_3}}\\&\Bigg[\frac{n_1}{1+\omega_1}\left(1-\frac{\eta_\psi}{d+1}\right)+\frac{n_2}{1+\omega_2}\left(1-\frac{\eta_{\phi_1}}{d+2}\right)+\frac{n_3}{1+\omega_3}\left(1-\frac{\eta_{\phi_2}}{d+2}\right)\Bigg],\\
\m{d}{n_1}{n_2}{\omega_1}{\omega_2}{\eta_\phi}=&\frac{1}{\left(1+\omega_1\right)^{n_1}\left(1+\omega_2\right)^{n_2}},\qquad\
\mTF{d}{\omega}{\eta_\psi}=\frac{1}{\left(1+\omega\right)^4},\\
\mFF{d}{\omega}{\eta_\psi}=&\frac{1}{\left(1+\omega\right)^4}+\frac{1-\eta_\psi}{d-2}\frac{1}{\left(1+\omega\right)^3}-\left(\frac{1-\eta_\psi}{2d-4}+\frac{1}{4}\right)\frac{1}{\left(1+\omega\right)^2},\\
\mFB{d}{n_1}{n_2}{\omega_1}{\omega_2}{\eta_\psi}{\eta_\phi}=&\left(1-\frac{\eta_\phi}{d+1}\right)\frac{1}{\left(1+\omega_1\right)^{n_1}\left(1+\omega_2\right)^{n_2}},\\
\mFBtilde{d}{1}{1}{\omega_1}{\omega_2}{\eta_\psi}{\eta_\phi}=&\frac{2}{d-1}\frac{1}{\left(1+\omega_1\right)\left(1+\omega_2\right)}\left(\frac{1-\frac{\eta_\phi}{d+1}}{1+\omega_2}+\frac{1-\frac{\eta_\psi}{d}}{1+\omega_1}-\frac{1}{2}+\frac{\eta_\psi}{2d}\right).
\end{aligned}
\end{equation*}
\end{widetext}

\section{Check of universality violations}
\label{app:universalityviolation}

Since the ungauged limit of our model, the Higgs-top-bottom model, does not 
exhibit a UV complete limit, we have to define it with an explicit UV cutoff 
$\Lambda$. This introduces a scheme dependence of our results which can be 
mapped onto a dependence of the choice of the bare couplings. Still, as long as 
we study flows which spend a sufficiently long ``RG time'' near the Gaussian 
weak-coupling fixed point, the dependence of IR observables on the initial 
conditions (or on the scheme) becomes suppressed by inverse powers of the 
cutoff scale $\Lambda$. In the case that the limit $\Lambda\to\infty$ can be 
taken, universality violations vanish exactly. 

In practice, we need find a compromise between a sufficiently large $\Lambda$ 
such that universality violations remain quantitatively irrelevant, and a 
conveniently small $\Lambda$ in order to keep the numerical effort manageable. 
The latter is dominated by the need to ``solve the fine-tuning problem'' in 
practice simply by fine-tuning the initial control parameter in order to arrive 
at a Fermi scale $v\lll \Lambda$. For our purposes, the choice 
$\Lambda=10^8$~GeV has turned out to represent such a suitable compromise.

\begin{figure}[b]
	\includegraphics[width=0.49\textwidth]{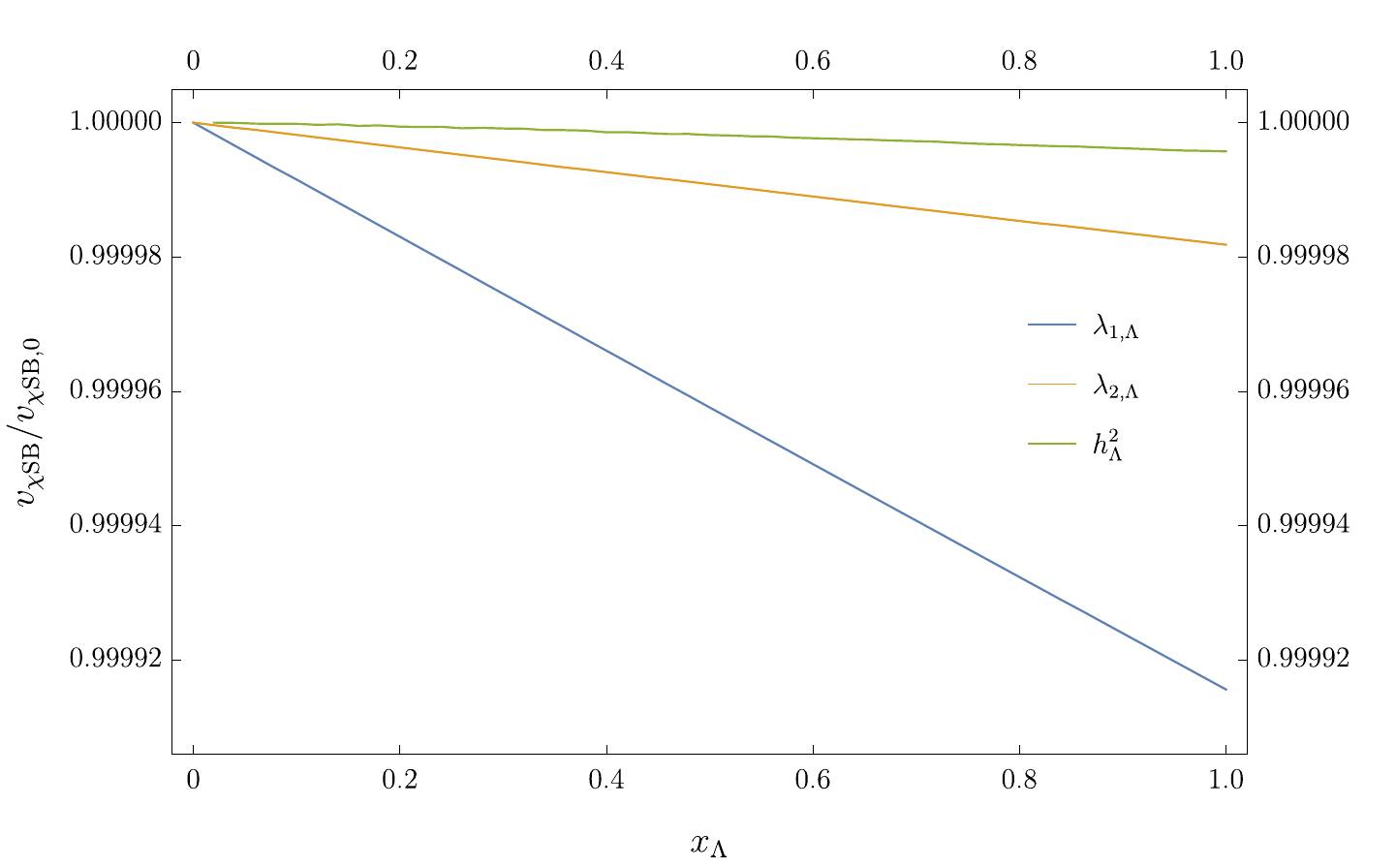}
	\caption{
		Test of universality violations:
		 normalized vacuum expectation 
		value in the deep QCD regime for different values of the marginal couplings at 
		the UV scale, 
		$x_\Lambda\in\{\lambda_{1,\Lambda},\lambda_{2,\Lambda},h_{\Lambda}^2\}$
		for the lower to upper curve respectively.
		The initial value of the gauge coupling $g_\Lambda$ is kept fixed for this 
		analysis. We observe that the vacuum expectation value changes at most on the 
		$0.01\%$ level for changes in the couplings studied in this work.}
	\label{fig:chiralIndependence}
\end{figure}

As an example of these universality violations, we show the influence of the 
marginal Yukawa and scalar self-interaction couplings on the vacuum 
expectation value $v$ in the deep QCD regime in Fig. 
\ref{fig:chiralIndependence}. Varying the initial conditions of these couplings 
at $\mathcal{O}(1)$, we observe that the vacuum expectation value $v$ varies 
merely on the sub-permille level at most. This illustrates the fact that the IR 
and the scale of all physical of physical observables is set only by the gauge 
coupling. 

\newpage
%
%
%
\bibliography{bibliography} 

\begin{thebibliography}{100}
\providecommand{\url}[1]{\texttt{#1}}
\providecommand{\urlprefix}{URL }
\providecommand{\eprint}[2][]{\url{#2}}

\bibitem{Nambu:1961tp}
Y.~Nambu and G.~Jona-Lasinio; {Dynamical Model of Elementary Particles Based on
  an Analogy with Superconductivity. 1.}; \emph{Phys. Rev.} (1961);
  \textbf{122}:345--358;
  \urlprefix\url{http://dx.doi.org/10.1103/PhysRev.122.345}.

\bibitem{Weinberg:1967tq}
S.~Weinberg; {A Model of Leptons}; \emph{Phys. Rev. Lett.} (1967);
  \textbf{19}:1264--1266;
  \urlprefix\url{http://dx.doi.org/10.1103/PhysRevLett.19.1264}.

\bibitem{Wilczek:2012sb}
F.~Wilczek; {Origins of Mass}; \emph{Central Eur. J. Phys.} (2012);
  \textbf{10}:1021--1037;
  \urlprefix\url{http://dx.doi.org/10.2478/s11534-012-0121-0};
  \eprint{1206.7114}.

\bibitem{BMW:2008jgk}
S.~Durr et~al.; {Ab-Initio Determination of Light Hadron Masses};
  \emph{Science} (2008); \textbf{322}:1224--1227;
  \urlprefix\url{http://dx.doi.org/10.1126/science.1163233};
  \eprint{0906.3599}.

\bibitem{ParticleDataGroup:2022pth}
R.~L. Workman et~al.; {Review of Particle Physics}; \emph{PTEP} (2022);
  \textbf{2022}:083C01; \urlprefix\url{http://dx.doi.org/10.1093/ptep/ptac097}.

\bibitem{Gerhold:2007yb}
P.~Gerhold and K.~Jansen; {The Phase structure of a chirally invariant lattice
  Higgs-Yukawa model for small and for large values of the Yukawa coupling
  constant}; \emph{JHEP} (2007); \textbf{09}:041;
  \urlprefix\url{http://dx.doi.org/10.1088/1126-6708/2007/09/041};
  \eprint{0705.2539}.

\bibitem{Gerhold:2007gx}
P.~Gerhold and K.~Jansen; {The Phase structure of a chirally invariant lattice
  Higgs-Yukawa model - numerical simulations}; \emph{JHEP} (2007);
  \textbf{10}:001;
  \urlprefix\url{http://dx.doi.org/10.1088/1126-6708/2007/10/001};
  \eprint{0707.3849}.

\bibitem{Gies:14}
H.~Gies and R.~Sondenheimer; {Higgs {M}ass {B}ounds from {R}enormalization
  {F}low for a {H}iggs-top-bottom model}; \emph{Eur. Phys. J. C} (2015);
  \textbf{75}(2):68;
  \urlprefix\url{http://dx.doi.org/10.1140/epjc/s10052-015-3284-1};
  \eprint{1407.8124}.

\bibitem{tHooft:1979rat}
G.~'t~Hooft; {Naturalness, chiral symmetry, and spontaneous chiral symmetry
  breaking}; \emph{NATO Sci. Ser. B} (1980); \textbf{59}:135--157;
  \urlprefix\url{http://dx.doi.org/10.1007/978-1-4684-7571-5_9}.

\bibitem{Giudice:2008bi}
G.~F. Giudice; {Naturally Speaking: The Naturalness Criterion and Physics at
  the LHC} (2008); 155--178;
  \urlprefix\url{http://dx.doi.org/10.1142/9789812779762_0010};
  \eprint{0801.2562}.

\bibitem{Grinbaum:2009sk}
A.~Grinbaum; {Which fine-tuning arguments are fine?}; \emph{Found. Phys.}
  (2012); \textbf{42}:615--631;
  \urlprefix\url{http://dx.doi.org/10.1007/s10701-012-9629-9};
  \eprint{0903.4055}.

\bibitem{Dine:2015xga}
M.~Dine; {Naturalness Under Stress}; \emph{Ann. Rev. Nucl. Part. Sci.} (2015);
  \textbf{65}:43--62;
  \urlprefix\url{http://dx.doi.org/10.1146/annurev-nucl-102014-022053};
  \eprint{1501.01035}.

\bibitem{Hossenfelder:2018ikr}
S.~Hossenfelder; {Screams for explanation: finetuning and naturalness in the
  foundations of physics}; \emph{Synthese} (2021); \textbf{198}(Suppl
  16):3727--3745; \urlprefix\url{http://dx.doi.org/10.1007/s11229-019-02377-5};
  \eprint{1801.02176}.

\bibitem{Quigg:2009xr}
C.~Quigg and R.~Shrock; {Gedanken Worlds without Higgs: QCD-Induced Electroweak
  Symmetry Breaking}; \emph{Phys. Rev. D} (2009); \textbf{79}:096002;
  \urlprefix\url{http://dx.doi.org/10.1103/PhysRevD.79.096002};
  \eprint{0901.3958}.

\bibitem{Quigg:2009vq}
C.~Quigg; {Unanswered Questions in the Electroweak Theory}; \emph{Ann. Rev.
  Nucl. Part. Sci.} (2009); \textbf{59}:505--555;
  \urlprefix\url{http://dx.doi.org/10.1146/annurev.nucl.010909.083126};
  \eprint{0905.3187}.

\bibitem{Iso:2017uuu}
S.~Iso, P.~D. Serpico and K.~Shimada; {QCD-Electroweak First-Order Phase
  Transition in a Supercooled Universe}; \emph{Phys. Rev. Lett.} (2017);
  \textbf{119}(14):141301;
  \urlprefix\url{http://dx.doi.org/10.1103/PhysRevLett.119.141301};
  \eprint{1704.04955}.

\bibitem{Frohlich:1980gj}
J.~Frohlich, G.~Morchio and F.~Strocchi; {Higgs phenomenon without a symmetry
  breaking order parameter}; \emph{Phys. Lett. B} (1980); \textbf{97}:249--252;
  \urlprefix\url{http://dx.doi.org/10.1016/0370-2693(80)90594-8}.

\bibitem{Maas:2017wzi}
A.~Maas; {Brout-Englert-Higgs physics: From foundations to phenomenology};
  \emph{Prog. Part. Nucl. Phys.} (2019); \textbf{106}:132--209;
  \urlprefix\url{http://dx.doi.org/10.1016/j.ppnp.2019.02.003};
  \eprint{1712.04721}.

\bibitem{Maas:2017xzh}
A.~Maas, R.~Sondenheimer and P.~T\"orek; {On the observable spectrum of
  theories with a Brout\textendash{}Englert\textendash{}Higgs effect};
  \emph{Annals Phys.} (2019); \textbf{402}:18--44;
  \urlprefix\url{http://dx.doi.org/10.1016/j.aop.2019.01.010};
  \eprint{1709.07477}.

\bibitem{Sondenheimer:2019idq}
R.~Sondenheimer; {Analytical relations for the bound state spectrum of gauge
  theories with a Brout-Englert-Higgs mechanism}; \emph{Phys. Rev. D} (2020);
  \textbf{101}(5):056006;
  \urlprefix\url{http://dx.doi.org/10.1103/PhysRevD.101.056006};
  \eprint{1912.08680}.

\bibitem{Maas:2020kda}
A.~Maas and R.~Sondenheimer; {Gauge-invariant description of the Higgs
  resonance and its phenomenological implications}; \emph{Phys. Rev. D} (2020);
  \textbf{102}:113001;
  \urlprefix\url{http://dx.doi.org/10.1103/PhysRevD.102.113001};
  \eprint{2009.06671}.

\bibitem{Lohitsiri:2019wpq}
N.~Lohitsiri and D.~Tong; {If the Weak Were Strong and the Strong Were Weak};
  \emph{SciPost Phys.} (2019); \textbf{7}(5):059;
  \urlprefix\url{http://dx.doi.org/10.21468/SciPostPhys.7.5.059};
  \eprint{1907.08221}.

\bibitem{Berger:2019yxb}
J.~Berger, A.~J. Long and J.~Turner; {Phase of confined electroweak force in
  the early Universe}; \emph{Phys. Rev. D} (2019); \textbf{100}(5):055005;
  \urlprefix\url{http://dx.doi.org/10.1103/PhysRevD.100.055005};
  \eprint{1906.05157}.

\bibitem{Schwinger:1957em}
J.~S. Schwinger; {A Theory of the Fundamental Interactions}; \emph{Annals
  Phys.} (1957); \textbf{2}:407--434;
  \urlprefix\url{http://dx.doi.org/10.1016/0003-4916(57)90015-5}.

\bibitem{Gell-Mann:1960mvl}
M.~Gell-Mann and M.~Levy; {The axial vector current in beta decay}; \emph{Nuovo
  Cim.} (1960); \textbf{16}:705;
  \urlprefix\url{http://dx.doi.org/10.1007/BF02859738}.

\bibitem{Jungnickel:1995fp}
D.~U. Jungnickel and C.~Wetterich; {Effective action for the chiral quark-meson
  model}; \emph{Phys. Rev. D} (1996); \textbf{53}:5142--5175;
  \urlprefix\url{http://dx.doi.org/10.1103/PhysRevD.53.5142};
  \eprint{hep-ph/9505267}.

\bibitem{Schaefer:1999em}
B.-J. Schaefer and H.-J. Pirner; {Renormalization group flow and equation of
  state of quarks and mesons}; \emph{Nucl. Phys. A} (1999);
  \textbf{660}:439--474;
  \urlprefix\url{http://dx.doi.org/10.1016/S0375-9474(99)00409-1};
  \eprint{nucl-th/9903003}.

\bibitem{Braun:2003ii}
J.~Braun, K.~Schwenzer and H.-J. Pirner; {Linking the quark meson model with
  QCD at high temperature}; \emph{Phys. Rev. D} (2004); \textbf{70}:085016;
  \urlprefix\url{http://dx.doi.org/10.1103/PhysRevD.70.085016};
  \eprint{hep-ph/0312277}.

\bibitem{Resch:2017vjs}
S.~Resch, F.~Rennecke and B.-J. Schaefer; {Mass sensitivity of the three-flavor
  chiral phase transition}; \emph{Phys. Rev. D} (2019); \textbf{99}(7):076005;
  \urlprefix\url{http://dx.doi.org/10.1103/PhysRevD.99.076005};
  \eprint{1712.07961}.

\bibitem{Eser:2018jqo}
J.~Eser, F.~Divotgey, M.~Mitter and D.~H. Rischke; {Low-energy limit of the
  $O(4)$ quark-meson model from the functional renormalization group approach};
  \emph{Phys. Rev. D} (2018); \textbf{98}(1):014024;
  \urlprefix\url{http://dx.doi.org/10.1103/PhysRevD.98.014024};
  \eprint{1804.01787}.

\bibitem{Divotgey:2019xea}
F.~Divotgey, J.~Eser and M.~Mitter; {Dynamical generation of low-energy
  couplings from quark-meson fluctuations}; \emph{Phys. Rev. D} (2019);
  \textbf{99}(5):054023;
  \urlprefix\url{http://dx.doi.org/10.1103/PhysRevD.99.054023};
  \eprint{1901.02472}.

\bibitem{Braun:2019aow}
J.~Braun, M.~Leonhardt and M.~Pospiech; {Fierz-complete NJL model study III:
  Emergence from quark-gluon dynamics}; \emph{Phys. Rev. D} (2020);
  \textbf{101}(3):036004;
  \urlprefix\url{http://dx.doi.org/10.1103/PhysRevD.101.036004};
  \eprint{1909.06298}.

\bibitem{Gross:73}
D.~J. Gross and F.~Wilczek; Ultraviolet {B}ehavior of {N}on-{A}belian {G}auge
  {T}heories; \emph{Phys. Rev. Lett.} (1973); \textbf{30}:1343--1346;
  \urlprefix\url{http://dx.doi.org/10.1103/PhysRevLett.30.1343}.

\bibitem{Politzer:73}
H.~D. Politzer; Reliable {P}erturbative {R}esults for {S}trong {I}nteractions?;
  \emph{Phys. Rev. Lett.} (1973); \textbf{30}:1346--1349;
  \urlprefix\url{http://dx.doi.org/10.1103/PhysRevLett.30.1346}.

\bibitem{Pawlowski:1996ch}
J.~M. Pawlowski; {Exact flow equations and the U(1) problem}; \emph{Phys. Rev.
  D} (1998); \textbf{58}:045011;
  \urlprefix\url{http://dx.doi.org/10.1103/PhysRevD.58.045011};
  \eprint{hep-th/9605037}.

\bibitem{Gies:2006nz}
H.~Gies, J.~Jaeckel, J.~M. Pawlowski and C.~Wetterich; {Do instantons like a
  colorful background?}; \emph{Eur. Phys. J. C} (2007); \textbf{49}:997--1010;
  \urlprefix\url{http://dx.doi.org/10.1140/epjc/s10052-006-0178-2};
  \eprint{hep-ph/0608171}.

\bibitem{Pisarski:2019upw}
R.~D. Pisarski and F.~Rennecke; {Multi-instanton contributions to anomalous
  quark interactions}; \emph{Phys. Rev. D} (2020); \textbf{101}(11):114019;
  \urlprefix\url{http://dx.doi.org/10.1103/PhysRevD.101.114019};
  \eprint{1910.14052}.

\bibitem{Fejos:2020pcg}
G.~Fej\H{o}s; {Perturbative RG analysis of the condensate dependence of the
  axial anomaly in the three flavor linear sigma model}; \emph{Symmetry}
  (2021); \textbf{13}(3):488;
  \urlprefix\url{http://dx.doi.org/10.3390/sym13030488}; \eprint{2012.08706}.

\bibitem{Braun:2020mhk}
J.~Braun, M.~Leonhardt, J.~M. Pawlowski and D.~Rosenbl\"uh; {Chiral and
  effective $U(1)_{\rm A}$ symmetry restoration in QCD} (2020);
  \eprint{2012.06231}.

\bibitem{Jungnickel:95}
D.~Jungnickel and C.~Wetterich; {Effective action for the chiral quark-meson
  model}; \emph{Phys. Rev. D} (1996); \textbf{53}:5142--5175;
  \urlprefix\url{http://dx.doi.org/10.1103/PhysRevD.53.5142};
  \eprint{hep-ph/9505267}.

\bibitem{Schreyer:2020}
S.~Schreyer; Global stability of scalar potentials in asymptotically safe
  particle models (2020);
  \urlprefix\url{https://nbn-resolving.org/urn:nbn:de:gbv:27-dbt-20201006-093946-005}.

\bibitem{Tornqvist:2005yp}
N.~A. Tornqvist; {Mixing the strong and E-W Higgs sectors}; \emph{Phys. Lett.
  B} (2005); \textbf{619}:145--148;
  \urlprefix\url{http://dx.doi.org/10.1016/j.physletb.2005.05.058};
  \eprint{hep-ph/0504204}.

\bibitem{Schumacher:2011gi}
M.~Schumacher; {Structure of Scalar Mesons and the Higgs Sector of Strong
  Interaction}; \emph{J. Phys. G} (2011); \textbf{38}:083001;
  \urlprefix\url{http://dx.doi.org/10.1088/0954-3899/38/8/083001};
  \eprint{1106.1015}.

\bibitem{Schmieden:Master}
R.~Schmieden; {{A}spects of the {G}auge {H}ierarchy of the {S}tandard {M}odel};
  \emph{Master Thesis} (2020); Jena;
  \urlprefix\url{http://dx.doi.org/10.22032/dbt.47615}.

\bibitem{Miransky:1988xi}
V.~A. Miransky, M.~Tanabashi and K.~Yamawaki; {Dynamical Electroweak Symmetry
  Breaking with Large Anomalous Dimension and t Quark Condensate}; \emph{Phys.
  Lett. B} (1989); \textbf{221}:177--183;
  \urlprefix\url{http://dx.doi.org/10.1016/0370-2693(89)91494-9}.

\bibitem{Miransky:1989ds}
V.~A. Miransky, M.~Tanabashi and K.~Yamawaki; {Is the t Quark Responsible for
  the Mass of W and Z Bosons?}; \emph{Mod. Phys. Lett. A} (1989);
  \textbf{4}:1043; \urlprefix\url{http://dx.doi.org/10.1142/S0217732389001210}.

\bibitem{Bardeen:1989ds}
W.~A. Bardeen, C.~T. Hill and M.~Lindner; {Minimal Dynamical Symmetry Breaking
  of the Standard Model}; \emph{Phys. Rev. D} (1990); \textbf{41}:1647;
  \urlprefix\url{http://dx.doi.org/10.1103/PhysRevD.41.1647}.

\bibitem{Hill:2002ap}
C.~T. Hill and E.~H. Simmons; {Strong Dynamics and Electroweak Symmetry
  Breaking}; \emph{Phys. Rept.} (2003); \textbf{381}:235--402;
  \urlprefix\url{http://dx.doi.org/10.1016/S0370-1573(03)00140-6}; [Erratum:
  Phys.Rept. 390, 553--554 (2004)]; \eprint{hep-ph/0203079}.

\bibitem{vonSmekal:1997ohs}
L.~von Smekal, R.~Alkofer and A.~Hauck; {The Infrared behavior of gluon and
  ghost propagators in Landau gauge QCD}; \emph{Phys. Rev. Lett.} (1997);
  \textbf{79}:3591--3594;
  \urlprefix\url{http://dx.doi.org/10.1103/PhysRevLett.79.3591};
  \eprint{hep-ph/9705242}.

\bibitem{Gies:02a}
H.~Gies; {Running coupling in {Y}ang-{M}ills theory: {A} flow equation study};
  \emph{Phys. Rev. D} (2002); \textbf{66}:025006;
  \urlprefix\url{http://dx.doi.org/10.1103/PhysRevD.66.025006};
  \eprint{hep-th/0202207}.

\bibitem{Fischer:2003rp}
C.~S. Fischer and R.~Alkofer; {Nonperturbative propagators, running coupling
  and dynamical quark mass of Landau gauge QCD}; \emph{Phys. Rev. D} (2003);
  \textbf{67}:094020;
  \urlprefix\url{http://dx.doi.org/10.1103/PhysRevD.67.094020};
  \eprint{hep-ph/0301094}.

\bibitem{Pawlowski:2003hq}
J.~M. Pawlowski, D.~F. Litim, S.~Nedelko and L.~von Smekal; {Infrared behavior
  and fixed points in Landau gauge QCD}; \emph{Phys. Rev. Lett.} (2004);
  \textbf{93}:152002;
  \urlprefix\url{http://dx.doi.org/10.1103/PhysRevLett.93.152002};
  \eprint{hep-th/0312324}.

\bibitem{Fischer:2008uz}
C.~S. Fischer, A.~Maas and J.~M. Pawlowski; {On the infrared behavior of Landau
  gauge Yang-Mills theory}; \emph{Annals Phys.} (2009);
  \textbf{324}:2408--2437;
  \urlprefix\url{http://dx.doi.org/10.1016/j.aop.2009.07.009};
  \eprint{0810.1987}.

\bibitem{Weber:2011nw}
A.~Weber; {Epsilon Expansion for Infrared Yang-Mills theory in Landau Gauge};
  \emph{Phys. Rev. D} (2012); \textbf{85}:125005;
  \urlprefix\url{http://dx.doi.org/10.1103/PhysRevD.85.125005};
  \eprint{1112.1157}.

\bibitem{Reinosa:2017qtf}
U.~Reinosa, J.~Serreau, M.~Tissier and N.~Wschebor; {How nonperturbative is the
  infrared regime of Landau gauge Yang-Mills correlators?}; \emph{Phys. Rev. D}
  (2017); \textbf{96}(1):014005;
  \urlprefix\url{http://dx.doi.org/10.1103/PhysRevD.96.014005};
  \eprint{1703.04041}.

\bibitem{Gies:04}
H.~Gies and C.~Wetterich; Universality of spontaneous chiral symmetry breaking
  in gauge theories; \emph{Physical Review D} (2004); \textbf{69}(2);
  \urlprefix\url{http://dx.doi.org/10.1103/physrevd.69.025001};
  \eprint{hep-th/0209183}.

\bibitem{Miransky:1984ef}
V.~A. Miransky; {Dynamics of Spontaneous Chiral Symmetry Breaking and Continuum
  Limit in Quantum Electrodynamics}; \emph{Nuovo Cim. A} (1985);
  \textbf{90}:149--170; \urlprefix\url{http://dx.doi.org/10.1007/BF02724229}.

\bibitem{Gies:2005as}
H.~Gies and J.~Jaeckel; {Chiral phase structure of QCD with many flavors};
  \emph{Eur. Phys. J. C} (2006); \textbf{46}:433--438;
  \urlprefix\url{http://dx.doi.org/10.1140/epjc/s2006-02475-0};
  \eprint{hep-ph/0507171}.

\bibitem{Braun:2006jd}
J.~Braun and H.~Gies; {Chiral phase boundary of QCD at finite temperature};
  \emph{JHEP} (2006); \textbf{06}:024;
  \urlprefix\url{http://dx.doi.org/10.1088/1126-6708/2006/06/024};
  \eprint{hep-ph/0602226}.

\bibitem{Braun:2010qs}
J.~Braun, C.~S. Fischer and H.~Gies; {Beyond Miransky Scaling}; \emph{Phys.
  Rev. D} (2011); \textbf{84}:034045;
  \urlprefix\url{http://dx.doi.org/10.1103/PhysRevD.84.034045};
  \eprint{1012.4279}.

\bibitem{Mitter:2014wpa}
M.~Mitter, J.~M. Pawlowski and N.~Strodthoff; {Chiral symmetry breaking in
  continuum QCD}; \emph{Phys. Rev. D} (2015); \textbf{91}:054035;
  \urlprefix\url{http://dx.doi.org/10.1103/PhysRevD.91.054035};
  \eprint{1411.7978}.

\bibitem{Braun:2014ata}
J.~Braun, L.~Fister, J.~M. Pawlowski and F.~Rennecke; {From Quarks and Gluons
  to Hadrons: Chiral Symmetry Breaking in Dynamical QCD}; \emph{Phys. Rev. D}
  (2016); \textbf{94}(3):034016;
  \urlprefix\url{http://dx.doi.org/10.1103/PhysRevD.94.034016};
  \eprint{1412.1045}.

\bibitem{Eichmann:2016yit}
G.~Eichmann, H.~Sanchis-Alepuz, R.~Williams, R.~Alkofer and C.~S. Fischer;
  {Baryons as relativistic three-quark bound states}; \emph{Prog. Part. Nucl.
  Phys.} (2016); \textbf{91}:1--100;
  \urlprefix\url{http://dx.doi.org/10.1016/j.ppnp.2016.07.001};
  \eprint{1606.09602}.

\bibitem{Binosi:2016wcx}
D.~Binosi, L.~Chang, J.~Papavassiliou, S.-X. Qin and C.~D. Roberts; {Natural
  constraints on the gluon-quark vertex}; \emph{Phys. Rev. D} (2017);
  \textbf{95}(3):031501;
  \urlprefix\url{http://dx.doi.org/10.1103/PhysRevD.95.031501};
  \eprint{1609.02568}.

\bibitem{Cyrol:2017ewj}
A.~K. Cyrol, M.~Mitter, J.~M. Pawlowski and N.~Strodthoff; {Nonperturbative
  quark, gluon, and meson correlators of unquenched QCD}; \emph{Phys. Rev. D}
  (2018); \textbf{97}(5):054006;
  \urlprefix\url{http://dx.doi.org/10.1103/PhysRevD.97.054006};
  \eprint{1706.06326}.

\bibitem{Wetterich:93}
C.~Wetterich; {Exact evolution equation for the effective potential};
  \emph{Phys. Lett. B} (1993); \textbf{301}:90--94;
  \urlprefix\url{http://dx.doi.org/10.1016/0370-2693(93)90726-X};
  \eprint{1710.05815}.

\bibitem{Ellwanger:1993mw}
U.~Ellwanger; {FLow equations for N point functions and bound states}; \emph{Z.
  Phys. C} (1994); \textbf{62}:503--510;
  \urlprefix\url{http://dx.doi.org/10.1007/BF01555911};
  \eprint{hep-ph/9308260}.

\bibitem{Morris:1993qb}
T.~R. Morris; {The Exact renormalization group and approximate solutions};
  \emph{Int. J. Mod. Phys. A} (1994); \textbf{9}:2411--2450;
  \urlprefix\url{http://dx.doi.org/10.1142/S0217751X94000972};
  \eprint{hep-ph/9308265}.

\bibitem{Bonini:1992vh}
M.~Bonini, M.~D'Attanasio and G.~Marchesini; {Perturbative renormalization and
  infrared finiteness in the Wilson renormalization group: The Massless scalar
  case}; \emph{Nucl. Phys. B} (1993); \textbf{409}:441--464;
  \urlprefix\url{http://dx.doi.org/10.1016/0550-3213(93)90588-G};
  \eprint{hep-th/9301114}.

\bibitem{Berges:02}
J.~Berges, N.~Tetradis and C.~Wetterich; Non-perturbative renormalization flow
  in quantum field theory and statistical physics; \emph{Physics Reports}
  (2002); \textbf{363}(4-6):223–386;
  \urlprefix\url{http://dx.doi.org/10.1016/s0370-1573(01)00098-9}.

\bibitem{Gies:06}
H.~Gies; {Introduction to the functional {RG} and applications to gauge
  theories}; \emph{Lect. Notes Phys.} (2012); \textbf{852}:287--348;
  \urlprefix\url{http://dx.doi.org/10.1007/978-3-642-27320-9\_6};
  \eprint{hep-ph/0611146}.

\bibitem{Pawlowski:2005xe}
J.~M. Pawlowski; {Aspects of the functional renormalisation group};
  \emph{Annals Phys.} (2007); \textbf{322}:2831--2915;
  \urlprefix\url{http://dx.doi.org/10.1016/j.aop.2007.01.007};
  \eprint{hep-th/0512261}.

\bibitem{Braun:2011pp}
J.~Braun; {Fermion Interactions and Universal Behavior in Strongly Interacting
  Theories}; \emph{J. Phys. G} (2012); \textbf{39}:033001;
  \urlprefix\url{http://dx.doi.org/10.1088/0954-3899/39/3/033001};
  \eprint{1108.4449}.

\bibitem{Kopietz:2010zz}
P.~Kopietz, L.~Bartosch and F.~Sch\"utz; \emph{{Introduction to the functional
  renormalization group}}; vol. 798 (2010);
  \urlprefix\url{http://dx.doi.org/10.1007/978-3-642-05094-7}.

\bibitem{Dupuis:2020fhh}
N.~Dupuis, L.~Canet, A.~Eichhorn, W.~Metzner, J.~M. Pawlowski, M.~Tissier and
  N.~Wschebor; {The nonperturbative functional renormalization group and its
  applications}; \emph{Phys. Rept.} (2021); \textbf{910}:1--114;
  \urlprefix\url{http://dx.doi.org/10.1016/j.physrep.2021.01.001};
  \eprint{2006.04853}.

\bibitem{Caswell:1974cj}
W.~E. Caswell and F.~Wilczek; {On the Gauge Dependence of Renormalization Group
  Parameters}; \emph{Phys. Lett. B} (1974); \textbf{49}:291--292;
  \urlprefix\url{http://dx.doi.org/10.1016/0370-2693(74)90437-7}.

\bibitem{Ellwanger:95}
U.~Ellwanger, M.~Hirsch and A.~Weber; {Flow equations for the relevant part of
  the pure {Y}ang-{M}ills action}; \emph{Z. Phys. C} (1996);
  \textbf{69}:687--698;
  \urlprefix\url{http://dx.doi.org/10.1007/s002880050073};
  \eprint{hep-th/9506019}.

\bibitem{Litim:98}
D.~F. Litim and J.~M. Pawlowski; Flow equations for Yang-Mills theories in
  general axial gauges; \emph{Physics Letters B} (1998);
  \textbf{435}(1-2):181--188;
  \urlprefix\url{http://dx.doi.org/10.1016/s0370-2693(98)00761-8}.

\bibitem{Litim1}
D.~F. Litim; Optimized renormalization group flows; \emph{Physical Review D}
  (2001); \textbf{64}(10);
  \urlprefix\url{http://dx.doi.org/10.1103/physrevd.64.105007}.

\bibitem{Borchardt:2016kco}
J.~Borchardt and A.~Eichhorn; {Universal behavior of coupled order parameters
  below three dimensions}; \emph{Phys. Rev. E} (2016); \textbf{94}(4):042105;
  \urlprefix\url{http://dx.doi.org/10.1103/PhysRevE.94.042105};
  \eprint{1606.07449}.

\bibitem{Beekman:2019sof}
A.~J. Beekman and G.~Fej\H{o}s; {Charged and neutral fixed points in the
  $O(N)\oplus O(N)$-model with Abelian gauge fields}; \emph{Phys. Rev. D}
  (2019); \textbf{100}(1):016005;
  \urlprefix\url{http://dx.doi.org/10.1103/PhysRevD.100.016005};
  \eprint{1903.05331}.

\bibitem{Gies:2001nw}
H.~Gies and C.~Wetterich; {Renormalization flow of bound states}; \emph{Phys.
  Rev. D} (2002); \textbf{65}:065001;
  \urlprefix\url{http://dx.doi.org/10.1103/PhysRevD.65.065001};
  \eprint{hep-th/0107221}.

\bibitem{Gies:2002kd}
H.~Gies and C.~Wetterich; {Renormalization flow from UV to IR degrees of
  freedom}; \emph{Acta Phys. Slov.} (2002); \textbf{52}(4):215--220;
  \eprint{hep-ph/0205226}.

\bibitem{Floerchinger:2009uf}
S.~Floerchinger and C.~Wetterich; {Exact flow equation for composite
  operators}; \emph{Phys. Lett. B} (2009); \textbf{680}:371--376;
  \urlprefix\url{http://dx.doi.org/10.1016/j.physletb.2009.09.014};
  \eprint{0905.0915}.

\bibitem{Baldazzi:2021ydj}
A.~Baldazzi, R.~B.~A. Zinati and K.~Falls; {Essential renormalisation group};
  \emph{SciPost Phys.} (2022); \textbf{13}(4):085;
  \urlprefix\url{http://dx.doi.org/10.21468/SciPostPhys.13.4.085};
  \eprint{2105.11482}.

\bibitem{Ihssen:2023nqd}
F.~Ihssen and J.~M. Pawlowski; {Flowing fields and optimal RG-flows} (2023);
  \eprint{2305.00816}.

\bibitem{Gies:03}
H.~Gies, J.~Jaeckel and C.~Wetterich; {Towards a renormalizable standard model
  without fundamental {H}iggs scalar}; \emph{Phys. Rev. D} (2004);
  \textbf{69}:105008;
  \urlprefix\url{http://dx.doi.org/10.1103/PhysRevD.69.105008};
  \eprint{hep-ph/0312034}.

\bibitem{Zinn-Justin:2002ecy}
J.~Zinn-Justin; {Quantum field theory and critical phenomena}; \emph{Int. Ser.
  Monogr. Phys.} (2002); \textbf{113}:1--1054.

\bibitem{Pendleton:1980as}
B.~Pendleton and G.~G. Ross; {Mass and Mixing Angle Predictions from Infrared
  Fixed Points}; \emph{Phys. Lett. B} (1981); \textbf{98}:291--294;
  \urlprefix\url{http://dx.doi.org/10.1016/0370-2693(81)90017-4}.

\bibitem{Chang:1974}
N.-P. Chang; Eigenvalue conditions and asymptotic freedom for Higgs-scalar
  gauge theories; \emph{Phys. Rev. D} (1974); \textbf{10}:2706--2709;
  \urlprefix\url{http://dx.doi.org/10.1103/PhysRevD.10.2706}.

\bibitem{Callaway:1988ya}
D.~J.~E. Callaway; {Triviality Pursuit: Can Elementary Scalar Particles
  Exist?}; \emph{Phys. Rept.} (1988); \textbf{167}:241;
  \urlprefix\url{http://dx.doi.org/10.1016/0370-1573(88)90008-7}.

\bibitem{Zimmermann:1984sx}
W.~Zimmermann; {Reduction in the Number of Coupling Parameters}; \emph{Commun.
  Math. Phys.} (1985); \textbf{97}:211;
  \urlprefix\url{http://dx.doi.org/10.1007/BF01206187}.

\bibitem{Heinemeyer:2014vxa}
S.~Heinemeyer, J.~Kubo, M.~Mondragon, O.~Piguet, K.~Sibold, W.~Zimmermann and
  G.~Zoupanos; {Reduction of couplings and its application in particle physics,
  Finite theories, Higgs and top mass predictions}; \emph{PoS} (2014);
  \textbf{Higgs and top}:1--339; \eprint{1411.7155}.

\bibitem{Bond:2016dvk}
A.~D. Bond and D.~F. Litim; {Theorems for Asymptotic Safety of Gauge Theories};
  \emph{Eur. Phys. J. C} (2017); \textbf{77}(6):429;
  \urlprefix\url{http://dx.doi.org/10.1140/epjc/s10052-017-4976-5}; [Erratum:
  Eur.Phys.J.C 77, 525 (2017)]; \eprint{1608.00519}.

\bibitem{Giudice:2014tma}
G.~F. Giudice, G.~Isidori, A.~Salvio and A.~Strumia; {Softened Gravity and the
  Extension of the Standard Model up to Infinite Energy}; \emph{JHEP} (2015);
  \textbf{02}:137; \urlprefix\url{http://dx.doi.org/10.1007/JHEP02(2015)137};
  \eprint{1412.2769}.

\bibitem{Gies:2015lia}
H.~Gies and L.~Zambelli; {Asymptotically free scaling solutions in non-Abelian
  Higgs models}; \emph{Phys. Rev. D} (2015); \textbf{92}(2):025016;
  \urlprefix\url{http://dx.doi.org/10.1103/PhysRevD.92.025016};
  \eprint{1502.05907}.

\bibitem{Gies:2016kkk}
H.~Gies and L.~Zambelli; {Non-Abelian Higgs models: Paving the way for
  asymptotic freedom}; \emph{Phys. Rev. D} (2017); \textbf{96}(2):025003;
  \urlprefix\url{http://dx.doi.org/10.1103/PhysRevD.96.025003};
  \eprint{1611.09147}.

\bibitem{Cheng:1973nv}
T.~P. Cheng, E.~Eichten and L.-F. Li; {Higgs Phenomena in Asymptotically Free
  Gauge Theories}; \emph{Phys. Rev. D} (1974); \textbf{9}:2259;
  \urlprefix\url{http://dx.doi.org/10.1103/PhysRevD.9.2259}.

\bibitem{Alessandro1}
H.~Gies, R.~Sondenheimer, A.~Ugolotti and L.~Zambelli; Asymptotic freedom in
  $\mathbb {Z}_2$-{Y}ukawa-{QCD} models; \emph{The European Physical Journal C}
  (2019); \textbf{79}(2);
  \urlprefix\url{http://dx.doi.org/10.1140/epjc/s10052-019-6604-z}.

\bibitem{Alessandro2}
H.~Gies, R.~Sondenheimer, A.~Ugolotti and L.~Zambelli; Scheme {D}ependence of
  {A}symptotically {F}ree {S}olutions; \emph{The European Physical Journal C}
  (2019); \textbf{79}(6);
  \urlprefix\url{http://dx.doi.org/10.1140/epjc/s10052-019-6956-4}.

\end{thebibliography}
\bibliographystyle{dimtest}
\end{document}